\documentclass[manuscript,screen]{acmart}

\makeatletter
\def\@ACM@checkaffil{%
    \if@ACM@instpresent\else
    \ClassWarningNoLine{\@classname}{No institution present for an affiliation}%
    \fi
    \if@ACM@citypresent\else
    \ClassWarningNoLine{\@classname}{No city present for an affiliation}%
    \fi
    \if@ACM@countrypresent\else
        \ClassWarningNoLine{\@classname}{No country present for an affiliation}%
    \fi
}
\makeatother

\setcopyright{acmlicensed}
\copyrightyear{2018}
\acmYear{2018}
\acmDOI{XXXXXXX.XXXXXXX}

\usepackage{amsmath,amsfonts}
\usepackage{pifont}%
\usepackage{algorithmic}
\usepackage{graphicx}
\usepackage{tabularx}
\usepackage{textcomp}
\usepackage[dvipsnames]{xcolor}
\usepackage{hhline}
\usepackage{xltabular}
\usepackage{listings}

\usepackage{hyperref}
\def\BibTeX{{\rm B\kern-.05em{\sc i\kern-.025em b}\kern-.08em
    T\kern-.1667em\lower.7ex\hbox{E}\kern-.125emX}}

\usepackage[linesnumbered,ruled,vlined]{algorithm2e}
\usepackage{hyperref}
\usepackage{tabulary}

\usepackage{xurl} %

\usepackage{float}

\usepackage{graphics}
\graphicspath{ {./figs/} }

\usepackage{caption}
\usepackage{subcaption}
\usepackage[inline]{enumitem}

\usepackage{amsthm}
\newtheorem{definition}{Definition}

\usepackage{booktabs}

\usepackage{makecell}
\usepackage{multirow}

\newcommand{\cmark}{\ding{51}}%
\newcommand{\xmark}{\ding{55}}%

\usepackage{mathtools,tikz,caption}
\DeclareRobustCommand\sampleline[1]{%
    \tikz\draw[#1] (0,0) (0,\the\dimexpr\fontdimen22\textfont2\relax)
    -- (2em,\the\dimexpr\fontdimen22\textfont2\relax);%
}

\usepackage[most]{tcolorbox}

\usepackage{soul}

\definecolor{color-up-to-date}{HTML}{8CC5E3}
\definecolor{color-tood}{HTML}{2066A8}
\definecolor{color-pfet-tood}{HTML}{F72B8F}

\newif\ifrevmarks
\revmarksfalse
\definecolor{revcol}{RGB}{0,128,0}
\ifrevmarks
  \DeclareRobustCommand{\rev}[1]{{\color{revcol}#1}}
  \newcommand{\revcolor}{\color{revcol}}
\else
  \DeclareRobustCommand{\rev}[1]{#1}
  \newcommand{\revcolor}{}
\fi
\newenvironment{revblock}{\par\revcolor}{\par}

\usepackage{lipsum}                     %
\usepackage{xargs}                      %
\usepackage[colorinlistoftodos,prependcaption,textsize=scriptsize]{todonotes}
\newcommandx{\unsure}[2][1=]{\todo[linecolor=red,backgroundcolor=red!25,bordercolor=red,#1]{#2}}
\newcommandx{\change}[2][1=]{\todo[linecolor=blue,backgroundcolor=blue!25,bordercolor=blue,#1]{#2}}
\newcommandx{\info}[2][1=]{\todo[linecolor=OliveGreen,backgroundcolor=OliveGreen!25,bordercolor=OliveGreen,#1]{#2}}
\newcommandx{\improvement}[2][1=]{\todo[linecolor=Plum,backgroundcolor=Plum!25,bordercolor=Plum,#1]{#2}}
\newcommandx{\thiswillnotshow}[2][1=]{\todo[disable,#1]{#2}}

\long\def\longcaption#1#2{\centering\begin{minipage}{#1}\vspace{-0.7\baselineskip}\footnotesize\noindent\emph{#2}\end{minipage}\vspace{0.2\baselineskip}}
\long\def\longcaptionfig#1#2{\centering\begin{minipage}
{#1}\vspace{-0.1\baselineskip}\footnotesize\noindent\emph{#2}\end{minipage}}

\SetCommentSty{mycommfont}
\newcommand{\highlight}[2]{%
    \vspace{.05\baselineskip}
    \colorlet{currentcolor}{.}%
    {\color{#1}%
    \noindent\fbox{\parbox{0.985\linewidth}{\color{currentcolor}#2}}}%
    \vspace{.05\baselineskip}
}
\newcommand{\sensapi}{security-sensitive API\xspace}
\newcommand{\sensapis}{security-sensitive APIs\xspace}
\newcommand{\Sensapi}{Security-sensitive API\xspace} %
\newcommand{\Sensapis}{Security-sensitive APIs\xspace} %
\newcommand{\SenSapi}{Security-Sensitive API\xspace}
\newcommand{\SenSapis}{Security-Sensitive APIs\xspace}
\newcommand{\goalstatement}{\textit{The goal of this study is to aid developers in understanding sensitive resources accessible to their dependencies by identifying \sensapis in packages through an empirical study of Java packages.}\xspace}

\SetKwInput{KwInput}{Input}                %
\SetKwInput{KwOutput}{Output}              %
\SetKwInput{KwNotation}{Notation}

\newcounter{rqcounter}
\newcommand{\newrq}[2]
{\vspace{0.5\baselineskip}
\noindent\refstepcounter{rqcounter}
\fbox{\parbox{0.97\columnwidth}{\textbf{RQ\arabic{rqcounter}:} {\em #2}\label{#1}}}
\vspace{0.5\baselineskip}
}
\newcommand{\rqref}[1]{\textbf{RQ\ref{#1}}}

\usepackage{adjustbox}
\usepackage{array}

\newcolumntype{R}[2]{%
    >{\adjustbox{angle=#1,lap=\width-(#2)}\bgroup}%
    l%
    <{\egroup}%
}
\newcommand*\rot{\multicolumn{1}{R{45}{1em}}}%

\newcommand\pkgname[1]{\textsf{\small #1}}
\newcommand\api[1]{\textit{#1}}
\newcommand\apicat[1]{\textsc{\footnotesize #1}}
\newcommand\qt[1]{\textit{``#1''}}
\newcommand\user[1]{$\langle #1\rangle$}

\begin{document}

\title{What's in a Package? Getting Visibility Into Package Security-Sensitive Resource Access
}

\author{Imranur~Rahman,$^{*\dagger}$ Ranindya Paramitha,$^{*\dagger}$  Henrik~Plate,$^\ddagger$ Dominik~Wermke,$^{*}$ Laurie~Williams,$^{*}$}

\def \authors{Imranur Rahman, Ranindya Paramitha, Henrik Plate, Dominik Wermke, Laurie Williams}

\affiliation{%
    \institution{ $^*$North Carolina State University, Raleigh, NC, USA}
}

\affiliation{%
    \institution{ $^\ddagger$Endor Labs, USA}
}

\begin{abstract}
The Log4Shell vulnerability occurred because the JNDI (Java Naming and Directory Interface) lookup feature was used, even though it was not needed for log4j's core functionality.
\rev{This case illustrates why developers may want visibility into the sensitive capabilities available through a dependency. The presence of a call alone does not establish that the call is unnecessary or risky.}
\Sensapis provide access to security-sensitive resources, such as filesystem, network, or process-related resources.
\rev{If developers know which types of sensitive resources a dependency can access, they can control their dependencies at a finer grain, and they can consider an alternative dependency when a dependency makes unnecessary or suspicious \sensapi calls.}
\rev{However, prior work has not provided a documented, repeatable procedure for deriving an ecosystem-specific \sensapi list, so developers lack systematic visibility into the sensitive capabilities of their dependencies.}
\goalstatement
\rev{In this study, we present a novel and repeatable methodology to construct a \sensapi list for an ecosystem, and we apply it to Java.
We then compare the \sensapi calls of functionally similar Java packages.
The number of \sensapi calls (call level) ranges from 0 to 368 in one capability and from 0 to 429 in total among packages with similar functionality.
These packages also differ in \emph{capability breadth}, the number of the 15 capabilities that a package reaches.
For example, \emph{capability breadth} ranges from 0 to 10 in the ``Dependency Injection'' category and from 0 to 13 in the ``Web Frameworks'' category.
A higher number of \sensapi calls does not by itself mean a higher security risk; the calls show which sensitive resources a dependency can reach.
In a survey of 110 developers (response rate 1.4\%), 44.6\% of the respondents said that they would consider \sensapi information when they select a dependency.
Based on our findings, we recommend adding \sensapi information to dependency management tools (e.g., npm cli, OpenSSF Scorecard), so that developers can better control their projects.}
\end{abstract}

\maketitle
{\renewcommand{\thefootnote}{}\footnotetext{$^\dagger$Imranur Rahman and Ranindya Paramitha contributed equally to this paper.}}

\section{Introduction}

\begin{table}
\revcolor
\centering
\caption{Security-sensitive APIs called by 5 Java dependency injection packages, at the API level.}
\longcaption{0.9\linewidth}{Each value is the number of unique \sensapis of a capability that the package calls in its inter-package call graph (API level, Section~\ref{sec:terminology}). The number in parentheses is the number of APIs of that capability in our list of 227 \sensapis. Capability breadth (capability level) is the number of the 15 capabilities that the package reaches. F: \apicat{\scriptsize Filesystem}, N: \apicat{\scriptsize Network}, P: \apicat{\scriptsize Process}.}
    \begin{tabular}{ l  rrrrrrrrrrrrrrrrr }
    Alternates & 
    \rot{F/Modification (7)} & 
    \rot{F/Read Network Env (27)} & 
    \rot{F/Input (12)} & 
    \rot{F/Output (20)} & 
    \rot{F/Miscellaneous (11)} & 
    \rot{F/Read Env (8)} & 
    \rot{N/Socket (18)} & 
    \rot{N/Naming Directory (11)} & 
    \rot{N/HTTP (16)} & 
    \rot{N/Connection (14)} & 
    \rot{P/Reflection (39)} & 
    \rot{P/Codec Crypto (23)} & 
    \rot{P/Scripting (12)} &
    \rot{P/Operating System (5)} & 
    \rot{P/Dependency (4)} &
    \rot{Total (227)} &
    \rot{\makecell[l]{Capability\\breadth (15)}}
    \\
    \midrule
    dagger & 0 & 0 & 0 & 0 & 0 & 0 & 0 & 0 & 0 & 0 & 0 & 0 & 0 & 0 & 0 & 0 & 0 \\
    guice & 0 & 0 & 1 & 1 & 0 & 3 & 0 & 0 & 0 & 1 & 15 & 1 & 0 & 0 & 0 & 22 & 6 \\
    jakarta.inject-api & 0 & 0 & 0 & 0 & 0 & 0 & 0 & 0 & 0 & 0 & 0 & 0 & 0 & 0 & 0 & 0 & 0 \\
    spring-beans & 0 & 0 & 2 & 1 & 4 & 5 & 0 & 0 & 0 & 3 & 23 & 2 & 0 & 0 & 0 & 40 & 7 \\
    spring-context & 1 & 0 & 2 & 4 & 3 & 5 & 0 & 4 & 0 & 2 & 25 & 2 & 5 & 0 & 0 & 53 & 10 \\
    \end{tabular}

\label{tab:dependency-injection}
\end{table}

The Log4Shell (CVE-2021-44228) vulnerability impacted most networked organization at the time of its disclosure~\cite{cisa-log4j}.
The vulnerable versions of log4j included a JNDI (Java Naming and Directory Interface) lookup feature that allowed the remote loading of an object from an attacker-controlled website.
This JNDI lookup was added to log4j as a feature request in 2013~\cite{eversonLog4shellRedefiningWeb2022}.
Most developers were unaware that log4j included a JNDI lookup feature that was enabled by default.
One mitigation for the Log4Shell vulnerability was to disable the feature by default~\cite{cisa-log4j-mitigation}.
This mitigation approach indicates that the JNDI lookup feature was not core log4j functionality and could be avoided if developers who had log4j as a dependency in their projects were aware that such a feature existed and was enabled by default.

The JNDI lookup can be categorized as a \sensapi, an API for accessing network resources.
\Sensapis provide access to security-sensitive resources, such as filesystem, network, or process-related resources~\cite{yuMaltrackerFineGrainedNPM2024}.
Developers would benefit from knowing which sensitive resources their dependencies can access during dependency selection.
By understanding the sensitive resources accessible to a dependency, developers can apply fine-grained permissions for each dependency to manage the software's attack surface~\cite{amusuoZTD$_JAVA$MitigatingSoftware2025}.
However, developers cannot manually review their projects' dependencies line by line due to limited time and the high number of dependencies present typically in a project (e.g., the mean dependency count for a Python project is 44~\cite{bifolco2024accuracyGitHubDependencyGraph}).

Prior work found that when selecting among alternative third‑party packages, developers consider a range of criteria beyond basic functionality, including security, performance, documentation quality, and maintainability~\cite{lariosvargasSelectingThirdpartyLibraries2020}.
Having a systematic method to quantify \sensapi calls made by a dependency, developers can make informed dependency selection for their expected functionality.
\rev{As shown in Table~\ref{tab:dependency-injection} (a preview of our analysis result), a developer choosing a ``dependency injection'' package may notice that \pkgname{dagger} calls no sensitive API, whereas \pkgname{spring-context} calls 53 different \sensapis and reaches 10 of the 15 capabilities, although they provide similar functionality.
Our intent is not to claim that every \sensapi call is dangerous or that a higher number of calls means a higher security risk.
The calls show which sensitive resources a dependency can reach, and developers can compare this information with the functionality that they expect from the dependency.}

\rev{However, to get this information, a tool first needs a list of the \sensapis of the ecosystem.
The current state of the art~\cite{hermann2015getting} relies on expert selection to identify \sensapis within an ecosystem, and others cannot repeat this selection.
We lack a repeatable methodology for identifying \sensapis in an ecosystem, so no tool gives developers \sensapi information, and developers cannot use it to evaluate packages when they select a dependency.}
\goalstatement

We conducted a mixed-method empirical study on the Java ecosystem to answer the following four research questions to reach our goal:

\newrq{rq:sec-sen-api-and-roles}{How can \sensapis be identified in an ecosystem?}

To identify \sensapis in an ecosystem, we provide a novel, repeatable methodology that uses the ecosystem's documentation, CVE fixes, and associated CWEs and then apply this methodology to construct a list of 227 Java \sensapis to answer \rqref{rq:sec-sen-api-and-roles}.
We then evaluate the quality of our \sensapi list through two approaches: comparison against a Quasi Golden Standard (QGS) API list, and verification of the APIs' presence in known vulnerable functions.

\newrq{rq:sen-api-usage-pattern}{How frequently do Java open-source packages call \sensapis?}

To understand the prevalence of \sensapi calls made by Java open-source packages, we first select a list of \emph{45 core packages} based on our package selection criteria, comprising \emph{4,183 package versions}.
These 45 core packages are the representatives of the Java ecosystem.
These \emph{45 core packages} have \emph{1,210 unique packages} in their dependency tree, totaling \emph{30,772 package versions}.
We then quantitatively analyzed the prevalence of \sensapi usage within the call graphs of these 45 core packages in \rqref{rq:sen-api-usage-pattern}.

\newrq{rq:sen-api-diff}{How do functionally similar packages differ in terms of their \sensapi calls and why?}

We divided the 45 core packages into eight categories after applying package inclusion and exclusion criteria.
\rev{In \rqref{rq:sen-api-diff}, we compare packages within seven of these categories; the packages in the eighth category, JDBC Driver, are not interchangeable, so we report it only as an exploratory case (Appendix~\ref{sec:jdbc-exploratory}).}
We then quantitatively compared packages within each category \rev{at three levels: the number of \sensapi calls (call level), the number of unique \sensapis (API level), and capability breadth (capability level)} to assess their similarity or dissimilarity in \rqref{rq:sen-api-diff}.
Additionally, we qualitatively analyzed the root causes of these similarities and dissimilarities in \sensapi calls by manually reviewing packages' source code, documentation, design patterns, dependency usage, implementation choices, and usage examples.
We found that packages with similar functionality often differ in the number of \sensapi calls they make \rev{(call level: from 0 to 368 in one capability and from 0 to 429 in total) and in capability breadth (capability level: for example, from 0 to 10 in ``Dependency Injection'')}.

\newrq{rq:usefulness-of-sen-api}{To what extent would prior knowledge of \sensapi information influence developers' dependency selection decisions?}

Finally, we conducted a survey with developers to understand their perceptions of using the \sensapi calls of functionally similar packages in their dependency selection process to answer (\rqref{rq:usefulness-of-sen-api}).
\rev{Of the 110 respondents (response rate 1.4\%), 44.6\% said that they would consider \sensapi information in their dependency selection process.}

In summary, our contributions in this paper are the following:
\begin{enumerate}
    \item an empirical study on the prevalence of \sensapi in the Java ecosystem, which shows the impact of \sensapis in developers' dependency selection through a quantitative and qualitative analysis\rev{, and an error analysis of its call graph results with a second open-source call graph generator};
    \item a repeatable methodology for constructing the \sensapi list in a software ecosystem and a case study for using the methodology in Java to construct a list of 227 \sensapis in Java;
    \item actionable recommendations towards an informed dependency selection for developers and supply chain security researchers.
    \item our data and replication package at Zenodo~\cite{zenodo-us}.
    Our replication package also contains manually verified vulnerable functions associated with 255 CVEs, which can be used to foster research in other directions (e.g., matching vulnerable functions to CVE fixes~\cite{dunlapPairingSecurityAdvisories2024}).
\end{enumerate}

We organized the rest of the paper as follows. We explain definitions and terminology in Section~\ref{sec:terminology}, followed by the research
methodology in Section~\ref{sec:method}.
Section~\ref{method:val:qgs} discusses how we build a Quasi Gold Standard (QGS) to evaluate our \sensapi list.
Section~\ref{sec:sensapi-list-construction} presents our novel and repeatable methodology to construct \sensapi list, its application to Java ecosystem and its quality evaluation process (\rqref{rq:sec-sen-api-and-roles}).
Section~\ref{sec:method} presents the analysis methodology of our research.
Section~\ref{sec:sen-api-usage-pattern} presents the prevalence of \sensapi usage in Java packages (\rqref{rq:sen-api-usage-pattern}).
Section~\ref{sec:sen-api-diff} presents the quantitative and qualitative comparison of functionally similar Java packages with respect to \sensapi calls (\rqref{rq:sen-api-diff}).
\rev{Section~\ref{sec:cg-validation} validates the call graph results of \rqref{rq:sen-api-usage-pattern} and \rqref{rq:sen-api-diff} with a second open-source call graph generator.}
Section~\ref{sec:dev-survey} presents our developers survey result (\rqref{rq:usefulness-of-sen-api}).
Detailed discussion and recommendations are presented in Section~\ref{sec:discuss} and ethical considerations in conducting this research in Section~\ref{sec:ethics}. 
Section~\ref{sec:threats}
covers threats to validity and Section~\ref{sec:related_works} covers related works.
Section~\ref{sec:concl} concludes this research paper.

\section{Terminology}
\label{sec:terminology}
In this section, we formally define a \sensapi, a call graph (with and without dependencies), \rev{the levels of granularity at which we report \sensapi calls, }and the Quasi Golden Standard (QGS), which are used throughout the study.

\begin{definition}[\textbf{Security-Sensitive API}]
    We define a security-sensitive API as an API that can be used to access security-sensitive resources: the filesystem, network, or system processes. 
\end{definition}

\api{java.io.File.createTempFile()}, \api{java.net.Socket.connect()}, \api{javax.script.ScriptEngine.eval()}, and \api{java.lang.Runtime.load()} are examples of \sensapis.

We adopted the categorization by Ferreira et al.~\cite{ferreiraContainingMaliciousPackage2021}:  \apicat{Filesystem}, \apicat{Network}, and \apicat{Process} as sensitive resource types.

\begin{description}[style=nextline]

\item[\apicat{Filesystem}:] 
related to interacting with the local filesystem, including opening, reading, modifying, creating, and writing from/ to files.
It also includes the APIs for setting, removing, and updating file permissions.

\item[\apicat{Network}:]
interact with network resources, including HTTP servlet/client, URL connection, socket, and naming directory services, which are required for communicating with remote servers.

\item[\apicat{Process}:]
expose operating-system process and subprocess-handling routines, including program loading and script execution.
\end{description}

\begin{definition}[\textbf{Intra-Package and Inter-Package Call Graph}]
    An intra-package call graph is created by analyzing the code of a single package and represents the \texttt{\small <caller, callee>} relationships between methods of types defined in the package.

\end{definition}

An inter-package call graph is created by stitching (using type hierarchy) the intra-package call graphs of a given package and its dependencies, meaning that callee types specified in the intra-package call graphs are resolved to all possible implementation types.
As a result, an inter-package call graph represents all \texttt{\small <caller, callee>} relationships between methods in a package and its dependencies that can be connected to a given package.
    
For example, \pkgname{log4j-core}'s method \api{ConfigurationSource.fromUri()} calls the static method \api{FileUtils.fileFromUri()} and instantiates \api{java.io.FileInputStream}, which is represented by corresponding call graph edges in the intra- and inter-package call graphs of \pkgname{log4j-core}. 

\begin{revblock}
\begin{definition}[\textbf{Levels of Granularity}]
    We report the \sensapi calls of a package at four levels of granularity, from coarse to fine: the resource level, the capability level, the API level, and the call level (Table~\ref{tab:levels}).
\end{definition}

\begin{table}[tbp]
\revcolor
\centering
\small
\caption{Levels of granularity at which we report the \sensapi calls of a package.}
\label{tab:levels}
\begin{tabular}{l p{0.42\linewidth} p{0.3\linewidth}}
\toprule
Level & What the level shows for a package & Measure \\
\midrule
Resource level & Which of the 3 sensitive resources (\apicat{Filesystem}, \apicat{Network}, \apicat{Process}) the package reaches & Number of resources reached (0--3) \\
Capability level & Which of the 15 capabilities the package reaches & Capability breadth (0--15) \\
API level & Which of the 227 \sensapis the package calls & Number of unique \sensapis called \\
Call level & How many call-graph edges reach those \sensapis & Number of \sensapi calls \\
\bottomrule
\end{tabular}
\end{table}

We report the API level and the call level either in total or per capability.
A finer level counts repeated use, and a coarser level does not.
Capability breadth is less sensitive to repeated calls: calling APIs within an already reached capability does not increase breadth. It still depends on the API list, the API categorization, and the call-graph analysis, and it does not measure security risk.
For example, \pkgname{guice} calls \api{java.util.logging.Logger.log()} 35 times: this counts as 35 at the call level, but as one API at the API level and as one capability (\apicat{Filesystem/Output}) at the capability level.
We use these labels throughout the paper.
\rqref{rq:sen-api-usage-pattern} reports the call level and the capability level (Section~\ref{sec:sen-api-usage-pattern}).
Table~\ref{tab:dependency-injection} and the heatmaps of \rqref{rq:sen-api-diff} report the API level per capability, together with the capability level (Section~\ref{sec:sen-api-diff}).
The call graph validation reports the API level and the capability level (Section~\ref{sec:cg-validation}).
\end{revblock}

\begin{definition}[\textbf{Quasi Gold Standard (QGS)}]
    A Quasi Gold Standard (QGS) is a list of \sensapis in an ecosystem from known venues and journals for a given time frame.
\end{definition}

To evaluate the coverage of our \sensapi list, we adopted (with some modifications) Quasi Gold Standard (QGS), which is a commonly used approach in evaluating the completeness of a systematic literature review as suggested by Zhang et al.~\cite{zhang2011empirical, zhang2011identifying}, Tran et al.~\cite{tran2024good}, and Kitchenham and Brereton~\cite{kitchenham2013systematic}.
According to Zhang et al.~\cite{zhang2011identifying}, a QGS is a list of work from known venues and journals for a given time frame.
\rev{Zhang et al. evaluate a search string by how much of the QGS it retrieves.
We use the QGS in the same way, to measure coverage only.}
This approach allows us to evaluate our search string and, if necessary, refine it until we achieve adequate coverage. 
In our study, instead of a set of papers, the QGS contains \sensapis from the studied ecosystem, extracted from known articles within a given time frame, to evaluate our identification process.

\section{Research Methodology:  Package Selection}
\label{sec:package-selection}

\begin{figure}
    \centering
    \includegraphics[width=0.45\linewidth]{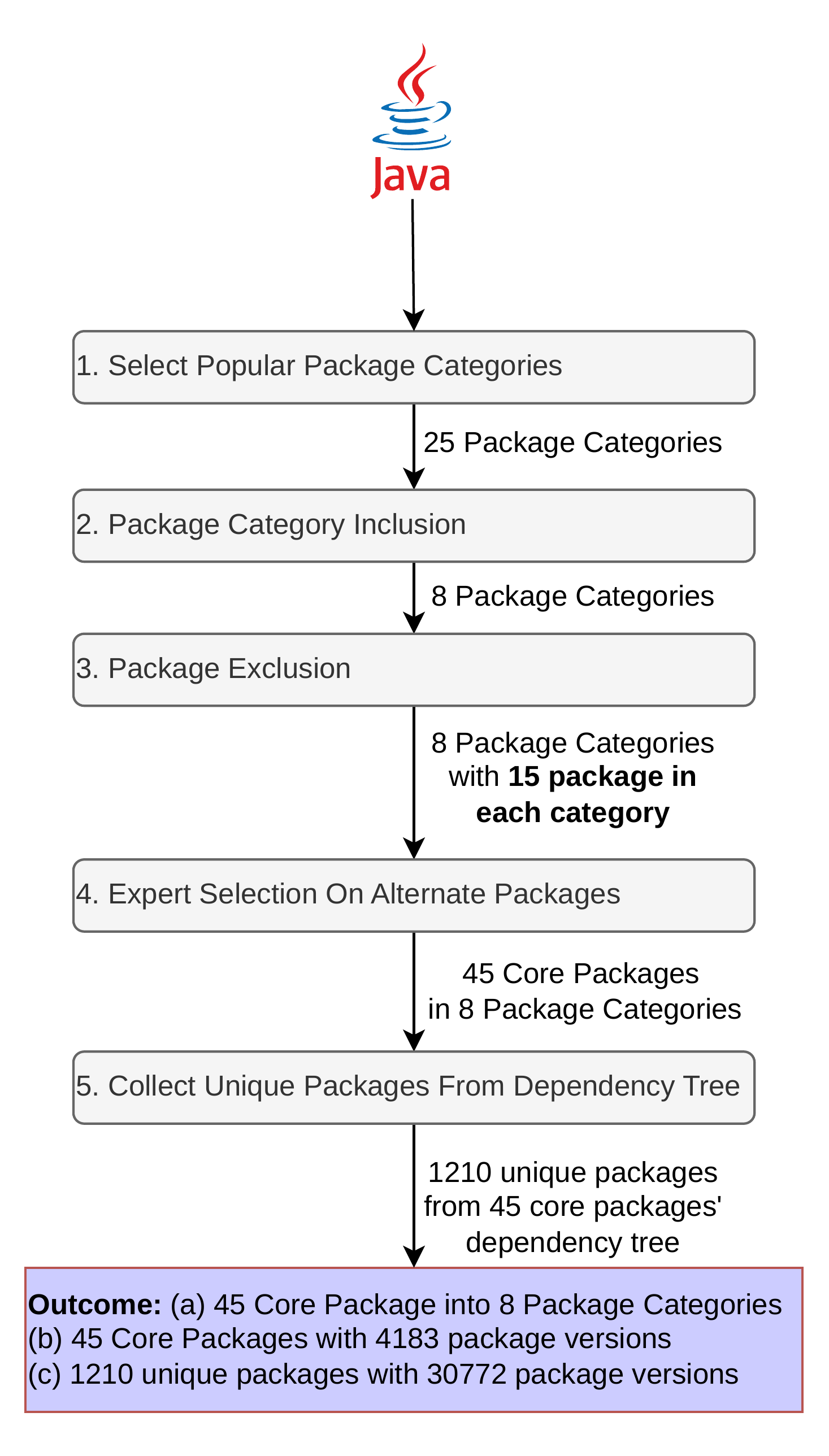}
    \caption{Our package selection process.}
    \label{fig:senapi-highlevel}
\end{figure}

In this section, we present the methodology for selecting the core packages and their corresponding package categories for our empirical analysis.
Developers often have multiple functionally similar packages to choose from when building their project.
Our goal in this research is to show developers the \sensapi calls made by the functionally similar packages.
So we need more than one functionally similar packages grouped together to understand the similarities and differences in \sensapi usage.
Our package selection process consists of five steps, as shown in Figure \ref{fig:senapi-highlevel}. 

\begin{description}
    \item[1. Select Popular Maven Package Categories.] MvnRepository~\cite{maven} provides a list of widely used open-source package categories, each representing functionally similar packages.
    Examples of package categories are HTTP Clients, JSON Libraries, and Logging Frameworks.
    We select the top 25 most-used package categories from MvnRepository to reflect what developers use in practice.
    To determine the main functionality of each package category, we review the descriptions of the top 15 packages in each category ($15\times25=375$ total packages) and how to use these packages as dependencies from GitHub.
    \item[2. Package Category Inclusion.] We then apply the following \textit{inclusion criteria} on the Maven package categories: we only use package categories with packages that
    \begin{enumerate}[leftmargin=*]
        \item \textbf{are used at runtime} that excludes anything related to building, testing, or mocking. We excluded test or build-time dependencies, which are not present in production, and included only runtime dependencies, since our goal is to assess vulnerability risks via sensitive API usage.
        \item \textbf{provide alternative functionality for a given task.} We wanted to compare similar-functioning packages with their respective sensitive API usage. If there is only one package that provides a unique functionality, we cannot compare it with any other package by design. For example, the category ``Java Specifications'' does not meet this criterion because its task range is too broad.
        \item \textbf{are used in ``typical'' server-side Java Web applications}, which makes it possible to create call graphs using a variety of open-source and proprietary tools, and excludes other JVM languages. With this criterion, we retain the functionalities needed in Java web backends (e.g., frameworks, parsers, database clients) and exclude unrelated areas (e.g., mobile, IoT), thereby keeping our analysis focused and relevant.
    \end{enumerate}

\begin{table}[tbp]
    \centering
    \caption{Chosen Core Packages}
    \rev{\longcaption{0.9\linewidth}{$^\dagger$ Exploratory case only (Appendix~\ref{sec:jdbc-exploratory}); the packages are not interchangeable, so \rqref{rq:sen-api-diff} does not compare them.}}
    \begin{tabular}{lp{0.7\linewidth}}
    \toprule
        Maven Categories & Package Names\\
        \midrule
        Dependency Injection & \textit{dagger, guice, jakarta.enterprise.cdi-api, jakarta.inject-api, javax.inject, spring-beans, spring-context} \\[13pt]
        HTTP Client & \textit{httpasyncclient, httpclient5, jetty-client, okhttp, retrofit} \\[3pt]
        I/O Utilities & \textit{commons-io, jetty-io, okio, plexus-io} \\[3pt]
        \rev{JDBC Driver$^\dagger$} & \textit{derby, derbyclient, mariadb-java-client, mssql-jdbc, mysql-connector-j, postgresql, sqlite-jdbc} \\[13pt]
        JSON Parsers & \textit{fastjson2, gson, jackson-core, json} \\[3pt]
        Logging & \textit{log4j-core, logback-classic, jboss-logging, timber} \\[3pt]
        Web Frameworks & \textit{jakarta.faces-api, spring-boot-starter-web, spring-web, spring-webflow, spring-webmvc, struts2-core, tapestry-core, vaadin, wicket-core} \\[13pt]
        XML Parsers & \textit{dom4j, jakarta.xml.bin-api, jaxb-api, xercesImpl, xstream} \\
        \bottomrule
    \end{tabular}
    \label{tab:packages}
\end{table}

    With these inclusion criteria, we identify eight categories, as shown in Table~\ref{tab:packages}.
    The JDBC driver category is an exception to inclusion criterion 2, as the JDBC driver choice is determined by the developers' underlying database, and packages in this category are not interchangeable.
    \rev{Therefore, we do not use this category in the comparison of alternatives in \rqref{rq:sen-api-diff} or in the comparison across package categories, and we report its heatmap only as an exploratory case in Appendix~\ref{sec:jdbc-exploratory}. We keep the case because the drivers are common runtime dependencies of Java web back ends, and their \sensapi calls are useful information even when a developer cannot choose an alternative.}
    \item[3. Package Inclusion Exclusion.] We then select the top 15 most-used packages in each category, based on total downloads from MvnRepository~\cite{maven}, to mitigate potential selection bias.
    We exclude packages that (1) are used for other JVM languages like Kotlin or Scala; and (2) were last updated prior to 2019.
    The rationale for this exclusion criterion is to include maintained Java packages in our analysis.
    With these inclusion criteria, we chose $8\times15=120$ packages.
    \item[4. Expert Selection for Alternative Packages.] We then use an expert selection process, following negotiator-agreement techniques~\cite{campbellCodingIndepthSemistructured2013}, to agree on a set of packages for each of the eight categories that can serve as alternatives to one another in terms of the core functionality of the package category. The three authors conduct this expert selection process: two authors review the top 15 package descriptions and select the functionally similar packages in each category, and the three authors reach consensus on the final selection.
    To minimize confirmation bias, the expert selection process involves authors with diverse backgrounds (two PhD students and an industry professional), and each expert must justify their choice.
    We used a structured framework to assess the packages against predefined criteria, ensuring consistency and objectivity in the selection process. After this process, we have 45 core packages, as shown in Table~\ref{tab:packages}. From these \emph{45 core packages}, we have \emph{4,183 package versions} after excluding unavailable versions, i.e., missing POM or JAR files in Maven Central. 
    \item[5. Collect Unique Packages from Dependency Trees.] At the end of this package selection process, we have collected the unique packages from these \emph{package versions' dependency} trees, yielding a total of \emph{1,210 packages}, corresponding to \emph{30,772 package versions}.
    Unique packages are the packages present in the dependency trees (direct and transitive) of the core packages.
    We compared the \emph{4,183 package versions} from \emph{45 core packages} with the \emph{30,772 package versions} from \emph{1,210 packages} with respect to \sensapi calls.
\end{description}

We will use the selected packages in our Analysis Methodology in Section~\ref{sec:method}.
To be specific, we use Outcome (b) and (c) to understand the prevalence of \sensapi usage (\rqref{rq:sen-api-usage-pattern}), Outcome (a) to understand the similarities and dissimilarities in functionally similar packages (\rqref{rq:sen-api-diff}), and Outcome (a) and (b) to understand whether having the \sensapi information would influence developers' dependency selection (\rqref{rq:usefulness-of-sen-api}).

\section{\rqref{rq:sec-sen-api-and-roles}: How Can \Sensapis Be Identified In An Ecosystem?}
\label{sec:sensapi-list-construction}
In this section, we answer \rqref{rq:sec-sen-api-and-roles}.
First, in Section~\ref{sec:rq1-method}, we present the methodology on how to construct the \sensapi list and \sensapi list evaluation.
Then, in Section~\ref{sec:sensapi-list-case-study}, we present the implementation of the method to construct the \sensapi list in Java as a case study and the evaluation result for the Java \sensapi list.

\subsection{Security-Sensitive API List Construction Process For An Ecosystem}
\label{sec:rq1-method}
In this section, we first present the CVE and CWE collection and mapping in Section~\ref{sec:cve-cwe-collection-mapping}, then the three-pronged approach to the \sensapi list construction process in Section~\ref{sec:api-list-construction-general}, and finally the methods to evaluate the constructed \sensapi list in Section~\ref{sec:api-list-eval}.
\subsubsection{CVE and CWE Collection and Mapping}
\label{sec:cve-cwe-collection-mapping}
We conduct a \textit{CVE-to-CWE-to-OWASP Top 10 Mapping} to support the identification of \sensapi entries from CWE examples.
For a given set of packages and their dependencies, CVEs are retrieved from the NVD and, where available, verified against their associated vulnerable functions by a domain expert or automated tooling.
The unique CWEs surfacing from these CVEs are then mapped to the OWASP Top 10~\cite{owasp}.
When a CWE has no direct mapping, its parent CWE is checked for a mapping and used accordingly; if no mapping exists at any level, the entry is classified as ``Others.''
The input of this step is a set of selected packages of an ecosystem (in our case, the 1210 unique packages in Outcome (c) from Figure~\ref{fig:senapi-highlevel}).
The output of this step is the CVEs for these selected packages mapped to CWEs.

\subsubsection{Security-Sensitive API List Construction Process}
\label{sec:api-list-construction-general}

We define a three-pronged approach to construct a \sensapi list for an ecosystem, as illustrated in Figure~\ref{fig:api-list-construction}:
\begin{description}
    \item[1. Ecosystem's API Documentation.] The Ecosystem API documentation provides a comprehensive list of APIs available to users without requiring additional dependencies. We argue that the ecosystem API documentation should be the main source to construct \sensapi list for an ecosystem. Additional inclusion criteria in this step are APIs that interact with system resources, e.g., the filesystem, network, or processes. This inclusion criterion ensures that the API list only contains APIs that access security-sensitive resources.
    Inter-rater agreement should be achieved at this step, and manual validation is required after the initial list is created.
    The input for this step is the API documentation for the ecosystem, and the output is a list of \sensapis for that ecosystem.
    \item[2. CVE Fixes.] 
    CVE fixes contain the added, removed, or modified code to address the vulnerability. We argue that if a certain ecosystem API is added, removed, or modified in the CVE fix, the API must be security-sensitive.
    The input to this step is the fixes for the CVEs from Section~\ref{sec:cve-cwe-collection-mapping}, and the output is the APIs added, removed, or modified in the CVE fixes.
    We include such APIs in our \sensapi list.
    \item[3. CWE Examples.]
    The MITRE CWE~\cite{mitre_cwe} definitions provide code snippets as ``Demonstrative Examples'' for most CWEs in various languages.
    CWE examples can complement the above two steps when no vulnerability exists in a given API, but the API can still be used to exploit a vulnerability.
    The input of this step is the mapped CWEs from Section~\ref{sec:cve-cwe-collection-mapping}, and the output is a list of \sensapi. 
\end{description}

After these three steps, an expert rule-based selection is conducted with multiple individuals to manually validate the constructed \sensapi list.
In this step, experts categorize the APIs into resource types and capability types for the ecosystem.
Categorization is done using card-sorting techniques (open, closed, or hybrid).
Card sorting is a qualitative technique to classify text into themes~\cite{zimmermannCardsortingTextThemes2016} and is commonly used in research to create informative categories~\cite{basakWhatChallengesDevelopers2023,rahmanWhatQuestionsProgrammers2018}.
Ecosystem-specific or language-specific inclusion-exclusion criteria are applied in this step if necessary (e.g., excluding packages written in Kotlin or Scala when creating \sensapi list for Java, although these three are JVM-based languages).
Disagreements between raters are resolved by the negotiated agreement technique~\cite{campbellCodingIndepthSemistructured2013}.
The final outcome is a list of \sensapis for an ecosystem with a categorization.

\subsubsection{\Sensapi List Evaluation}
\label{sec:api-list-eval}
To evaluate the quality of the process to identify our \sensapi list, we performed two approaches: (1) developed a Quasi-Gold Standard and \rev{measured how much of it our list covers (sensitivity)};
and (2) measured the prevalence of the APIs in our list in real-world vulnerable functions.

\paragraph{Methodology to Create Quasi Gold Standard (QGS) for evaluating \sensapi list construction process}
\label{method:val:qgs}

\begin{figure}
    \centering
    \includegraphics[width=\linewidth]{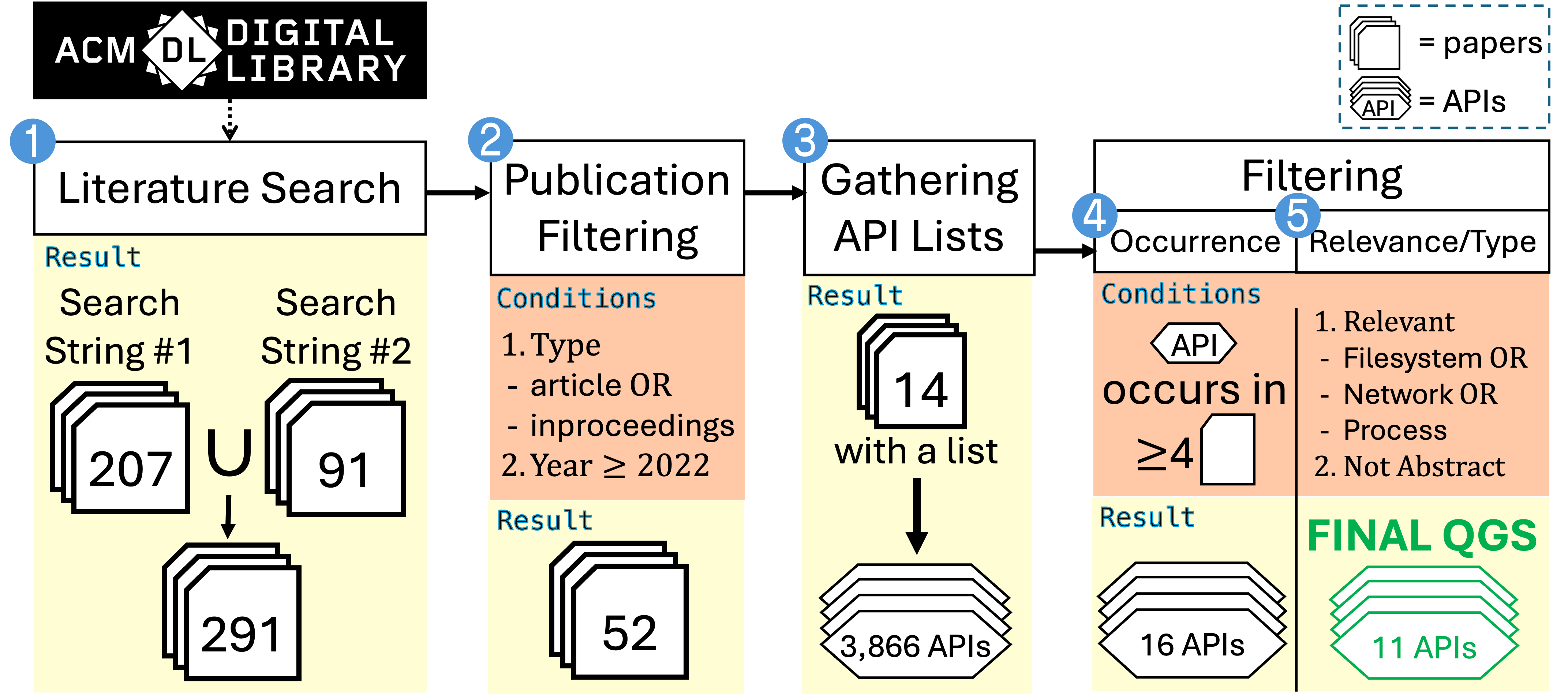}
    \caption{Quasi-Gold Standard Creation}
    \label{fig:qgs-generation}
\end{figure}

To develop the QGS for our \sensapi list to evaluate on, we adapted the QGS construction from Tran et al.~\cite{tran2024good}.
An illustration of our QGS creation is depicted in Figure~\ref{fig:qgs-generation}.
The steps are the following:
\begin{description}
    \item[1. Perform Literature Search on Databases.] We performed a literature search on ACM Digital Library~\cite{acmdigitallib} with the following search strings.

    \vspace{0.5\baselineskip}
    \fbox{\parbox{0.92\columnwidth}{\textbf{Search String \#1:} {\texttt{[["security" AND "sensitive api*"] OR "security api*"] AND [java]}}\label{search1}}}\vspace{0.5\baselineskip}
    
    \fbox{\parbox{0.92\columnwidth}{\textbf{Search String \#2:} {\texttt{[["capability model" OR "capability analysis" OR "capability inference"] AND [api] AND [java]]}}\label{search2}}}\vspace{0.5\baselineskip}
    
    We then merged the results from both search strings and removed duplicate entries.
    We also used the IEEE Xplore Digital Library and Scopus.

        \textbf{\emph{Result:}} Search string \#1 returned 207 papers and search string \#2 returned 91 papers. After merging and removing duplicate entries, we have a list of 291 papers.

    \item[2. Inclusion Exclusion Criteria (Publication Filtering).] We limit the result to publication type \texttt{article} and \texttt{inproceedings} only, as \texttt{proceedings} usually only contains a table of contents and a list of \texttt{inproceedings}. We also limit the results to work from the last four years (2022-2025) to maintain recency and make sure of the relevance of the APIs.
    
        \textbf{\emph{Result:}} After filtering by publication type and year, we are left with 52 papers. 
    
    \item[3. Additional Filtering (Gathering the API List).]  We manually reviewed the published works to see whether they define a list of ``\sensapi'' in Java (different papers may use a different name for the list). For the works that have a defined API list, we went to their paper/artifact and collected the list until we had a list of lists.
    
        \textbf{\emph{Result:}} Among the 52 papers, only 14 papers have a defined sensitive Java API list. The 14 API lists contain, on average, 317 APIs (std dev: 948), ranging from 2 to 3,585 APIs (median: 13). The total number of unique APIs across all lists is 3,866.
    
    \item[4. Additional Filtering (API List Filter based on Occurrence).] For each API in the lists, we counted its occurrences across the lists; i.e., if an API occurs in 5 lists (5 papers defined it as sensitive), we assigned it an occurrence count of 5. We included only APIs that appear in more than 3 papers (i.e., more than 3 papers agreed that this API is sensitive). This filter is also used to limit the number of APIs in our QGS to no more than 100.

        \textbf{\emph{Result:}} From the unique API list, only 16 APIs are considered sensitive in more than 3 papers.
    
    \item[5. Additional Filtering (API List Filter based on Relevance and Type).] We manually went through each of the 16 APIs' descriptions, and we only kept the APIs that (1) are included in our resource types (Section~\ref{sec:terminology}: filesystem, network, and process); and (2) are not abstract methods.

        \textbf{\emph{Result:}} Our filter reduces our QGS to 11 APIs \textbf{(FINAL QGS)}, and the \textbf{FINAL QGS} is present in Table~\ref{tab:qgs}.

\end{description}

\begin{table}[htbp]
    \centering
    \caption{Quasi-Gold Standard of Java \Sensapi List}
    \label{tab:qgs}
    \begin{tabular}{l|rl}
        \toprule
        APIs & Occur in \# lists & Relevant capability \\
        \midrule
        java.net.URL.openConnection() & 9 & \apicat{Network}\\
        java.net.URL.openStream() & 7 & \apicat{Network}\\
        java.net.URL.getContent() & 6 & \apicat{Network}\\
        java.lang.ProcessBuilder.start() & 6 & \apicat{Process}\\
        java.io.FileOutputStream.write() & 6 & \apicat{Filesystem}\\
        java.net.ServerSocket.bind() & 5 & \apicat{Network}\\
        java.net.URLConnection.getInputStream()	& 5 & \apicat{Network}\\
        java.lang.Runtime.exec() & 5 & \apicat{Process}\\
        java.net.URLConnection.getOutputStream() & 4 & \apicat{Network}\\
        java.net.Socket.connect() & 4 & \apicat{Network}\\
        java.io.RandomAccessFile.write() & 4 & \apicat{Filesystem}\\
        \bottomrule
    \end{tabular}
\end{table}

Following these five steps, we obtained a set of 11 APIs as our QGS, shown in Table~\ref{tab:qgs}, which we deemed sufficient based on the ranges reported by Tran et al.~\cite{tran2024good}.

\noindent \textbf{Evaluation Metrics for Our \Sensapi List With Respect To QGS.}
\rev{We use this QGS to evaluate the coverage of our \sensapi list.
Equation~\ref{eq:sensitivity} presents our adapted formula for sensitivity from Zhang et al.~\cite{zhang2011identifying}, and Equation~\ref{eq:overlap} presents the overlap ratio.}

\begin{equation}
    \label{eq:sensitivity}
    \text{Sensitivity} = \frac{\text{The number of APIs in the API list that are also in the QGS}}{\text{Total APIs in the QGS}} \times 100\%
\end{equation}

\begin{equation}
    \label{eq:overlap}
    \text{\rev{Overlap ratio}} = \frac{\text{The number of APIs in the API list that are also in the QGS}}{\text{Total APIs in the API list}} \times 100\%
\end{equation}
\rev{The overlap ratio measures only how much of our list the QGS accounts for.
The QGS is small and non-exhaustive by design, so most APIs in our list cannot appear in it, and the overlap ratio does not bound the false positives of our list.
We therefore use the QGS to support a coverage claim only.
We do not quantify the precision of our list: a sound estimate needs a reference list of Java \sensapis that is independent of ours and complete enough to serve as ground truth, and no such list exists.}

\paragraph{Prevalence in Vulnerable Functions}
\label{method:val:prev_vuln_func}
To further validate our \sensapi list, we analyzed the prevalence of the APIs in the vulnerable functions across our set of 45 core packages, comprising 4,183 package versions.
The rationale for this validation is that a \sensapi should contribute to vulnerabilities, which would cause it to appear in vulnerable functions.
For this validation, we performed a manual qualitative analysis to assess the occurrence of the \sensapis in vulnerable functions associated with the 255 CVEs from Section~\ref{sec:cve-cwe-collection-mapping}.

\subsection{Implementation: Case study of constructing the \sensapi list using the process in Java ecosystem}
\label{sec:sensapi-list-case-study}
In this section, we present the implementation of the methodology presented in Section~\ref{sec:sensapi-list-construction} for Java as a case study.
We present the CVE and CWE collection and mapping in Section~\ref{sec:cve-cwe-collection-mapping-impl}, the implementation of \sensapi list construction process in Section~\ref{sec:sen-api-list-impl}, and the implementation of the evaluation of \sensapi list in Section~\ref{sec:sen-api-list-eval-impl}.

\subsubsection{CVE and CWE Collection and Mapping}
\label{sec:cve-cwe-collection-mapping-impl}
Following the methodology presented in Section~\ref{sec:cve-cwe-collection-mapping}, we conduct a \textit{CVE-to-CWE-to-OWASP Top 10 Mapping} so that it can be used later in our process of identifying \sensapi list from the CWE examples in Section~\ref{sec:sen-api-list-impl}.
Our 45 core packages and their dependencies had 255 CVEs in the NVD as of June 2024, recorded by our industry partner, Endor Labs.
Endor Labs manually verified the vulnerable functions associated with the 255 CVEs, covering 71 unique CWEs.
We then mapped the CWEs of these CVEs to OWASP's Top 10 (2025)~\cite{owasp}. %
If the CWE has no mapping to OWASP's Top 10, we check whether its parent has a mapping and map it accordingly; otherwise, we map it as ``Other''.
The input of this step is a set of selected packages of an ecosystem (in our case, the 1210 unique packages in Outcome (c) from Figure~\ref{fig:senapi-highlevel}).
The output of this step is the CVEs for these selected packages mapped to CWEs.

\subsubsection{\Sensapi List Construction}
\label{sec:sen-api-list-impl}
We apply the methodology present in Section~\ref{sec:api-list-construction-general} to construct the \sensapi list for Java using a three-pronged approach.
For categorizing the \sensapi list, we adopt the categorization by Ferreira et al.~\cite{ferreiraContainingMaliciousPackage2021}:  \apicat{Filesystem}, \apicat{Network}, and \apicat{Process}.

\begin{description}[style=nextline]

\item[\apicat{Filesystem}:] 
related to interacting with the filesystem, including opening, reading, modifying, creating, and writing from/ to files.
It also includes the APIs for file permissions.

\item[\apicat{Network}:]
interact with network resources, including HTTP servlet/client, URL connection, socket, and naming directory services.

\item[\apicat{Process}:]
manage encoding/ encryption, program loading, and script execution.
\end{description}

\begin{figure}[tbp]
    \centering
    \includegraphics[width=0.7\columnwidth]{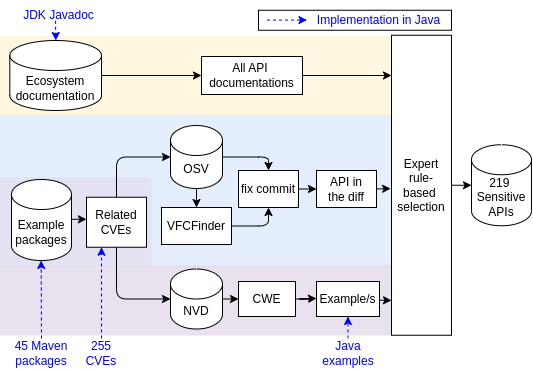}
    \caption{Security-sensitive API List Construction Process}
    \label{fig:api-list-construction}
\end{figure}

To construct a \sensapi list in Java, we use a three-pronged approach, as illustrated in Figure~\ref{fig:api-list-construction}:
\begin{description}
    \item[1. Ecosystem's API documentation (JDK).] \rev{One author went through all the API documentation (Java 11 JDK) and selected a preliminary sensitive API list using two rules: either (1) the API accesses sensitive resources based on the defined resource types: file system, network, and process, or (2) the API can perform the operation that a weakness in a fixed list of CWEs exploits, i.e., CWE-22 (Path Traversal), CWE-78 (OS Command Injection), CWE-94 (Code Injection, which can lead to remote code execution). An API meets rule (2) when its Javadoc describes this operation, for example, an API that opens a file at a given path for CWE-22.} This preliminary list contains 127 \sensapis. Two other authors manually validated this list.
    The input for this step is the JDK Javadoc, and the output is a list of 127 Java \sensapis.
    \item[2. CVE fixes.] 
    One author analyzed the 255 CVE fixes \rev{ associated with our 45 core packages (from Endor Labs)} using VFCFinder~\cite{dunlapVFCFinderPairingSecurity2024} to identify APIs associated with system or information compromise if the vulnerability were exploited.
    For APIs affected by the vulnerability fixes, we check whether the API call is added, removed, or modified in the fix.
    This process resulted in 76 \sensapis, of which 66 are new (not on the preliminary list).
    The input to this step is the CVE fixes from Section~\ref{sec:cve-cwe-collection-mapping-impl}, and the output is the 76 APIs that were added, removed, or modified in those fixes.
    We analyze the CWEs associated with these CVEs in the next section.
    \item[3. CWE examples.] 
    One author reviewed the MITRE CWE~\cite{mitre_cwe} for each of the 71 unique CWEs identified in the 255 CVEs to find common exploitation examples in Java.
    \rev{If no example was found in MITRE CWE, we used two LLM-based tools to propose examples: Gemini integrated with Google search, with the query ``What is the example of CWE-XXX in Java?'', and ChatGPT (GPT 4.0), with the prompt ``Give me an example of CWE-XXX in JAVA.''.
    The tools only proposed candidate APIs through generated examples: we kept only the APIs that both tools returned, and we checked every candidate manually against the JDK documentation for hallucinations before it entered the list, so the tools could not add an API that does not exist.
    This step applied to 12 of the 71 CWEs and contributed 26 APIs.
    To measure the stability of this step, we repeated the prompts five times for each of the 12 CWEs with both tools: both tools returned all 26 APIs in every run, and the raw outputs contained no API that does not exist in the JDK.}
    In this way, we compile a list of APIs present in the CWE examples. From this process, we got 42 \sensapis of which 26 are new (not in the list of other 2 steps).
    The input of this step is the mapped CWEs from Section~\ref{sec:cve-cwe-collection-mapping-impl}, and the output is 42 Java \sensapi.
\end{description}

\begin{table*}[htbp]
    \centering
    \caption{Java \sensapi Categorization}
    \longcaption{\textwidth}{}
    \begin{tabularx}{\textwidth}{llp{0.43\textwidth}rp{0.20\textwidth}}
        \toprule
         Resource & Capability & Description & \#API & Related CWE IDs\\
         \midrule
         & \apicat{Input} & Opening or reading files. & 12 & 73, 91, 319, 552, 576, 611 \\
         & \apicat{Output} & Creating a new file or writing to an existing file. & 20 & 73, 116, 117, 552 \\
         & \apicat{Modification} & Modifying files, such as deleting and copying files. & 7 & 73, 552\\
         & \apicat{Miscellaneous} & Any other actions related to the file system, including APIs that can read \emph{and} write files, APIs related to paths, and file permissions. & 11 & 22, 73, 367, 552, 732 \\
         & \apicat{Read\_env} & Reading environment variables. & 8 & 214, 526\\
         \multirow{-7}{*}{\apicat{Filesystem}} & \apicat{Read\_network\_env} & Reading environment variables related to the network. & 27 & 214, 291, 526, 706, 755, 1327\\
        \midrule
         & \apicat{Connection} & Creating and managing connections such as URL connections, web connections, and dispatching requests. & 14 & 89, 404, 444, 523, 600, 601, 772, 830, 918, 943, 1072\\
         & \apicat{Http} & Creating and managing HTTP requests, responses, cookies, and client operations. & 16 & 20, 79, 116, 213, 295, 352, 384, 444, 600, 601, 602, 614, 754, 918 \\
         & \apicat{Socket} & Creating and managing sockets as endpoints of communications between two machines. & 18 & 246, 577, 602, 923, 941, 1385\\
         \multirow{-7}{*}{\apicat{Network}}& \apicat{Naming\_directory} & Providing naming and directory functionality to Java applications. & 11 & 502 \\
         \midrule
         & \apicat{Codec\_crypto} & Encoding, decoding, encrypting, and decrypting APIs. & 23 & 84, 177, 261, 327, 1385\\
         & \apicat{Dependency} & Loading packages as dependencies. &  4 & 111, 114\\
         & \apicat{Reflection} & Dynamic loading of accessible objects: classes, methods, constructors, etc. & 39 & 470, 578, 749, 917\\ 
         & \apicat{Operating\_system} & Executing OS programs. & 5 & 78\\
         \multirow{-6}{*}{\apicat{Process}} & \apicat{Scripting} & Building, loading, and executing scripts. & 12 & 79\\
         \bottomrule
    \end{tabularx}
    \label{tab:categories}
\end{table*}

Another author reviewed the results from all approaches, and three authors reached consensus on the final \sensapi list, comprising 227 APIs.
During this construction, we also have two \rev{criteria}:
\begin{enumerate}[label=(\alph*),leftmargin=*]
\item If a class has one or more \sensapis (methods), the class's constructor is added to the \sensapi list, except if: (1) the class cannot be instantiated, (2) there is no constructor available, or (3) the constructor is not public.
The rationale is that including the constructor provides forward compatibility \textemdash{} if a new \sensapi is added to the class in the future, the constructor will already be tracked in the list.

\item We also exclude interfaces that do not declare any methods with default implementations.
\item We exclude the abstract classes since these cannot be instantiated with an implementation.

\end{enumerate}

Two authors independently applied hybrid card sorting (sorting APIs into resource types, with the flexibility to add capability types as well) to the selected APIs.
Card sorting is a qualitative technique to classify text into themes~\cite{zimmermannCardsortingTextThemes2016}.
Card sorting is commonly used in research to create informative categories~\cite{basakWhatChallengesDevelopers2023,rahmanWhatQuestionsProgrammers2018}.
We followed Zimmermann et al.~\cite{zimmermannCardsortingTextThemes2016}'s three-phase card-sorting technique.
We resolved disagreements among raters using the negotiated agreement technique~\cite{campbellCodingIndepthSemistructured2013}.
Using these techniques, the three resource types are refined to 15 more specific capabilities. One special case raised in the negotiated agreement process is the \apicat{codec\_crypto} capability; intuitively, it is not directly related to any of the three resource types.
However, both the raters then agreed that \apicat{codec\_crypto} surely does not fall into the \apicat{filesystem} or \apicat{network} resources, and so classified it as \apicat{process}~\cite{gisevInterraterAgreementInterrater2013}.
\apicat{codec\_crypto} APIs were included because the insecure use of codec/crypto libraries can lead to vulnerabilities~\cite{krugerSecuringYourCryptoAPI2023}.

After categorization, we mapped the \sensapis to CWEs to increase developers' awareness of the risks (vulnerability types that may arise) associated with adopting certain dependencies.
Table~\ref{tab:categories} describes the \sensapis, their resource types, their functionality, how many \sensapis there are in each capability, and the CWEs related to these APIs and their accessed sensitive resources.
We put the full list of APIs in the replication package.

Each resource and capability in Table~\ref{tab:categories} corresponds to a set of sensitive APIs in Java.
Our intent is not to claim that all such API calls are \emph{always} dangerous, but rather that the presence may signal elevated risk, depending on the context.
For example, the rationale for including Sockets and Reflection as sensitive APIs is that \emph{insecure} use of Sockets or Reflection could lead to vulnerabilities (as also supported by prior work~\cite{ferreiraContainingMaliciousPackage2021}).
Some examples of CWEs associated with Sockets and Reflections are CWE-246: J2EE Bad Practices: Direct Use of Sockets~\cite{cwe-246}, 
CWE-577: EJB Bad Practices: Use of Sockets~\cite{cwe-577}, 
CWE-470: Use of Externally-Controlled Input to Select Classes or Code ('Unsafe Reflection')~\cite{cwe-470}.
Some recent CVEs associated with these CWEs are CVE-2024-8048~\cite{cve-2024-8048} and CVE-2018-1000613~\cite{cve-2018-1000613}, both of which resulted from unsafe use of reflection.

\subsubsection{\Sensapi List Evaluation}
\label{sec:sen-api-list-eval-impl}
We applied both methodologies mentioned in Section~\ref{sec:api-list-eval} to understand the quality of our defined process for identifying the \sensapi list for an ecosystem.

\begin{revblock}
\paragraph{\SenSapi List's Coverage of the QGS}
To assess coverage of the QGS, we applied Equation~\ref{eq:sensitivity} to our list (the QGS is defined in Section~\ref{method:val:qgs}).
Our list coverage of the QGS is shown in Table~\ref{tab:qgs-eval}.
We report the QGS figures at two points.
Before the QGS evaluation, our three-pronged process produced a list of 226 APIs that contained 9 of the 11 QGS APIs, a sensitivity of 81.8\% (9/11).
After the evaluation, we added \texttt{java.net.URL.getContent()} from the QGS (explained below), so the final list of 227 APIs contains 10 of the 11 QGS APIs, a sensitivity of 90.9\% (10/11).
The figure 81.8\% measures the process; the figure 90.9\% includes one API that we took from the QGS itself.
We do not compare these values with the sensitivity thresholds that Zhang et al.~\cite{zhang2011identifying} give for literature search strings, because these thresholds come from a different task, and our QGS has only 11 APIs, so one API changes the sensitivity by 9.1 percentage points.
The sensitivity shows only that our process found 9 of the 11 APIs that several prior lists agree on.
The overlap ratio (Equation~\ref{eq:overlap}) is 4.0\% (9/226) before and 4.4\% (10/227) after the evaluation; as explained in Section~\ref{method:val:qgs}, this ratio does not bound the false positives of our list.
Our construction process prioritizes coverage: an omitted API gives developers an incomplete picture, while an extra API can be inspected by a developer. These checks provide evidence that the list covers APIs discussed in prior work and includes APIs used in known vulnerable functions; they do not establish that the list is complete or quantify its precision.
\end{revblock}

\begin{table}[htbp]
    \centering
    \caption{Our \SenSapi List Evaluation with the QGS}
    \rev{\longcaption{0.5\textwidth}{* Not in the list produced by our process; we added it to the final list after this evaluation.}}
    \label{tab:qgs-eval}
    \begin{tabular}{l|c}
        \toprule
        APIs & In Our List? \\
        \midrule
        java.net.URL.openConnection() & \textcolor{PineGreen}{\cmark} \\
        java.net.URL.openStream() & \textcolor{PineGreen}{\cmark}\\
        java.net.URL.getContent() & \textcolor{red}{\xmark}*\\
        java.lang.ProcessBuilder.start() & \textcolor{PineGreen}{\cmark}\\
        java.io.FileOutputStream.write() & \textcolor{PineGreen}{\cmark}\\
        java.net.ServerSocket.bind() & \textcolor{red}{\xmark}\\
        java.net.URLConnection.getInputStream()	& \textcolor{PineGreen}{\cmark}\\
        java.lang.Runtime.exec() & \textcolor{PineGreen}{\cmark}\\
        java.net.URLConnection.getOutputStream() & \textcolor{PineGreen}{\cmark}\\
        java.net.Socket.connect() & \textcolor{PineGreen}{\cmark}\\
        java.io.RandomAccessFile.write() & \textcolor{PineGreen}{\cmark}\\
        \midrule
        \rev{Sensitivity before the evaluation (226 APIs)} & \rev{81.8\% (9/11)} \\
        \rev{Sensitivity of the final list (227 APIs)} & \rev{90.9\% (10/11)} \\
        \bottomrule
    \end{tabular}
\end{table}

Based on this evaluation, we decided to add \texttt{java.net.URL.getContent()} to our final list, as merging QGS into the search result is also a recommended practice~\cite{zhang2011identifying}.
However, we decided \emph{not} to add \texttt{java.net.ServerSocket.bind()} because, according to the Java 11 JDK documentation, this API is only used to bind a ServerSocket to an IP address, but if a user wants to listen to the connection (and accept it to get the input information), they have to use \texttt{java.net.ServerSocket.accept()} API, which is included in our list.
Therefore, after adding \texttt{java.net.URL.getContent()}, the list now contains 227 APIs.

\paragraph{\SenSapis in Vulnerable Functions.}
To examine the prevalence of \sensapis in past vulnerabilities, we analyze vulnerable functions across the 45 core packages we selected, which are related to 255 CVEs.
Overall, 72 of 227 APIs in our list are directly called in vulnerable functions of these 255 CVEs.
These CVEs are associated with the same selected package set used in the list-construction study, so this result shows a link to known vulnerabilities in our sample; it is not independent external validation of the list.
The vulnerable functions in our sample are spread across 2806 package versions.
Given the granularity of the package versions, the number of \sensapi calls in the vulnerable functions of those versions has a median of 1.

\begin{figure}[htbp]
    \centering
    \includegraphics[width=0.7\linewidth]{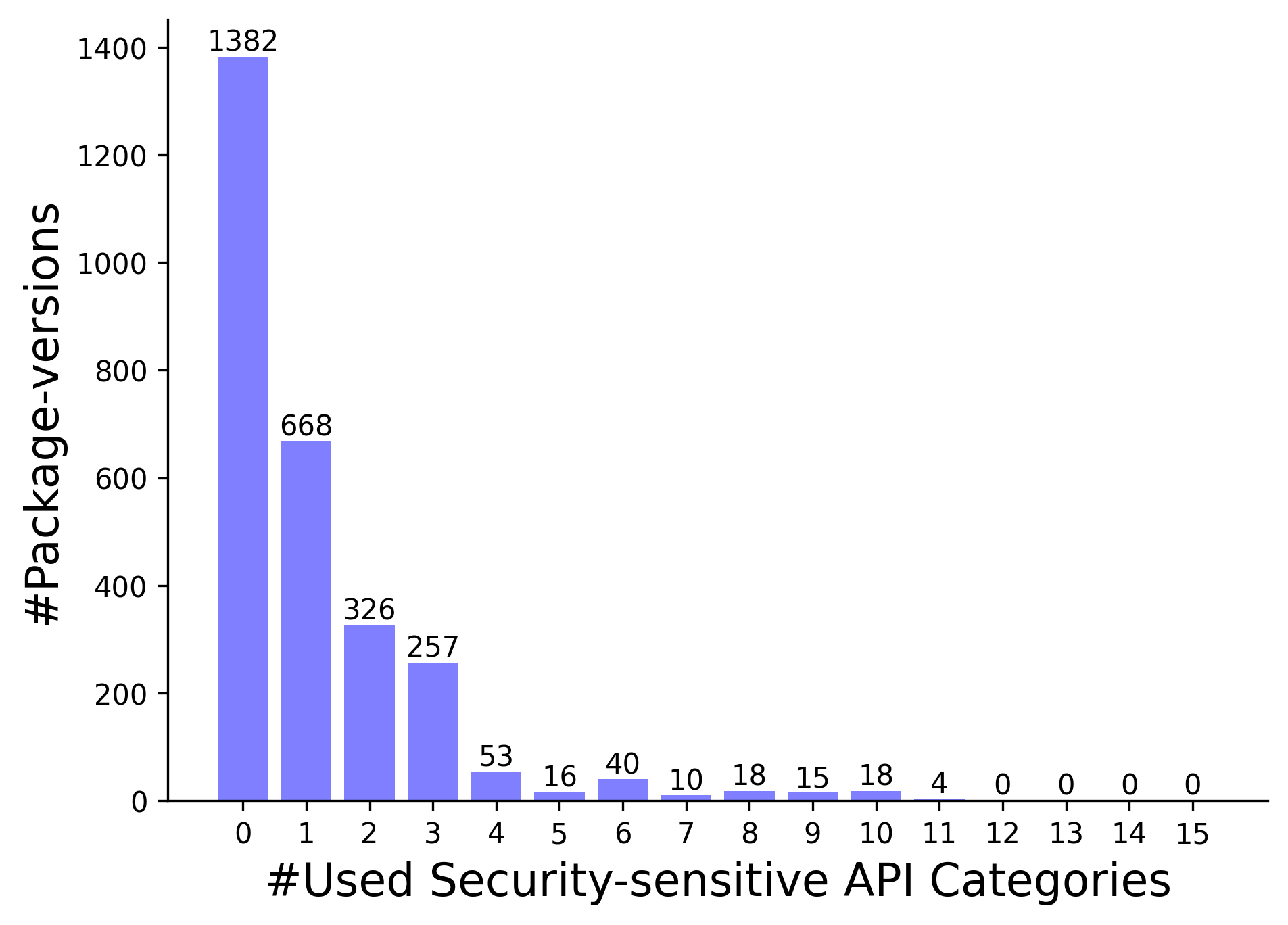}
    \longcaptionfig{\linewidth}{\textbf{Observation}: More than two-thirds of the vulnerable functions in our set have at least one \sensapi call.}
    \caption{Security-sensitive API usage in vulnerable functions from 4,183 package versions of 45 core packages.}
    \label{fig:vuln_func__sen_api_bin_cat_cgstore}
\end{figure}

Fig.~\ref{fig:vuln_func__sen_api_bin_cat_cgstore} shows the number of unique Security-Sensitive resource types used by package versions in their vulnerable functions.
49.2\% of package versions have no \sensapi call in their vulnerable function, while the rest (50.8\%) have at least one \sensapi call. 
We then analyze the generalizability of our findings when scaling to a larger number of packages. To achieve this, we use the confidence interval from Wilson-Agresti-Coull~\cite{agresti2016statistics} as used in other empirical papers~\cite{bui2024apr4vul}. 
The formula is shown in Equation~\ref{eq:1}.
Using this formula, we got a confidence interval of $(49.28\%, 52.31\%)$ for the population's proportion.
We show the calculation of the upper and lower bounds of the confidence interval in Equation~\ref{eq:1}. This equation provides the lower and upper bounds for the population proportion of packages/package versions with at least one security-sensitive API. In Equation \ref{eq:1}, $p$ is the population proportion, $\hat{p}$ is the sampled proportion packages/package versions with at least one security-sensitive API in our sample, $z_{95\%}$ is the two-sided 95\% confidence interval, in this case, $z$ has a value of $1.96$, and $n$ is the sample size, i.e., the total number of packages/package versions in the sample. 

\begin{equation} 
	\label{eq:1}
	|\hat{p}-p| = z_{95\%}\sqrt{\frac{p(1-p)}{n}}
\end{equation}

However, at the package level, 25 out of 45 packages (56.82\%) have at least one \sensapi call in one of their vulnerable functions.
In this analysis, we only identified the directly called APIs from vulnerable functions, which is an underestimation.
To assess the generalizability of this proportion, we again applied Equation~\ref{eq:1} and obtained a confidence interval of $(42.38\%, 70.19\%)$ for the population proportion.

\highlight{black}{\textbf{Finding \#1: } We identify a list of 227 Java \sensapis (categorized into 3 resource types and 15 capability types, related to 57 CWE categories), of which 72 are found in vulnerable Java functions. These \sensapis are directly called at least once in 50.8\% of our set's vulnerable functions.}

\section{Analysis Methodology}
\label{sec:method}

\begin{figure}
    \centering
    \includegraphics[width=\linewidth]{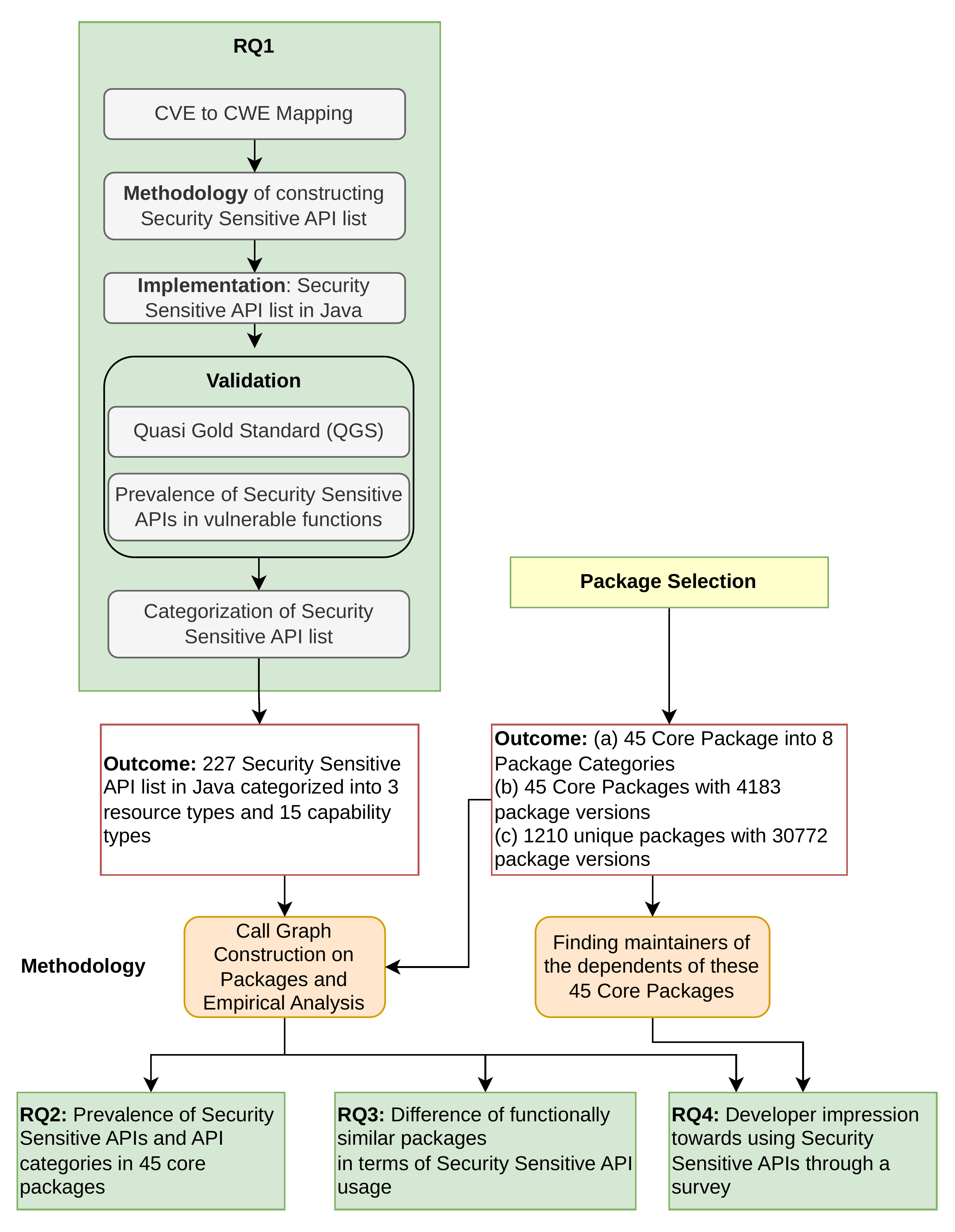}
    \caption{High level overview of our research.}
    \label{fig:senapi-workflow}
\end{figure}

This section describes our methodology for empirical call graph analysis and a developer survey (110 respondents) to quantitatively and qualitatively analyze \sensapi usage in Java packages. The API-list construction procedure is described in Section~\ref{sec:sensapi-list-construction}.
A high-level overview of our research methodology is presented in Figure~\ref{fig:senapi-highlevel}.
\rev{Table~\ref{tab:study-design} summarizes, for each research question, the sample, the scope of the call graph, and the level at which we report the results (Section~\ref{sec:terminology}).}

\begin{table}[tbp]
\revcolor
\centering
\small
\caption{Study design: the sample, the call graph scope, and the measure of each research question and of the call graph validation.}
\label{tab:study-design}
\begin{tabular}{>{\raggedright\arraybackslash}p{0.1\linewidth} >{\raggedright\arraybackslash}p{0.3\linewidth} >{\raggedright\arraybackslash}p{0.17\linewidth} >{\raggedright\arraybackslash}p{0.33\linewidth}}
\toprule
Part & Sample & Scope & Measure (level) \\
\midrule
\rqref{rq:sec-sen-api-and-roles} & JDK~11 documentation; 255 CVE fixes and 71 CWEs of the 45 core packages and their dependencies & -- & List of 227 \sensapis; coverage of an 11-API QGS; APIs in vulnerable functions \\
\addlinespace
\rqref{rq:sen-api-usage-pattern} & 30,772 versions of 1,210 packages; 4,183 versions of the 45 core packages & Intra-package; intra vs.\ inter-package (4,183 versions) & Number of calls (call level); number of capabilities used (capability level) \\
\addlinespace
\rqref{rq:sen-api-diff} & Latest versions of 34 packages in seven categories (JDBC Driver: exploratory case only) & Inter-package & Unique \sensapis per capability (API level); capability breadth (capability level); ranges of calls (call level); qualitative review \\
\addlinespace
Validation & The same 34 package versions as \rqref{rq:sen-api-diff} & Intra- and inter-package & Package--API pairs (API level); capability breadth (capability level) \\
\addlinespace
\rqref{rq:usefulness-of-sen-api} & 110 respondents of 7,785 contacted developers & -- & Shares of respondents \\
\bottomrule
\end{tabular}
\end{table}

We selected Java as our evaluation environment because 35.4\% of professional developers use Java, according to the 2021 Stack Overflow Developer Survey~\cite{stackoverflow-2021}, and according to the JetBrains ``State of Developer Ecosystem 2023'' report, 74\% of surveyed Java developers regularly use Maven as their build system.

\subsection{Call Graph Generation on Packages and Empirical Analysis (\rqref{rq:sen-api-usage-pattern} and \rqref{rq:sen-api-diff})}

In this section, we present our methodology for constructing the call graphs on our selected packages and for the empirical analysis.

We opted to use the specific call graph generated by Endor Labs Call Graph Tool because it can accumulate the reachable APIs of dependencies.
However, this can also be done using other open-source Java call graph generation tools.
\rev{The Endor Labs tool is proprietary.
So that others can recompute RQ2 and RQ3 without it, our replication package contains the \sensapi call-graph edges and the \sensapi call tables for each package version.}
The tool generated both intra-pkg call graphs and stitched them together to form inter-pkg call graphs.
To generate the inter-pkg call graph for a package, the tool examines all call sites in the package and connects them to potential callees in packages in its dependency graph.
The call graph construction process includes building a type hierarchy for all types across all packages in the dependency graph, which is important for resolving dynamically dispatched calls.
In addition, the call graph construction process allows, e.g., to stitch an interface call across all implementations of this interface, regardless of which package in the dependency graph contains them.
Using this call graph generation tool, we generated a call graph for each package version (intra-pkg call graph) of our set of 30,772 package versions.
\rev{The tool uses Class Hierarchy Analysis (CHA)~\cite{deanOptimizationObjectorientedPrograms1995}: at a virtual call site, it adds an edge to every matching method implementation in the package's dependency graph.
Section~\ref{sec:cg-validation-method} describes how we validated these call graphs with a second, open-source call graph generator.}

After receiving the call graphs in JSON format, we traversed them using a pre-order breadth-first search (BFS) to determine whether any of our \sensapis were called in the package version.
Our pre-order breadth-first-traversal is similar to Mir et al.~\cite{mirEffectTransitivityGranularity2023}'s analysis of reachable vulnerable call chains.
\emph{Reachability analysis} is a program analysis concept to determine whether functions containing vulnerable code are called within dependencies~\cite{dunlapPairingSecurityAdvisories2024,pontaDetectionAssessmentMitigation2020}.
After this \emph{reachable} method identification step, we can find the reachable functions and sensitive APIs from the vulnerable functions.
We present our empirical findings on the prevalence of \sensapi use in these 30,772 package versions (\rqref{rq:sen-api-usage-pattern}).

After obtaining the CGs, we systematically and qualitatively compare the Maven package categories of functionally similar packages to understand why they exhibit similar or different \sensapi patterns for \rqref{rq:sen-api-diff}.
\rev{For this analysis, we used the latest versions of 34 of the 45 core packages: the packages of the seven categories that \rqref{rq:sen-api-diff} compares.
We excluded two groups of packages.
First, we excluded the seven JDBC Driver packages, because these packages are not interchangeable (Section~\ref{sec:package-selection}, step 2); Appendix~\ref{sec:jdbc-exploratory} shows five of these drivers only as an exploratory case, so the results with JDBC cover 39 packages.
Second, we excluded four packages that have no dependencies (\pkgname{javax.inject}, \pkgname{json}, \pkgname{jboss-logging}, and \pkgname{xercesImpl}), because their intra- and inter-package call graphs are the same, and \rqref{rq:sen-api-diff} compares the inter-package call graphs.
The quantitative comparison (the heatmaps), the qualitative review, and the call graph validation (Section~\ref{sec:cg-validation-method}) therefore use the same 34 packages.}
First, we gathered each package's source code from GitHub and documentation from GitHub, Maven Central Repository, and the project's homepage.
We further collect API usage examples for those packages by analyzing the top 10 dependents of each package.
To be specific, we extract the call sites where a dependent package invokes an API from the target package, with the API name, its arguments, and the surrounding code context.
We also collect dependency information for these 34 packages from deps.dev.
Next, one author carried out a uniform review of the collected information: 
(1) \textit{Intended usage patterns}\textemdash{}by examining project documentation (e.g., READMEs, wikis) to identify the typical sequence of API calls a client is expected to make (e.g., initialize $\rightarrow$ configure $\rightarrow$ execute);
(2) \textit{Internal design choices} \textemdash{} by assessing the architectural and implementation decisions made by package authors, such as whether the package uses an interface, callback-based vs. promise-based async handling, or a plugin/extension model; we identified these patterns inductively from the source code;
(3) \textit{Real-world usage} \textemdash{} by reviewing example code and tutorials linked from the repository or homepage, and examining how the top dependents invoke the package, to capture how client code concretely uses the package in practice (e.g., which APIs are called together, which are rarely used).
Finally, we aggregated our qualitative findings for each package category and compared them across package categories.
For example, one qualitative finding for HTTP client packages were that some HTTP clients used other HTTP client packages as a dependency to inherit basic HTTP functionality and provide additional features.

\begin{revblock}
\subsection{Call Graph Validation}
\label{sec:cg-validation-method}
A static call graph can be wrong in two directions.
The call graph \emph{over-approximates} when it contains an edge that no execution of the program takes.
The call graph \emph{under-approximates} when it misses an edge that an execution can take.
For our counts, an over-approximation can add a \sensapi that a package does not call, and an under-approximation can hide a \sensapi that a package calls.
Our call graphs can contain both kinds of error.
CHA over-approximates at virtual call sites: CHA connects a virtual call to every matching implementation, also to implementations that the program never creates.
A static call graph under-approximates the calls through reflection: a package can reach a \sensapi only as the \emph{target} of a reflective call (e.g., \api{Method.invoke()} on a method name read from a configuration file), and a static call graph has no edge for such a call.
In addition, our call graphs come from a proprietary tool, so the counts can depend on how this one tool works.
We therefore validated the call graphs with a second, open-source call graph generator, OPAL~\cite{helmUnimocgModularCallgraph2024}, to answer two questions:
(1)~how much does CHA over-approximate the \sensapi calls, and
(2)~do the results depend on the call graph generator?
Table~\ref{tab:cg-design} shows the two comparisons that answer these questions.
The validation does not measure the under-approximation through reflection; Section~\ref{sec:threats} discusses this limitation.

\begin{table}[tbp]
\revcolor
\centering
\small
\caption{The two comparisons of the call graph validation. In each comparison, only one factor differs between the two call graphs.}
\label{tab:cg-design}
\begin{tabular}{>{\raggedright\arraybackslash}p{0.19\linewidth} >{\raggedright\arraybackslash}p{0.16\linewidth} >{\raggedright\arraybackslash}p{0.2\linewidth} >{\raggedright\arraybackslash}p{0.14\linewidth} >{\raggedright\arraybackslash}p{0.16\linewidth}}
\toprule
Comparison & Call graphs & Factor that differs & Scope & Results \\
\midrule
(1) Over-approximation & OPAL CHA vs.\ OPAL RTA & The algorithm that resolves virtual calls & Intra- and inter-package & Section~\ref{sec:cg-validation-rta}, Table~\ref{tab:cg-capability} \\
\addlinespace
(2) Tool agreement & Endor Labs tool vs.\ OPAL CHA & The tool (both use CHA) & Intra-package & Section~\ref{sec:cg-validation-tools}, Table~\ref{tab:cg-packages} \\
\bottomrule
\end{tabular}
\end{table}

\noindent\textbf{Sample.}
We analyzed the latest version of each of the 34 packages in the seven RQ3 categories (Figure~\ref{fig:dependency-injection-heatmap}, Figure~\ref{fig:httpclient-heatmap}, and the Appendix), because RQ3 compares exactly these versions; we excluded the JDBC Driver packages of the exploratory case.

\noindent\textbf{Configurations.}
We ran OPAL 5.0.0 with two algorithms.
The first is CHA, the algorithm that the Endor Labs tool also uses.
The second is Rapid Type Analysis (RTA)~\cite{baconFastStaticAnalysis1996}, which resolves a virtual call only to the implementations whose class the reachable code instantiates.
RTA therefore keeps a subset of the CHA edges, and it is more precise than CHA.
RTA is still a static over-approximation, so the CHA edges that RTA removes are a lower bound on the over-approximation of CHA, not its full size.
Both configurations used the same inputs: the package JAR, the resolved runtime dependency JARs, and the same JDK (OpenJDK 21).
Both used the library setting of Reif et al.~\cite{reifCallGraphConstruction2016}: every public and protected method of the analyzed JARs is an entry point, and a client can instantiate any public type.
In the inter-package scope, the analyzed JARs include the dependency JARs.
Both also ran the same analyses that find the entry points and the instantiated types in library code, so the two configurations differ only in how they resolve virtual calls.
OPAL did not analyze the JDK implementation.
OPAL records a call from a package to a JDK method, but OPAL does not follow the calls inside the JDK, and OPAL does not resolve a call on a JDK interface (e.g., \api{ScriptEngine.eval()}) to an implementation.
For each call graph, we extracted the \sensapis that the package calls, in two scopes: \emph{intra-package} (callers in the package's own JAR) and \emph{inter-package} (callers in the package or in its dependencies).

\noindent\textbf{Units of comparison.}
We compared the call graphs at the API level and the capability level (Section~\ref{sec:terminology}): for each package, the set of unique \sensapis and the set of capabilities that the call graph reaches.
We count the results in \emph{package--API pairs}: a pair $(p, a)$ means that the call graph of package $p$ contains at least one edge to \sensapi $a$.
For example, if 10 packages each call \api{File.delete()}, this API gives 10 pairs.
The \sensapi counts that RQ2 and RQ3 report come from the call graphs of the Endor Labs tool.

\noindent\textbf{Comparison 1: over-approximation (OPAL CHA vs.\ OPAL RTA).}
The tool and the inputs are the same, so every difference comes from the algorithm.
For each scope and each capability, we report the share of the CHA package--API pairs that RTA keeps, with a 95\% bootstrap confidence interval that resamples packages.

\noindent\textbf{Comparison 2: tool agreement (Endor Labs tool vs.\ OPAL CHA).}
We compared the two tools in the intra-package scope only, where both tools analyze the same code.
We did not compare the inter-package scope, because the two tools define it differently: the inter-package call graph of the Endor Labs tool contains only the dependency code that the package's own code can reach, while OPAL treats every public method of the dependencies as an entry point.
We report three statistics:
\begin{itemize}
    \item \emph{Spearman's $\rho$} of the numbers of unique \sensapis per package. The coefficient shows whether the two tools rank the packages in the same order ($\rho = 1$: the same order).
    \item \emph{Cohen's $\kappa$} for the presence of each capability in each package ($34 \times 15$ decisions). The coefficient shows whether the tools agree on which capabilities a package reaches, corrected for agreement by chance ($\kappa = 1$: full agreement).
    \item The median \emph{Jaccard index} of the API sets. For one package, the Jaccard index is the number of \sensapis that both tools report, divided by the number of \sensapis that at least one tool reports ($1$: the same set).
\end{itemize}
To explain the differences between the tools, we separate two kinds of edges to a \sensapi.
A \emph{direct edge} goes to the method that the call instruction in the bytecode names.
A \emph{virtual edge} goes to a different method, which the call graph generator selects when it resolves a virtual call on a supertype to a method of a subtype.
We gave every package--API pair that only one tool reports one of three labels (the short names are the column names of Table~\ref{tab:cg-packages}):
\begin{itemize}
    \item \emph{OPAL over-approx.} (over-approximation by OPAL CHA): only OPAL reports the API, and all its edges to the API are virtual edges. For example, the package calls \api{OutputStream.write()}, and OPAL resolves this call to \api{FileOutputStream.write()}, although the stream may not be a file.
    \item \emph{Endor under-approx.} (under-approximation by the Endor Labs tool): only OPAL reports the API, and at least one OPAL edge to the API is not a virtual edge. The call is a direct call, or a call on the package's own subclass that inherits the JDK method (e.g., a class loader of the package that calls the inherited \api{ClassLoader.defineClass()}).
    \item \emph{OPAL under-approx.} (under-approximation by OPAL): only the Endor Labs tool reports the API, and the package's bytecode contains a call instruction to the API.
\end{itemize}
To assign these labels, we inspected the bytecode of each package with \texttt{javap}.
For each pair, we checked whether the bytecode contains a call instruction to the API, and whether the instruction names the API itself or a method of a supertype.
The labels classify the differences between the outputs of the two tools.
The labels do not show that a call executes at runtime.

\noindent\textbf{Effect on the findings.}
Finally, we recomputed the RQ2 and RQ3 findings at the API level under each call graph: the capability with the most package--API pairs, the packages with no \sensapi, and, for each package category, the ranking of packages and the packages with the most and the fewest \sensapis.
\end{revblock}

\subsection{Experimental Design of User Study (\rqref{rq:usefulness-of-sen-api})}
To understand if having \sensapi information would have influenced developers' dependency selection, we conducted an online survey of developers who maintain projects that use the 45 core packages we selected.

\subsubsection{Participant Selection:}
To find the respondents for our survey, we first used \texttt{\small deps.dev}~\cite{depsdev} to identify the dependents of our chosen 45 core packages, yielding 37,192 dependent packages (and 118,775 package versions).
Only 29\% (10,785/37,192) of dependent packages had a valid GitHub URL.
We used the GitHub API to find the top 20 contributors to each package (12,565 people) and pulled their email addresses, yielding 11,703 available email addresses.
We then filtered out users who had not contributed to the project in the last 2 years, leaving 7,785 contributors.
We made sure to send only one email per person, even though they may contribute to multiple packages on our list.
We sent an email with the link to a personalized survey built in Qualtrics\footnote{https://www.qualtrics.com/} to these email addresses.
We conducted the survey between June and July 2024.
As a thank you for their participation, we offered a \$20 Amazon gift card to five randomly selected participants if they wished to participate in the lottery.
We discussed the IRB approval and other ethical considerations in Section~\ref{sec:ethics}.

\subsubsection{Feedback Elicitation:}
\begin{figure}[hbp]
    \centering
    \includegraphics[width=0.6\linewidth]{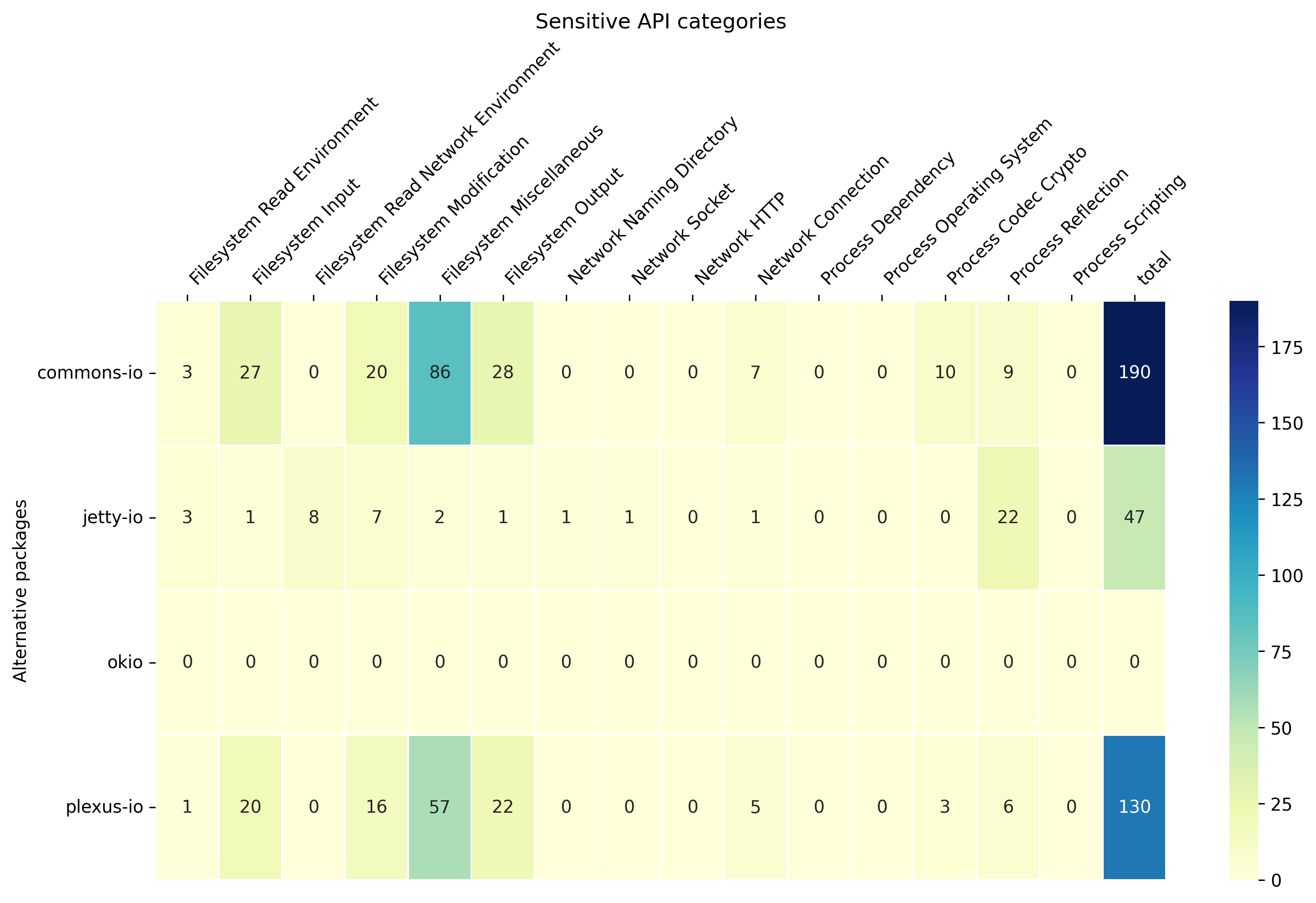}
    \caption{Security-sensitive API usage in ``I/O Utilities'' package category.}
    \label{fig:io-utilities-heatmap}
\end{figure}

To elicit feedback from the developers, we showed them a personalized heatmap visualization of the \sensapi calls of their chosen dependency and other alternative packages in the same category.
An example is shown in Figure~\ref{fig:io-utilities-heatmap}.
We designed the survey so participants would require as little effort as possible to complete it; i.e., it was self-contained, included all relevant \sensapi information, our motivation, and the study's context, and would not take more than 15 minutes to complete.

\subsubsection{Visualization:}
We conducted a pilot with different visualizations and ultimately chose a heatmap because it was more intuitive and clearer to the pilot participants.
We used an accessible palette for the heatmap to ensure it is clear for colorblind people.
We also randomized the order of the Security-Sensitive resource types in the visualization to avoid bias in our results~\cite{gerosaShiftingSandsMotivation2021,guerraHowAnnotationsAffect2024}.

\subsubsection{Survey Design:}
In our survey, we asked developers: (1) 
whether they would have decided differently if they had known this information before (yes/no) and why (open-ended); and (2) what resource types they think should be considered as security-sensitive in general, along with the degree of sensitivity (checkbox and multiple-choice).
Our survey design, including the participant selection process, was inspired by Mujahid et al.~\cite{mujahidWhereGoNow2023}.

\subsubsection{Analysis and Coding:}
We used thematic analysis and an iterative open-coding approach to analyze the responses in the open-ended part of question 1~\cite{charmazConstructingGroundedTheory2014,corbinGroundedTheoryPractice1997,corbinGroundedTheoryResearch1990}.
Initially, two authors developed a codebook based on their impressions of the survey responses (inductive open-coding).
Then, the same authors iteratively coded the responses in multiple rounds~\cite{birksGroundedTheoryPractical2022,urquhartGroundedTheoryQualitative2022}.
After each iteration, we resolve the conflicts by discussion.
At the end, we got a hypothetical final agreement of 100\%~\cite{mcdonaldReliabilityInterraterReliability2019,wermke2022committed,wermkeAlwaysContributeBack2023}.
We include the full survey questionnaire and codebook in the supplementary material and the artifact.

\section{\rqref{rq:sen-api-usage-pattern}: How Frequently Do Java Open-Source Packages Call \Sensapis?}
\label{sec:sen-api-usage-pattern}
\rev{\noindent\textbf{How to read the results of \rqref{rq:sen-api-usage-pattern} and \rqref{rq:sen-api-diff}.}
The results in Sections~\ref{sec:sen-api-usage-pattern} and~\ref{sec:sen-api-diff} come from the call graphs of the Endor Labs tool, which uses CHA.
These call graphs report the \sensapis that the static analysis finds, not the calls that an execution makes.
Section~\ref{sec:cg-validation} tests these results in two ways.
First, a more precise algorithm (RTA) shows how much the over-approximation of CHA changes the reported \sensapis: RTA keeps 96.9\% (intra-package) and 97.4\% (inter-package) of the package--API pairs.
Second, a second tool (OPAL) shows that the comparisons between packages do not depend on the Endor Labs tool.
Neither test measures the \sensapis that a package reaches only as the target of a reflective call (Section~\ref{sec:threats}).}

To answer this RQ, we first analyze the prevalence of our \sensapis calls across the chosen packages, with and without dependencies.
For analysis, we create a tool that takes a package call graph as input and generates comparative visualizations of \sensapi calls for those packages.
Then, we qualitatively analyze each Maven package category (Table~\ref{tab:packages}) and compare the \sensapi calls among packages in each Maven category.
Functionally similar packages are grouped together in a package category.

\begin{figure}
    \centering
    \includegraphics[width=0.7\linewidth]{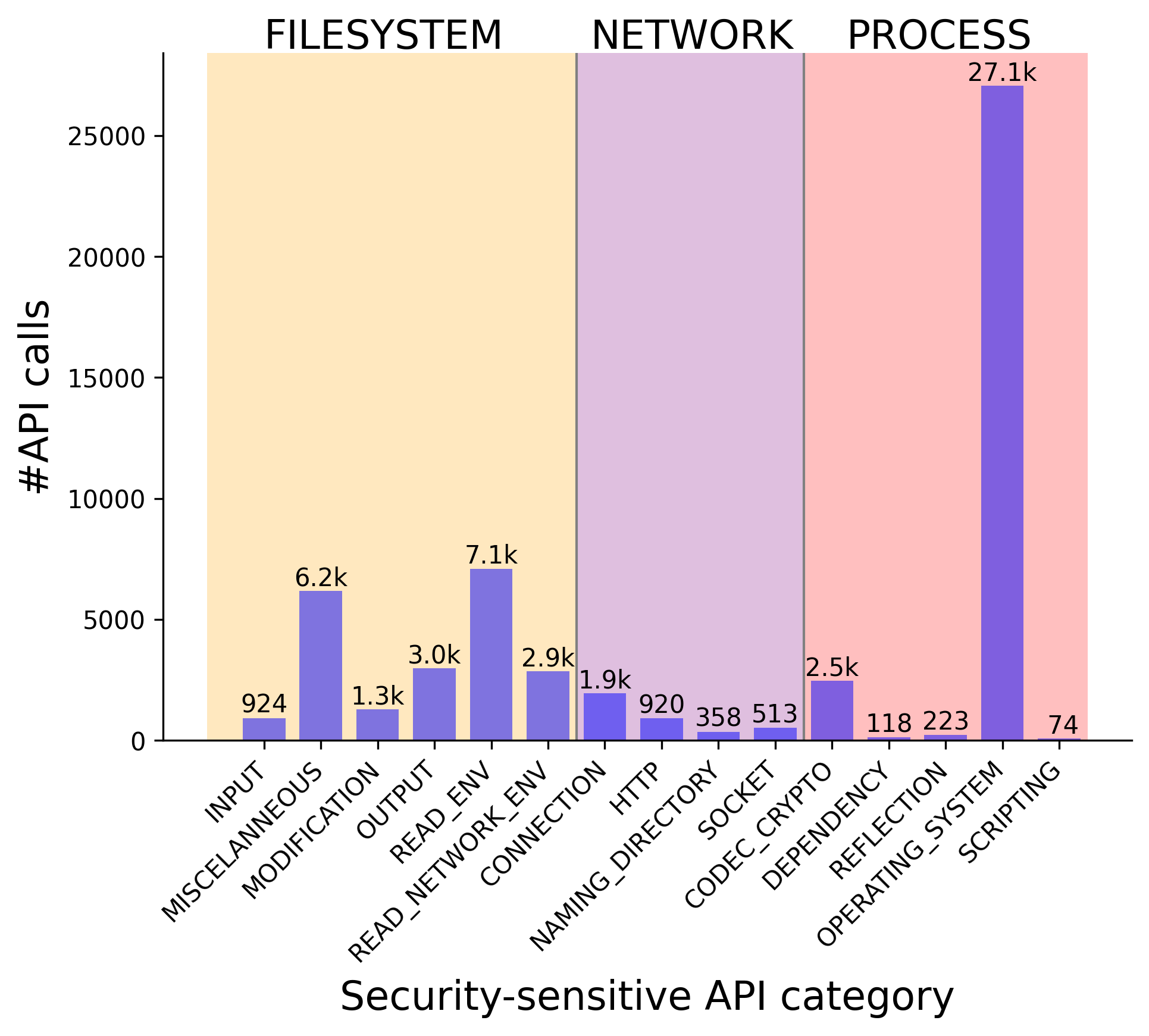}
    \caption{Sensitive API usage in 1,210 open-source packages\rev{ (call level, per capability)}.}
    \label{fig:sen_api_freq_in_cat_cgstore}
\end{figure}

\begin{table}
    \centering
    \setlength\tabcolsep{2.5pt}
    \caption{Top 10 \sensapi Call\rev{ (call level, per API)}}
    \longcaption{\columnwidth}{These top 10 API calls are from 1,210 packages. We first average the calls on each package and then sum for each category to avoid bias. \apicat{f} is from \apicat{filesystem} resource and \apicat{p} is from \apicat{process}.}
    \begin{tabular}{llrr}
        \toprule
        Capability & API & \# & \%  \\
        \midrule
        \apicat{P/Reflection} & java.lang.Class.forName() & 7243 & 14.80\%\\
        \apicat{P/Reflection} & java.lang.reflect.Method.invoke() & 3688 & 7.54\%\\
        \apicat{F/Read\_env} & java.lang.System.getProperty() & 3319 & 6.78\%\\
        \apicat{F/Miscellaneous} & java.io.File.exists() & 2693 & 5.50\%\\
        \apicat{F/Miscellaneous} & java.io.File.getAbsolutePath() & 2016 & 4.12\%\\
        \apicat{P/Reflection} & java.lang.Class.newInstance() & 1767 & 3.61\%\\
        \apicat{P/Reflection} & java.lang.Class.getMethod() & 1711 & 3.50\%\\
        \apicat{F/Output} & java.util.logging.Logger.log() & 1548 & 3.16\%\\
        \apicat{F/Read\_env} & java.util.Properties.getProperty() & 1527 & 3.12\%\\
        \apicat{P/Reflection} & java.lang.reflect.Constructor.newInstance() & 1440 & 2.94\%\\
        \bottomrule
    \end{tabular}
    \label{tab:top10}
\end{table}

\noindent \subsection{What Is The Most Frequently Invoked \Sensapis and Most Frequently Invoked \Sensapi in 1,210 Packages?}
Fig.~\ref{fig:sen_api_freq_in_cat_cgstore} shows the \sensapi call in 1,210 open-source packages.
These 1,210 packages are the unique packages in the dependency tree of our chosen 45 packages.
The most used \sensapi capability is \apicat{process/reflection}.
Table~\ref{tab:top10} shows the top 10 \sensapi calls in the observed packages, which shows that the \apicat{process/reflection} capability takes the first, second, sixth, seventh, and tenth place (32.4\% of total calls).
\rev{These five APIs are direct calls to the reflection API: \api{Class.forName()} is static, and \api{Class}, \api{Method}, and \api{Constructor} are final classes.
Each of these call sites therefore has exactly one target, and CHA resolves each call site with a direct edge.
In our call graph validation (Section~\ref{sec:cg-validation}), none of the 3,844 inter-package edges to these five APIs was a virtual edge.
The count measures calls \emph{to} reflection; it does not include the \sensapis that a package reaches only as the \emph{target} of a reflective call (Section~\ref{sec:threats}).}

\noindent \subsection{How Many Security-Sensitive Resources Are Called At The Package-Version Granularity In 30,772 Package Versions?}
\begin{figure}
    \centering
    \includegraphics[width=0.8\linewidth]{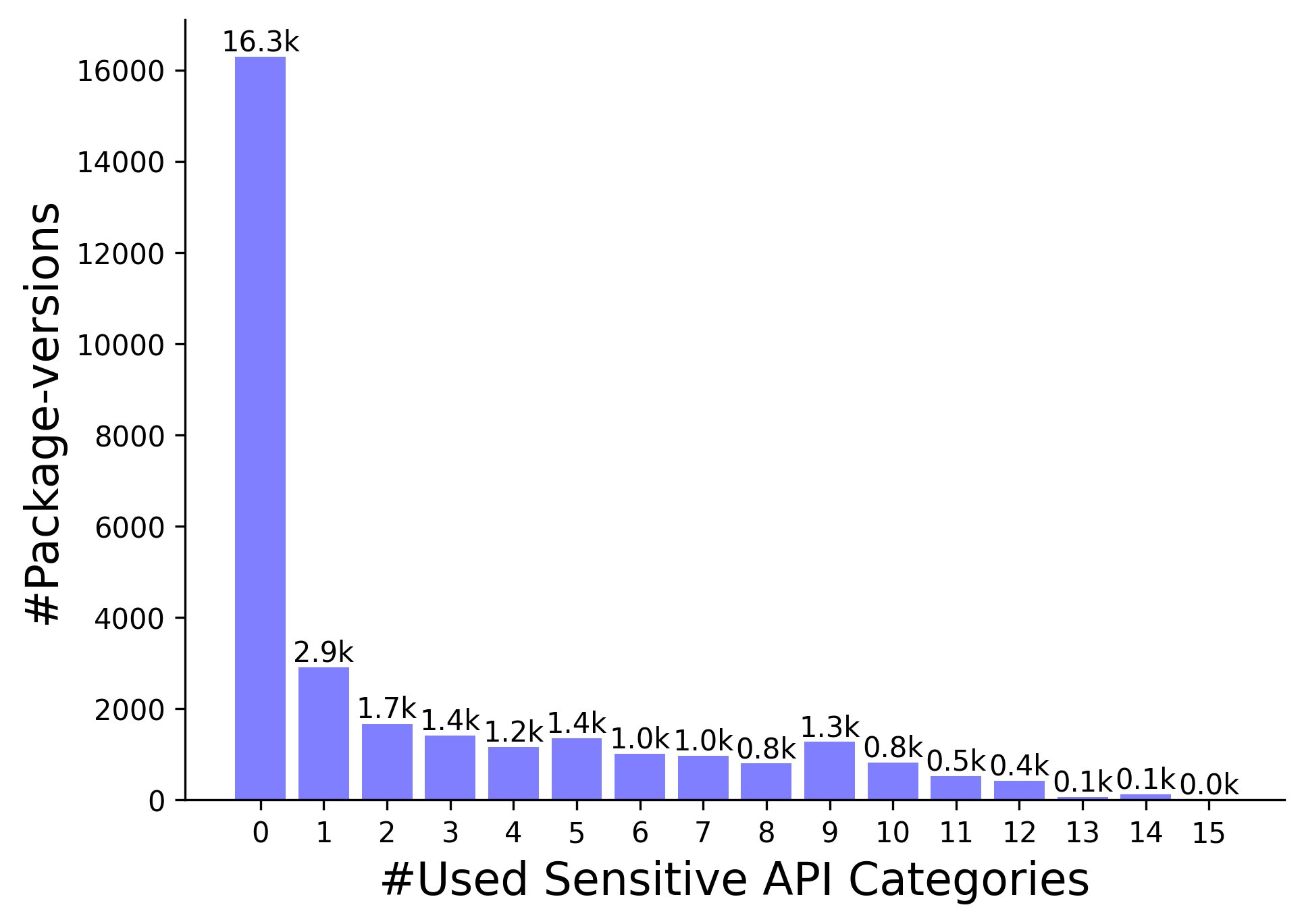}
    \caption{Sensitive API usage in 30,772 open-source package versions \rev{ (capability level: number of capabilities used)}.}
    \label{fig:sen_api_bin_cat_cgstore}
\end{figure}

While developers may first choose a package, their project would use a \textit{version} (or more) of that package in production. Using a different version may result in different \sensapis being included and thus different possible risks.
Fig.~\ref{fig:sen_api_bin_cat_cgstore} shows the number of Security-Sensitive resources each package version uses.
Half of the package versions (50.2\%) use no \sensapi (consecutively no \sensapi capability), and only 2.6\% package versions use more than 10 security-sensitive capabilities.

\noindent \subsection{How Do \Sensapi Calls In Intra Package Call Graphs Differ From Inter Package Call Graphs In 4,183 Package Versions?}
\begin{figure}[tbp]
    \centering
    \includegraphics[width=0.7\linewidth]{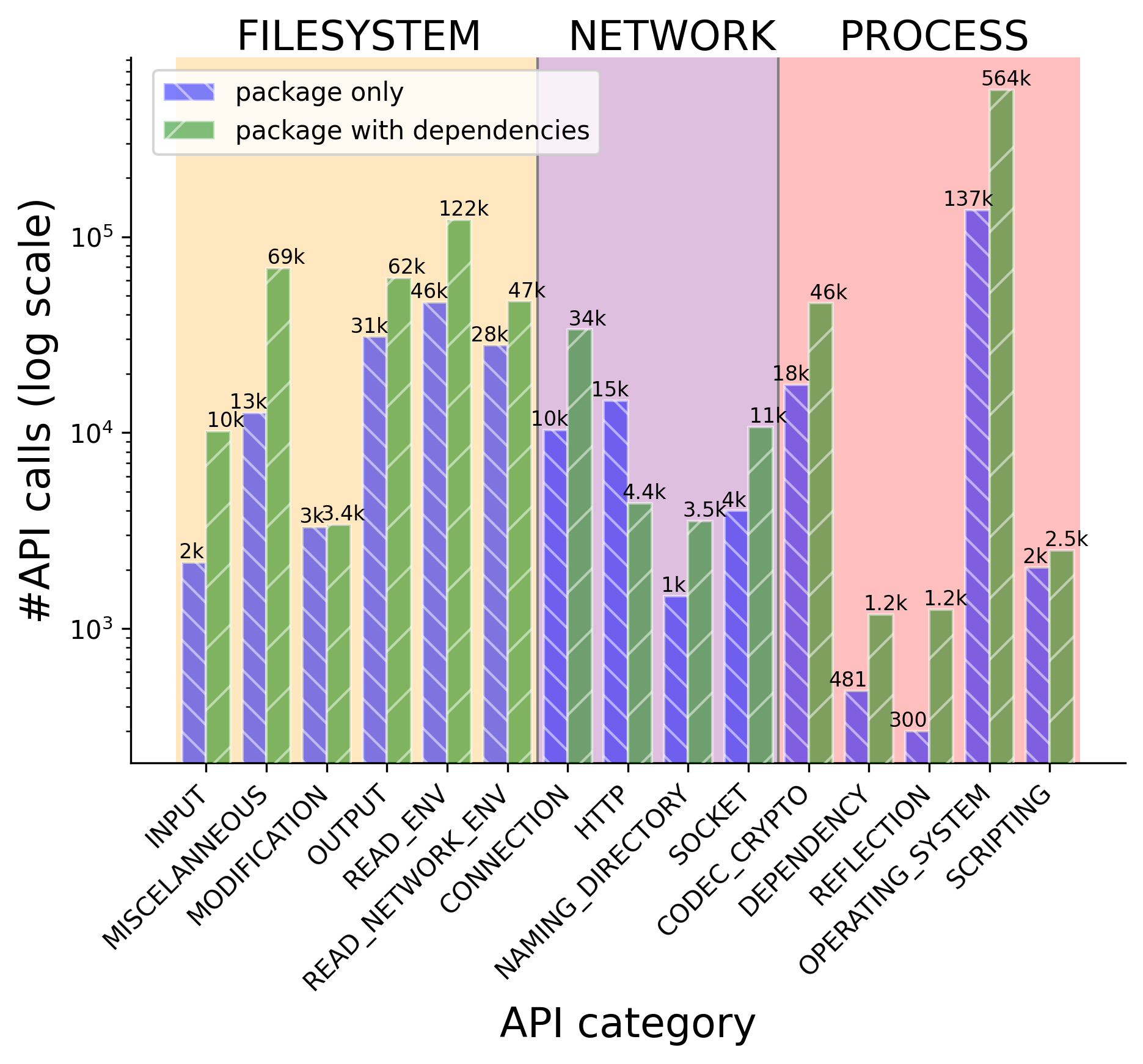} 
    \caption{Security-sensitive API call increase in 4183 open-source package versions without vs. with dependencies \rev{(call level)}.}
    \label{fig:combined_sensitive_api_usage_freq}
\end{figure}

Fig.~\ref{fig:combined_sensitive_api_usage_freq} shows the comparison of \sensapi call frequency between intra- (without dependencies) and inter- (with dependencies) package call graphs. 
On average, each package version has 168 additional \sensapi calls \rev{(call level)} from its dependencies.
The most increase happens in capability \apicat{process/reflection} (on average, 109 additional calls per package version).
\apicat{Filesystem/miscellaneous} calls are increased 5.3 times (from 13k to 69k), the highest multiply factor in a capability. This phenomenon can be explained by the nature of the \apicat{filesystem} APIs, which are used in tandem with other functionalities; e.g., checking whether a file exists is a standard practice before creating or reading from a file.
\apicat{Process/scripting} has the lowest multiply factor (1.19 times) with 2.5k to 2.1k calls.

However, \apicat{network/http} is the only capability where the number of calls is reduced (15k to 4.4k) when dependencies are included.
The number of calls in inter-package call graphs is smaller than in intra-package call graphs if there is no suitable implementation type of an interface found in the package or its dependencies.
In such cases, Endor Labs' call graph generator does not create any edge because no concrete \texttt{<caller, callee>} relationship can be determined.
One example is the calls to the interface method \api{HttpSession.getAttribute()} in \pkgname{spring-webflow@2.2.1}.
Due to a lack of interface implementation in the package and its dependencies, no corresponding edge was created in the inter-package call graph of \pkgname{spring-webflow@2.2.1}.
One possible explanation is that implementations of the servlet API are commonly provided by a servlet container such as Apache Tomcat.
The choice of a specific servlet container, however, is not prescribed by \pkgname{spring-webflow@2.2.1}.
This requirement can be satisfied, for example, by additional dependencies of \pkgname{spring-webflow@2.2.1}.

\begin{figure}[tbp]
    \centering
    \includegraphics[width=0.6\linewidth]{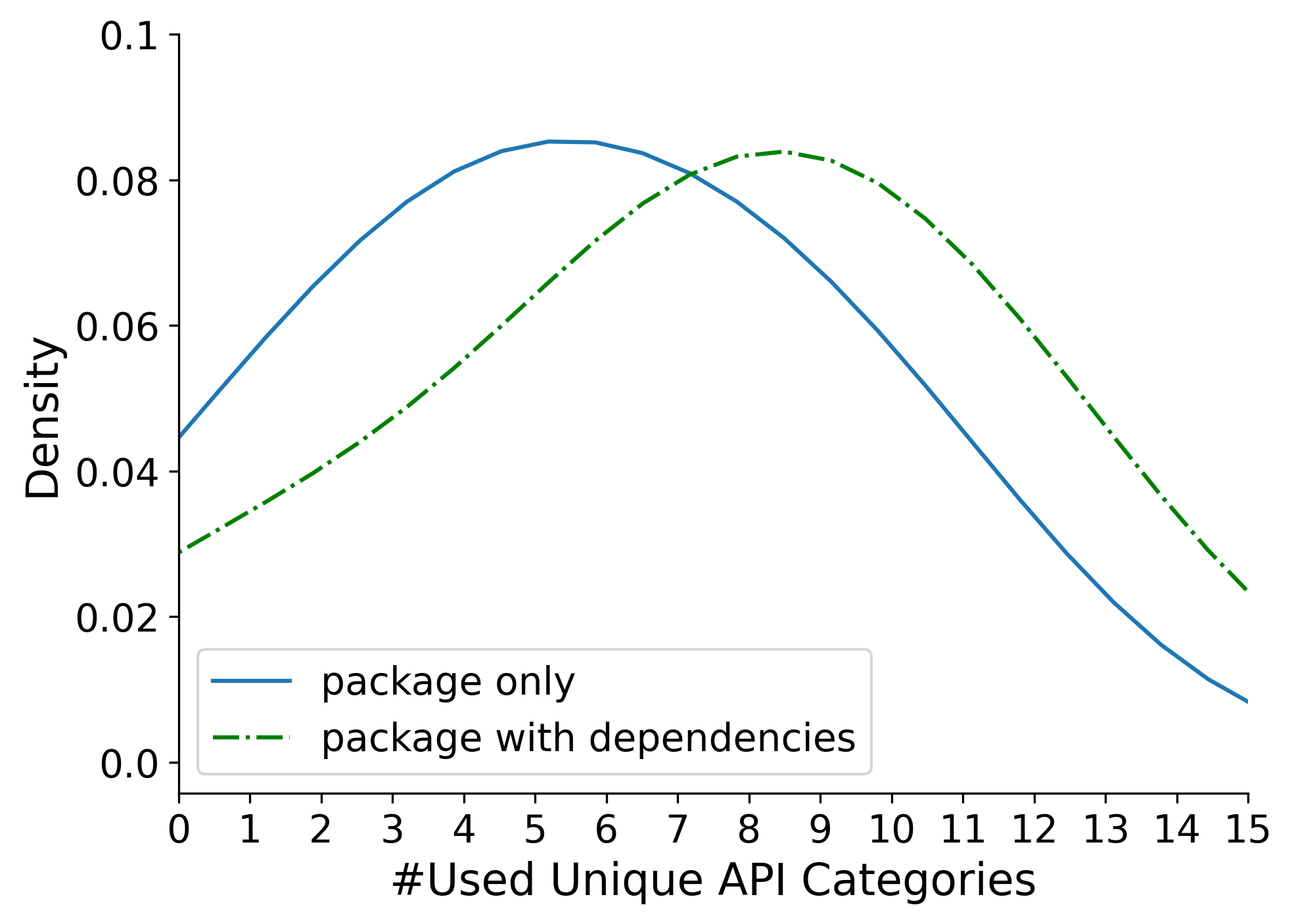}
    \caption{Security-sensitive API usage in 4183 open-source package versions \rev{ (capability level: number of capabilities used)}.}
    \label{fig:combined_unique_api_categories_used}
\end{figure}

To compare the use of security-sensitive resources, we show the density functions with and without dependencies in Fig.~\ref{fig:combined_unique_api_categories_used}.
The rightward shift in the density function indicates that dependencies increase the number of resource and capability types used.

\highlight{black}{\textbf{Finding \#2: } On average, each package version makes 72 \sensapi calls \rev{(call level)} in its own code. This average increases by a factor of 2.3 when we consider dependencies, indicating that dependencies play a major role in increasing the number of \sensapi calls in a program. \rev{The call graph validation (Section~\ref{sec:cg-validation}) measures a different quantity, the number of unique \sensapis per package (API level), on the latest versions of the 34 RQ3 packages. With this measure, the increase is between 1.8 and 2.6 times, depending on the call graph. This range does not replace the call-level factor of 2.3, because the API level does not count repeated calls, and the validation covers a smaller sample. The range shows that dependencies increase the number of \sensapis under every call graph that we used.}}

 \begin{figure}[bp]
    \centering
    \includegraphics[width=\linewidth]{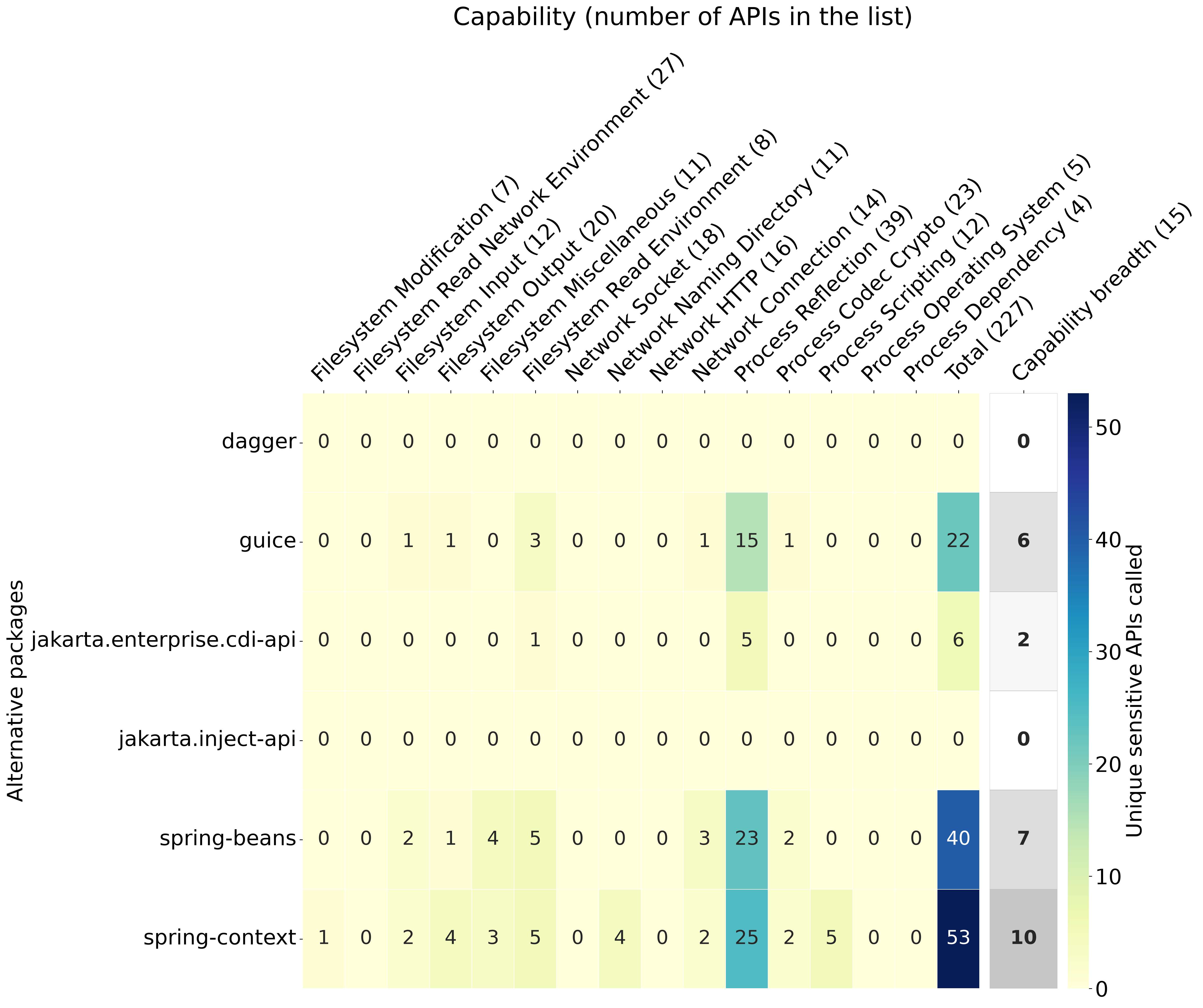}
    \caption{\rev{Security-sensitive API usage in ``Dependency Injection'' package category, at the API level. Each cell is the number of unique \sensapis of a capability that the package calls in its inter-package call graph; the number in parentheses in a column header is the number of APIs of that capability in our list. The last column shows the capability breadth (capability level).}}
    \label{fig:dependency-injection-heatmap}
\end{figure}
 \begin{figure}[tbp]
    \centering
    \includegraphics[width=\linewidth]{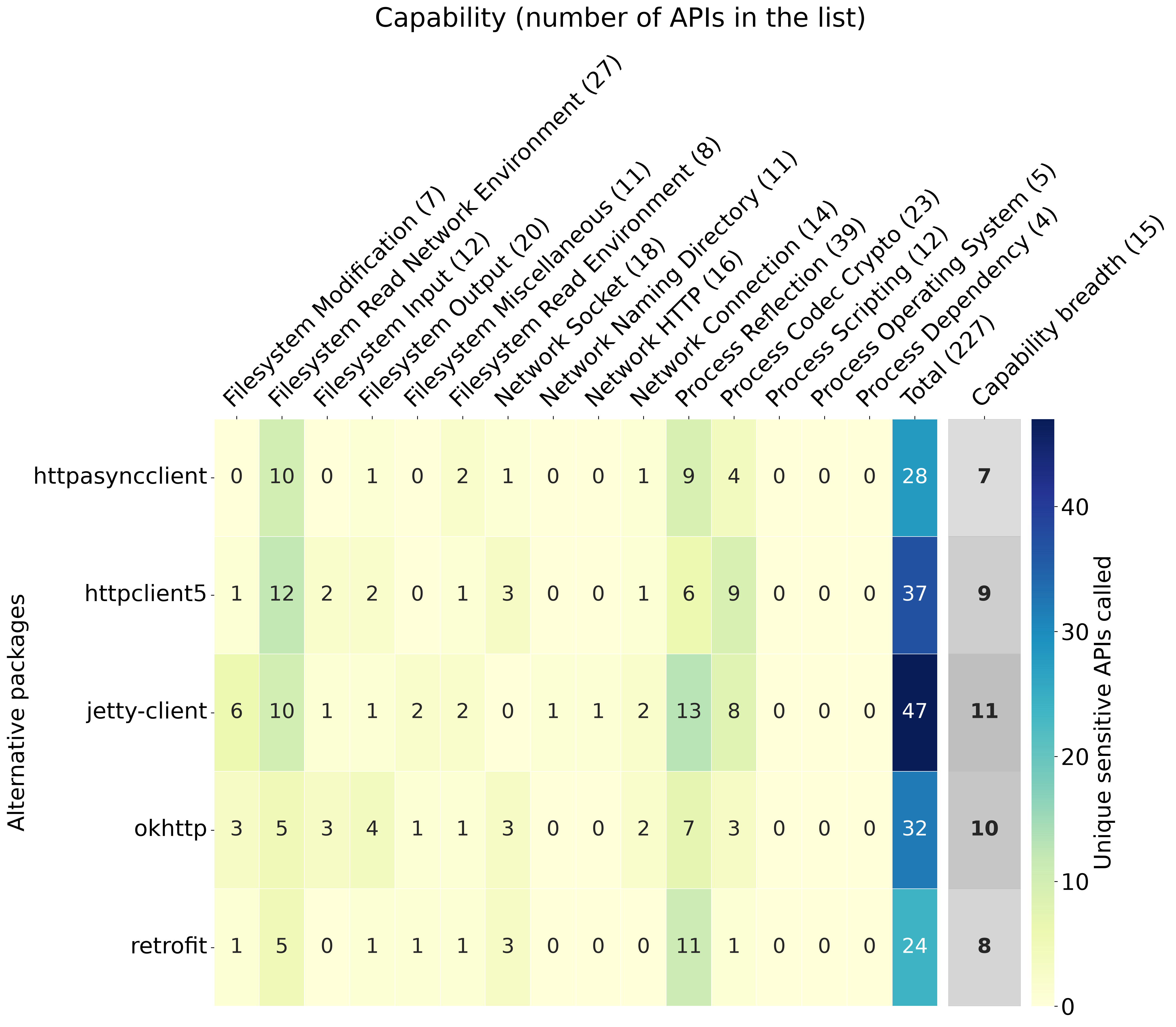}
    \caption{\rev{Security-sensitive API usage in ``HTTP Client'' package category, at the API level. Each cell is the number of unique \sensapis of a capability that the package calls in its inter-package call graph; the number in parentheses in a column header is the number of APIs of that capability in our list. The last column shows the capability breadth (capability level).}}
    \label{fig:httpclient-heatmap}
\end{figure}

\section{\rqref{rq:sen-api-diff}: How Do functionally similar Packages Differ In Terms Of Their \Sensapi Calls And Why?}
\label{sec:sen-api-diff}
In this section, we first present a heatmap of \sensapi calls made by package categories to visualize differences in \sensapi calls across similar-functioning packages.
\rev{Rather than showing a heatmap of all seven package categories, we show only two in this section, and the Appendix contains the others.
We exclude the JDBC Driver category from this section because its packages are not interchangeable; the Appendix shows it as an exploratory case.}
Figure~\ref{fig:dependency-injection-heatmap} contains the comparison for the ``Dependency Injection'' package category, and Figure~\ref{fig:httpclient-heatmap} contains the comparison for the ``HTTP Client'' package category.
We can see that for the ``Dependency Injection'' category (Figure~\ref{fig:dependency-injection-heatmap}), \pkgname{dagger} and \pkgname{jakarta.inject.api} do not make any \sensapi calls, whereas \pkgname{spring.context} \rev{calls 53 unique \sensapis (API level, as the heatmap shows) and} makes 429 calls in total \rev{(call level)}, although they provide a similar core functionality.
\rev{The last column of each heatmap shows the capability breadth, which is the number of the 15 capabilities that a package reaches.
Capability breadth separates a package that calls one capability many times from a package that can reach many different kinds of sensitive resources, and routine filesystem calls do not inflate it.
In the ``Dependency Injection'' category, \pkgname{dagger} reaches none of the 15 capabilities, while \pkgname{spring.context} reaches 10.
Most of the difference is in one capability: \pkgname{spring.context} makes 368 \apicat{process/reflection} calls (call level), and \pkgname{dagger} makes none.}
Similarly, for the ``HTTP Client'' package category (Figure~\ref{fig:httpclient-heatmap}), the total number of \sensapi calls \rev{(call level)} varies from 44 to 88\rev{, and the capability breadth (capability level) varies from 7 to 11}.
\rev{The two measures do not always agree: \pkgname{jetty-client} reaches the most capabilities (11) but makes fewer calls (73) than \pkgname{httpclient5} (88 calls, 9 capabilities).}
Moreover, \pkgname{jetty-client} makes one call to each \apicat{NETWORK/HTTP} and \apicat{NETWORK/NAMING DIRECTORY} whereas none of the other packages need to call these two capabilities.
In addition, although \pkgname{retrofit} has the lower total number of API calls in this package category (44), \pkgname{retrofit} calls \apicat{PROCESS/REFLECTION} category 31 times \rev{(call level)}, which is the highest in this package category.

We then qualitatively compare functionally similar packages (Table~\ref{tab:categories}) with respect to \sensapi calls and discuss each package category's findings as a case study.
For this analysis, we consider only the \sensapi calls from the inter-package CG of the latest version of the packages.
One author reviewed the source code, documentation, design patterns, dependency usage, implementation choices, usage examples, and overall API usage of each of the 34 packages \rev{in the seven categories (Section~\ref{sec:method}); the review does not include the JDBC Driver packages}.
The full analysis is present in the supplementary material.
Our observation of qualitative analysis on the individual packages in each category can be summarized into four patterns:

\begin{description}
    \item[1. Same Functionality, Different Implementation.]
    The expected functionality of a package is fixed (eg, HTTP spec~\cite{http2_rfc}).
    But \pkgname{dagger}, for example, has a completely different implementation, resulting in different \sensapi calls from the rest of the ``dependency injection'' packages.
   \item[2. Different Features.]
    At a high level, all web framework packages are supposed to provide the functionality of building web applications.
    However, each web framework has a slightly different feature set, making it unique.
    Similarly, JSON library \pkgname{fastjson2} provides additional features than the \pkgname{gson} and \pkgname{jackson-core}.
    Having additional features is reflected in the differences in \sensapi usage.
    \item[3. Modularity And Reuse.]
    HTTP client packages have a low usage of \apicat{network} APIs.
    The implementations are very modular.
    For example, rather than having three \apicat{network} calls for GET, POST, and PUT, \pkgname{httpclient5} uses one \api{java.lang.Socket.connect()} method in \api{ProxyClient.java} and reuse the \api{ProxyClient} class to implement other methods.
    \item[4. Using A Similar Package As Dependency.]
    Even with another HTTP client package \pkgname{okhttp} as a dependency, \pkgname{retrofit} uses fewer \sensapi calls than \pkgname{okhttp}.
    \pkgname{Retrofit} is a high-level REST abstraction built on top of \pkgname{okhttp}.
    In this setting, \pkgname{okhttp} provides low-level HTTP functionality, whereas \pkgname{retrofit} provides \textit{additional features}, e.g., URL manipulation.
    A similar pattern can be found in web frameworks as well:
    \pkgname{spring-webflow} and \pkgname{spring-webmvc} both use \pkgname{spring-web} for different use cases from the high-level. 
\end{description}

\noindent \textbf{Comparison Across Package Categories.}
\rev{This comparison covers the latest versions of the 34 packages in the seven categories of alternatives (the counts in the heatmaps); the comparison excludes the JDBC Driver category.
\apicat{Process/reflection} is the most used capability, both by the number of calls (3,044, call level) and by the number of packages that call it (28 of 34), and every package category calls it.
\apicat{Process/dependency} is the least used capability: none of the 34 packages calls it.
With the JDBC Driver packages included (39 packages), the result is the same: \apicat{process/reflection} is the most used (3,494 calls, 33 packages), and \apicat{process/dependency} is the least used (6 calls), called only by four of the five JDBC drivers.}

\highlight{black}{\textbf{Finding \#3: } Packages have varying \sensapi calls (call level: from 0 to 368 in one capability and from 0 to 429 in total), although they provide similar functionality. \rev{Capability breadth (capability level) also varies, for example from 0 to 10 in ``Dependency Injection''.} \rev{These ranges are the same with and without the exploratory JDBC Driver category.}
The differences in \sensapi calls in a package category stem from diverse implementations, additional features, modularity, and abstraction levels.
}

\begin{revblock}
\section{Robustness of the Call Graph Results}
\label{sec:cg-validation}
This section reports the call graph validation described in Section~\ref{sec:cg-validation-method}, at the API level and the capability level.
The validation covers the latest versions of the 34 packages in the seven RQ3 categories; 28 of them call at least one \sensapi in their own code, and 31 when dependencies are included.

\subsection{Over-Approximation: CHA versus RTA}
\label{sec:cg-validation-rta}
RTA is much more precise than CHA on the call graph as a whole, but this precision barely changes which \sensapis a package calls.
Compared with OPAL CHA, OPAL RTA reached 28.3\% fewer methods and produced 39.3\% fewer call edges.
However, RTA kept 96.9\% of the package--API pairs of CHA intra-package (95\% CI: 95.3--98.3\%) and 97.4\% inter-package (95\% CI: 96.9--98.0\%), and every RTA pair was also a CHA pair.
Table~\ref{tab:cg-capability} shows the result for each capability.

\begin{table}[tbp]
\revcolor
\centering
\small
\caption{Package--API pairs (API level) in the OPAL CHA and OPAL RTA call graphs of the 34 RQ3 packages (latest version of each), and the share of CHA pairs that RTA keeps. Kept (\%) is the number of RTA pairs divided by the number of CHA pairs, times 100; every RTA pair is also a CHA pair. The number in parentheses is the number of APIs of the capability in our list.}
\label{tab:cg-capability}
\begin{tabular}{l r rrr rrr}
\toprule
 & & \multicolumn{3}{c}{Intra-package} & \multicolumn{3}{c}{Inter-package} \\
\cmidrule(lr){3-5}\cmidrule(lr){6-8}
Capability (\#APIs) & & CHA & RTA & Kept (\%) & CHA & RTA & Kept (\%) \\
\midrule
F/Input & (12) & 15 & 12 & 80.0 & 80 & 72 & 90.0 \\
F/Miscellaneous & (11) & 56 & 56 & 100.0 & 153 & 153 & 100.0 \\
F/Modification & (7) & 21 & 18 & 85.7 & 97 & 97 & 100.0 \\
F/Output & (20) & 57 & 56 & 98.2 & 205 & 199 & 97.1 \\
F/Read Env & (8) & 56 & 55 & 98.2 & 119 & 118 & 99.2 \\
F/Read Network Env & (27) & 67 & 67 & 100.0 & 162 & 160 & 98.8 \\
N/Connection & (14) & 46 & 39 & 84.8 & 100 & 80 & 80.0 \\
N/Http & (16) & 24 & 23 & 95.8 & 45 & 43 & 95.6 \\
N/Naming Directory & (11) & 2 & 2 & 100.0 & 16 & 16 & 100.0 \\
N/Socket & (18) & 10 & 10 & 100.0 & 43 & 43 & 100.0 \\
P/Codec Crypto & (23) & 50 & 49 & 98.0 & 136 & 135 & 99.3 \\
P/Dependency & (4) & 0 & 0 & -- & 4 & 4 & 100.0 \\
P/Operating System & (5) & 0 & 0 & -- & 7 & 7 & 100.0 \\
P/Reflection & (39) & 262 & 258 & 98.5 & 538 & 534 & 99.3 \\
P/Scripting & (12) & 5 & 5 & 100.0 & 20 & 20 & 100.0 \\
\midrule
All & (227) & 671 & 650 & 96.9 & 1725 & 1681 & 97.4 \\
\bottomrule
\end{tabular}
\end{table}

RTA keeps a virtual edge only when the reachable code instantiates the class of the target method, and RTA keeps no edge from a method that it does not reach.
Every RTA edge is therefore also a CHA edge, and every difference between the two algorithms is an edge that only CHA has: an over-approximation by CHA.
The removed pairs concentrate on a few APIs.
\api{URLConnection.getInputStream()} accounts for 7 of the 21 removed pairs intra-package and 20 of the 44 inter-package: \api{URLConnection} is an abstract class, and RTA removes the call when no subclass is instantiated in the reachable code.
The same reason explains \api{DocumentBuilder.parse()}, which RTA removes in \pkgname{spring-web}, \pkgname{spring-webflow}, and \pkgname{tapestry-core}.
RTA also removes \api{AccessibleObject.setAccessible()} in \pkgname{fastjson2}, \pkgname{log4j-core}, and \pkgname{struts2-core}, and four file APIs in \pkgname{spring-web}, such as \api{Files.copy()} and \api{Files.delete()}.
For the five reflection APIs in Table~\ref{tab:top10}, no edge was a virtual edge (789 edges intra-package, 3,844 inter-package).
This result is consistent with the argument in Section~\ref{sec:sen-api-usage-pattern}: each of these calls has one target, so CHA resolves it exactly.
RTA can remove such an edge only when it does not reach the calling method.
RTA removed 1 and 16 of these edges, respectively, and in every case the calling method had no other \sensapi edge under RTA, which is consistent with a calling method that RTA does not reach.

RTA did not change the comparison of packages.
In every package category, the packages with the most and the fewest \sensapis were the same under CHA and RTA, in both scopes.
Spearman's $\rho$ between the numbers of unique \sensapis under CHA and RTA was 1.00 in every category, except Web Frameworks inter-package ($\rho = 0.996$).
Capability breadth changed for four packages intra-package (\pkgname{log4j-core} from 11 to 10, \pkgname{spring-web} from 10 to 8, \pkgname{spring-webflow} from 5 to 4, and \pkgname{tapestry-core} from 9 to 8) and for one package inter-package (\pkgname{log4j-core}).
\apicat{Process/reflection} remained the capability with the most package--API pairs (39.0\% under CHA and 39.7\% under RTA, intra-package).

\subsection{Tool Agreement: Endor Labs versus OPAL}
\label{sec:cg-validation-tools}
Table~\ref{tab:cg-packages} compares, for each package, the unique \sensapis that the two tools find in the package's own code.
The tools agree closely at the API level and at the capability level.
\begin{itemize}
    \item \emph{Packages without \sensapis.} The two tools agree on which packages call no \sensapi in their own code. Both tools report zero \sensapis for the same six packages (\pkgname{dagger}, \pkgname{jakarta.inject-api}, \pkgname{okio}, \pkgname{spring-boot-starter-web}, \pkgname{timber}, \pkgname{vaadin}), and both tools report at least one \sensapi for each of the other 28 packages.
    \item \emph{API sets and rankings.} The Endor Labs tool reports 583 package--API pairs and OPAL 671; 551 pairs are shared, so 32 pairs appear only under the Endor Labs tool and 120 only under OPAL. The median Jaccard index of the per-package API sets is 0.78. Separately, Spearman's $\rho$ for per-package counts of unique \sensapis is 0.99 (95\% CI 0.97--1.00), showing that package rankings are similar even though the exact API sets sometimes differ. In every category, the package with the most unique \sensapis is the same under both tools.
    \item \emph{Capabilities.} Across all $34 \times 15$ package--capability pairs, including pairs where both tools report absence, the tools agree in 95.9\% of cases (Cohen's $\kappa = 0.90$). For \apicat{process/reflection}, they agree for every package ($\kappa = 1.00$).
\end{itemize}

\begin{table}[tbp]
\revcolor
\centering
\scriptsize
\caption{Unique \sensapis (API level) and capability breadth (capability level) of the 34 RQ3 packages (latest version of each), intra-package, under the Endor Labs tool and under OPAL. The disagreement columns count the package--API pairs that only one tool reports, by label (Section~\ref{sec:cg-validation-method}): OPAL under-approx., under-approximation by OPAL; OPAL over-approx., over-approximation by OPAL CHA (virtual edges only); Endor under-approx., under-approximation by the Endor Labs tool (at least one edge that is not a virtual edge). The last two rows give the total and the mean over the 34 packages.}
\label{tab:cg-packages}
\resizebox{\linewidth}{!}{%
\begin{tabular}{ll rrr r rr rr}
\toprule
 & & \multicolumn{3}{c}{Unique APIs} & Endor only & \multicolumn{2}{c}{OPAL only} & \multicolumn{2}{c}{Capability breadth} \\
\cmidrule(lr){3-5}\cmidrule(lr){6-6}\cmidrule(lr){7-8}\cmidrule(lr){9-10}
Category & Package & Endor & OPAL CHA & OPAL RTA & OPAL under-approx. & OPAL over-approx. & Endor under-approx. & Endor & OPAL \\
\midrule
Dependency Injection & dagger & 0 & 0 & 0 & 0 & 0 & 0 & 0 & 0 \\
 & guice & 19 & 21 & 20 & 0 & 0 & 2 & 3 & 3 \\
 & jakarta.enterprise.cdi-api & 6 & 7 & 7 & 0 & 0 & 1 & 2 & 2 \\
 & jakarta.inject-api & 0 & 0 & 0 & 0 & 0 & 0 & 0 & 0 \\
 & spring-beans & 29 & 33 & 33 & 1 & 0 & 5 & 6 & 6 \\
 & spring-context & 28 & 26 & 24 & 8 & 1 & 5 & 6 & 5 \\
\midrule
HTTPClient & httpasyncclient & 4 & 6 & 6 & 0 & 0 & 2 & 2 & 3 \\
 & httpclient5 & 22 & 25 & 25 & 1 & 1 & 3 & 6 & 7 \\
 & jetty-client & 15 & 16 & 16 & 0 & 0 & 1 & 3 & 3 \\
 & okhttp & 16 & 20 & 20 & 1 & 0 & 5 & 5 & 6 \\
 & retrofit & 8 & 9 & 9 & 0 & 0 & 1 & 2 & 2 \\
\midrule
I/O Utilities & commons-io & 45 & 48 & 47 & 2 & 1 & 4 & 8 & 8 \\
 & jetty-io & 5 & 5 & 5 & 1 & 1 & 0 & 5 & 5 \\
 & okio & 0 & 0 & 0 & 0 & 0 & 0 & 0 & 0 \\
 & plexus-io & 12 & 14 & 13 & 0 & 0 & 2 & 6 & 6 \\
\midrule
JSON Libraries & fastjson2 & 30 & 40 & 39 & 0 & 1 & 9 & 7 & 8 \\
 & gson & 12 & 15 & 15 & 0 & 0 & 3 & 3 & 4 \\
 & jackson-core & 7 & 11 & 11 & 0 & 2 & 2 & 3 & 4 \\
\midrule
Logging Framework & log4j-core & 55 & 61 & 57 & 5 & 1 & 10 & 11 & 11 \\
 & logback-classic & 18 & 20 & 20 & 2 & 1 & 3 & 6 & 7 \\
 & timber & 0 & 0 & 0 & 0 & 0 & 0 & 0 & 0 \\
\midrule
Web Frameworks & jakarta.faces-api & 14 & 14 & 13 & 1 & 0 & 1 & 3 & 3 \\
 & spring-boot-starter-web & 0 & 0 & 0 & 0 & 0 & 0 & 0 & 0 \\
 & spring-web & 38 & 39 & 34 & 4 & 1 & 4 & 10 & 10 \\
 & spring-webflow & 12 & 15 & 14 & 0 & 0 & 3 & 4 & 5 \\
 & spring-webmvc & 13 & 16 & 16 & 1 & 1 & 3 & 5 & 6 \\
 & struts2-core & 58 & 64 & 62 & 2 & 1 & 7 & 10 & 10 \\
 & tapestry-core & 21 & 25 & 24 & 1 & 1 & 4 & 9 & 9 \\
 & vaadin & 0 & 0 & 0 & 0 & 0 & 0 & 0 & 0 \\
 & wicket-core & 37 & 46 & 45 & 1 & 1 & 9 & 10 & 10 \\
\midrule
XML Processing & dom4j & 10 & 14 & 14 & 0 & 0 & 4 & 4 & 5 \\
 & jakarta.xml.bind-api & 14 & 18 & 18 & 0 & 0 & 4 & 4 & 6 \\
 & jaxb-api & 15 & 20 & 20 & 0 & 0 & 5 & 4 & 6 \\
 & xstream & 20 & 23 & 23 & 1 & 0 & 4 & 4 & 4 \\
\midrule
\multicolumn{2}{l}{Total (34 packages)} & 583 & 671 & 650 & 32 & 14 & 106 & -- & -- \\
\multicolumn{2}{l}{Mean per package} & 17.1 & 19.7 & 19.1 & 0.9 & 0.4 & 3.1 & 4.4 & 4.8 \\
\bottomrule
\end{tabular}%
}
\end{table}

We labeled the 152 package--API pairs that only one tool reports (32 + 120; Table~\ref{tab:cg-packages}). As Section~\ref{sec:cg-validation-method} describes, we inspected the package bytecode with \texttt{javap} to check whether the call instruction is present. The labels classify the differences between the tool outputs, but the labels do not show that the call executes at runtime.
\begin{itemize}
    \item \emph{OPAL over-approx.: over-approximation by OPAL CHA (14 pairs).} OPAL's only edges to the API are virtual edges. In 12 packages, OPAL resolves a call to \api{OutputStream.write()} to \api{FileOutputStream.write()}, so it reports \apicat{Filesystem/Output} although the code writes to a stream of unknown type. The other two cases are \api{DataOutput.write()} resolved to \api{RandomAccessFile.write()} in \pkgname{jackson-core}, and \api{Context.lookup()} resolved to \api{InitialContext.lookup()} in \pkgname{spring-context}.
    \item \emph{Endor under-approx.: under-approximation by the Endor Labs tool (106 pairs).} OPAL reports a call in the package's bytecode that reaches the API without a virtual edge, but the Endor Labs tool does not report it. Most of these pairs (86) are constructor calls, such as \api{new File(...)} and \api{new URL(...)}. Four are calls on the package's own class loader that inherit a JDK method, such as \api{defineClass()} in \pkgname{guice}, \pkgname{spring-context}, and \pkgname{fastjson2}. The others include \api{Class.getMethod()} in \pkgname{logback-classic} and \pkgname{tapestry-core}, and \api{Paths.get()} in \pkgname{commons-io}, \pkgname{log4j-core}, and \pkgname{wicket-core}.
    \item \emph{OPAL under-approx.: under-approximation by OPAL (32 pairs).} Only the Endor Labs tool reports the API, and in all 32 cases the package's bytecode contains the call instruction, but OPAL has no edge for it. Examples are \api{Context.bind()}, \api{Context.lookup()}, and \api{ScriptEngine.eval()} in \pkgname{spring-context}, \api{HttpURLConnection.connect()} in \pkgname{log4j-core} and \pkgname{spring-web}, and \api{DocumentBuilder.parse()} in \pkgname{spring-beans}, \pkgname{log4j-core}, and \pkgname{xstream}. Eight of the 32 calls target JDK interface methods, such as \api{Context.bind()} and \api{ScriptEngine.eval()}, which OPAL does not resolve because it does not analyze the JDK implementation.
\end{itemize}
No pair is an over-approximation by the Endor Labs tool: every API that only the Endor Labs tool reports is called in the bytecode of the package.

These labels explain every difference in capability breadth (the two capability breadth columns of Table~\ref{tab:cg-packages}).
Capability breadth is the same under both tools for 21 of the 34 packages.
Capability breadth is greater under the Endor Labs tool for one package: \pkgname{spring-context} reaches \apicat{Filesystem/Output} only under the Endor Labs tool, through \api{Files.createFile()} and \api{Files.writeString()}, which OPAL misses (under-approximation by OPAL).
Capability breadth is smaller under the Endor Labs tool for 12 packages, for two reasons:
\begin{itemize}
    \item \emph{Over-approximation by OPAL CHA.} The virtual edge to \api{FileOutputStream.write()} adds \apicat{Filesystem/Output} in \pkgname{httpclient5}, \pkgname{logback-classic}, and \pkgname{spring-webmvc}.
    \item \emph{Under-approximation by the Endor Labs tool.} A constructor call adds a capability in \pkgname{httpasyncclient} (\api{new RandomAccessFile(...)}, \apicat{Filesystem/Output}), \pkgname{okhttp} and \pkgname{gson} (\api{new URL(...)}, \apicat{Network/Connection}), \pkgname{spring-webflow} (\api{new File(...)}, \apicat{Filesystem/Miscellaneous}), \pkgname{dom4j} (\api{new FileOutputStream(...)}, \apicat{Filesystem/Output}), and \pkgname{jakarta.xml.bind-api} and \pkgname{jaxb-api} (\api{new File(...)} and \api{new URI(...)}).
\end{itemize}
In \pkgname{fastjson2} and \pkgname{jackson-core}, both reasons add \apicat{Filesystem/Output}.

\subsection{Effect on the Findings}
The validation supports the findings of RQ2 and RQ3 as statements about which \sensapis and capabilities the analyzed call graphs report and how functionally similar packages compare. The validation does not establish runtime reachability for every reported API.
\begin{itemize}
    \item \emph{The over-approximation that RTA removes is small.} RTA, a more precise algorithm, removes 3.1\% of the package--API pairs intra-package and 2.6\% inter-package, and it does not change any comparison between packages. RTA is also an over-approximation, so these figures are a lower bound on the over-approximation of CHA. The over-approximation that RTA finds is therefore too small to change any finding of RQ2 or RQ3. The validation cannot measure the over-approximation that RTA does not find.
    \item \emph{Effect of dependencies (Finding \#2).} Under every call graph, dependencies increase the number of \sensapis. On the 34 packages, at the API level, the number of package--API pairs increases from intra-package to inter-package by a factor of 1.8 under the Endor Labs tool (583 to 1,065) and 2.6 under OPAL CHA (671 to 1,725) and OPAL RTA (650 to 1,681). The OPAL factor is greater partly because OPAL treats every public method of the dependencies as an entry point (Section~\ref{sec:cg-validation-method}).
    \item \emph{Reflection.} \apicat{Process/reflection} is the capability with the most package--API pairs under every call graph (Endor Labs, OPAL CHA, and OPAL RTA, in both scopes), with a share of all package--API pairs between 31.2\% and 41.7\%.
    \item \emph{Differences between alternatives (Finding \#3).} Under every call graph, packages in the same category differ substantially, and the packages without any \sensapi are the same. For example, in Dependency Injection, the number of unique \sensapis ranges from 0 to 29 under the Endor Labs tool and from 0 to 33 under OPAL CHA (intra-package).
\end{itemize}

\highlight{black}{\textbf{Call graph validation: } At the API level, replacing CHA with RTA keeps 96.9\% (intra-package) and 97.4\% (inter-package) of the package--API pairs, and it does not change the ranking of packages in any category. An independent, open-source call graph generator agrees with the Endor Labs tool on which \sensapis ($\rho = 0.99$) and which capabilities ($\kappa = 0.90$) each package reaches. We labeled every remaining difference as an over-approximation by OPAL CHA or as an under-approximation by one of the two tools.}
\end{revblock}

\section{\rqref{rq:usefulness-of-sen-api}: To what extent would prior knowledge of \sensapi information influence developers' dependency selection decisions?}
\label{sec:dev-survey}
Our survey aims to understand how \sensapi information would influence developers' dependency selection.
We received valid responses from 110 developers (response rate: 1.4\%).
\rev{Because of the low response rate and possible self-selection (Section~\ref{sec:threats}), the results in this section describe our respondents, not the developer population.}

\noindent \textbf{If you had known this \sensapi information before choosing the dependency, how would it affect your decision?}
We found that 44.6\% of respondents said they would consider this information in a dependency selection process: 20\% said they would keep using the chosen dependency but factor \sensapi information into future decisions, 19.1\% said they would select the dependency with the lowest \sensapi call count, and 5.5\% said they would select based on calls to a specific capability. These are self-reported responses to the survey scenario, not observed choices.

We also found that 41.8\% of respondents said they would not have decided differently based solely on the \sensapi information. Most of them either think that functionality is still much more important or that the \sensapis in the package are necessary for it to do what it does, as \user{P7} mentioned: \qt{I would assume it does what it does for a reason.}
The rest (13.6\%) answered ``Others''.
Of these answers, 7 out of 15 still indicated that they would consider the \sensapi information in their decision-making process.
\rev{If we also count these 7 answers, which we recoded as indicating consideration, the share is 50.9\% (56 of 110); the directly measured value is 44.6\%.
The replication package contains the 15 ``Others'' answers, anonymized, together with our recoding decisions.}

To understand the reasons behind developers' perspectives, we qualitatively coded the open-ended survey responses.
In this section, the given themes are reported in \textbf{bold}, with the percentage of respondents' answers corresponding to the code.
We present the four high-level themes that appeared in our analysis:

\begin{description}
    \item[1. Security and Risk.] Participants are generally aware of \textbf{security context} ($45\%$) when responding to \sensapis.
    Participants are also aware of the \textbf{security risk with \sensapi} usage in dependencies ($27\%$).
    Few responses (9\%) mentioned \textbf{actual vulnerability}, giving more context to the necessity of the \sensapi information.
    Participants are also aware that \textbf{less API is better} ($12\%$).
    \user{P1} mentioned, \qt{Obviously[,] I want to minimize the attack vector as much as possible[;] therefore[,] the dependency with the least security sensitivity wins.}
    On the other hand, a few participants mentioned they \textbf{trust their dependencies} ($4\%$).
    \user{P2} mentioned, \qt{I would assume it does what it does for a reason.}
    Other participants suggested using \textbf{access control or sandboxing or containerizing} on the utility of \sensapi information. 
 
    \item[2. Reported decision and action.] Some participants described an action or decision prompted by the survey ($9\%$), including noticing an unused dependency. These reports illustrate possible uses of dependency information; they do not show that API information changed a real dependency choice.
    \user{P3} described, \qt{This survey also prompted me to realize that the dependency on \textit{guice} can be removed from this project. It doesn't use \textit{guice} anymore, but we forgot to remove the dependency from the build.}
    Participants mentioned to \textbf{use \sensapi in future decision} ($14\%$).
    \user{P4} mentioned, \qt{It would be another factor considered in the choice of the library.}
    However, few participants mentioned \textbf{\sensapis in choosing dependency, not changing}.
    Few participants mentioned they are not in the position of \textbf{making dependencies decision} ($5\%$).

    \item[3. Technical or Implementation.] Participants mentioned their concerns with \textbf{reachability}, \textbf{compatibility}, \textbf{functionality}, \textbf{specific use case} or \textbf{lack of alternatives} in regards of \sensapis.
    \user{P5} mentioned, \qt{I would not choose only because of security-sensitive API information. Functionality is very great factor to consider. I may choose the the one which has higher security-sensitive API if I don't use those parts of it and rely for something else. So it depends on the usage and the sensitivity.}
    \user{P6} described compatibility, \qt{[\ldots] Though in the case of Spring, it would be weird to have chosen another. Given one goal of Spring is its consistency across packages.}
    \user{P7} mentioned, \qt{Dependency selection is often a tradeoff between capabilities and concerns. Sometimes there are not many alternatives, or the alternatives have significant implementation challenges. However, understanding the security impact of a dependency is a vital part of the selection process.}
    \user{P8} described unavailability, \qt{Sensitive API calls information is not so broadly available, so I am not planning to factor it.}

    \item[4. Reaction or Attitude.] Few participants were \textbf{surprised} to see our result, \user{P15} \qt{[\ldots] I never thought jackson-core was interacting with the network for instance.}
    In addition, a few participants were negative towards \sensapis, \user{P9} \qt{Because I trust the authors of the package are doing what needs to get done to accomplish the functionality the package is supposed to deliver on. [\ldots]}
\end{description}

\highlight{black}{\textbf{Finding \#4: } 
\rev{44.6\% of the 110 respondents said that they would consider \sensapi information in their dependency selection process.}}

\section{Implications}
\label{sec:discuss}

We discuss the implications of our study's results in this section.
\rev{Note that, the goal of our study is to show the importance of visibility into dependencies through security sensitive API calls.}

\paragraph{For Practitioners}
\rev{Our respondents} appear to be generally aware of risks with \sensapis and risks with vulnerabilities in dependencies.
However, greater awareness of \sensapis and their potential implications will help developers in their dependency selection process.
As evidenced by the survey, our study can complement ${ZTD}_{JAVA}$~\cite{amusuoZTD$_JAVA$MitigatingSoftware2025}.
\textbf{Our results can give developers visibility into potentially sensitive resource use by a dependency (\rqref{rq:sen-api-usage-pattern} and~\rqref{rq:sen-api-diff}). If they use a dependency whose capabilities are unnecessary for its role, they could use a permission system such as ${ZTD}_{JAVA}$ to restrict access.}

\rev{In our survey, 44.6\% of respondents said they would consider \sensapi calls when selecting packages (\rqref{rq:usefulness-of-sen-api}); the survey measured stated interest, not observed selection outcomes.}
Incorporating this data alongside traditional selection criteria (e.g., functionality, popularity, ease of use) can help developers \rev{see which sensitive resources a dependency can reach} during the dependency selection process.
\rev{Our results show that packages in the same functional category can differ by up to 429 call-graph edges to listed APIs (call level) and by up to 13 of the 15 capabilities (capability level) (\rqref{rq:sen-api-diff}). These comparisons give developers information to consider alongside functionality and other selection criteria; the comparisons do not establish that a package with fewer calls or capabilities is safer.}
If developers opt for a dependency that makes excessive calls to any resource or capability due to the specific use case, they should be aware of the potential vulnerability classes associated with the \sensapi (our API-to-CWE mapping).
Knowing the potential risk might help developers mitigate it more quickly if such a vulnerability arises.
Overall, \textbf{we advocate developers to use \sensapis for dependency selection}.

\paragraph{For Tool Builders}
Our motivation for the lack of tools providing \sensapi information is reflected in the developer's response.
To bridge the gap, we contribute an open-source tool that takes a package call graph as input and generates comparative visualizations of \sensapi calls for those packages.
Regardless of how the call graph is generated (e.g., with or without dependencies), our tool can compare the \sensapi calls across the provided packages.
Our tool can be integrated in four ways: (a) as a standalone tool, (b) as a plugin in IDEs, (c) as a plugin in package management tools (e.g., Maven, Gradle), and (d) as a part of a dependency management/checking tool (e.g., OpenSSF Scorecard~\cite{zahanOpenSSFScorecardPath2023}).
Several developers stated that the way (c) and (d) would be more useful (\rqref{rq:usefulness-of-sen-api}), as stated by \user{P9} \qt{I would not run a tool to decide this manually\textemdash{}too much effort\textemdash{}but if it was auto-provided in the PR say\textemdash{}then it's a small nice to have.}
\textbf{We recommend integrating the tool into package management tools (e.g., npm cli, OpenSSF Scorecard) so developers can quickly compare and identify the most secure option.}

\paragraph{For Researchers}
To ensure that dependencies remain secure throughout the development lifecycle, \sensapi analysis should be integrated into CI/CD pipelines.
Regular security audits can help track and mitigate any introduction of risky dependencies or changes in \sensapi usage over time, especially with transitive dependencies.
\textbf{Our tool can be used to automatically audit the delta in used \sensapis between two versions of the same dependency, which can be helpful in an already-deployed system.}
We envision future research that further explores this type of longitudinal analysis of packages' \sensapi calls to chart the evolution of the attack surface over time.
In addition, future research should also explore whether such a delta can be used to identify malicious package updates.

\section{Ethical Considerations}
\label{sec:ethics}
We have followed the ACM Code of Ethics and Professional Conduct~\cite{acm-code-of-ethics} in our research.
We obtained IRB approval from our institution (\textit{North Carolina State University IRB 27712}) before conducting the survey.
We did not collect any personally identifiable information or demographic information from the survey.
We semi-automate email sending, ensuring that people do not receive more than one email or an email about a project they did not contribute to.
We attached the consent form with our initial email to the participants and informed them that the survey is voluntary and they can withdraw at any time.
No new vulnerabilities were introduced during our research, so no disclosures were made.
Our contact procedure focused on informing the responsible parties about the \sensapi calls in their chosen dependencies.
We mostly received positive responses and stopped further recruitment after receiving the first negative response.
After this study, we became aware of the extent of GitHub's terms of service~\cite{github_tos}, which restrict the collection of users' email addresses for academic research, and we chose to discontinue this type of recruitment for future research.

\section{Threats To Validity}
\label{sec:threats}
In this section, we present the threats to validity to our study.

\paragraph{Generalizability and Coverage}
We use CVEs and CWEs from the 45 packages in our chosen Maven repositories as a reference for our analysis to determine a list of \sensapis.
We acknowledge that this list of \sensapis may not reflect the whole population of (possibly) \sensapis.
However, during our analysis, we found that (1) not all CVE patches and CWEs have related \sensapis, and (2) the more CVEs and CWEs we analyzed, the more our list gets saturated with repeating \sensapi.
Our list of 227 \sensapis provides an overview of these API calls in the studied Java packages and can inform developers' consideration of open-source dependencies.
In addition, we did not consider CVE severity (e.g., CVSS score) in our analysis, but CVE severity may affect developers' decisions on selecting a dependency.

\paragraph{API List Construction Method.}
We opt not to consider only the APIs used by the vulnerable functions (from our chosen packages) sensitive.
Considering all APIs called from vulnerable functions as sensitive would be an overapproximation.
Thus, we constructed our \sensapi list through expert selection (based on manual inspection of JavaDoc, CVE patches, and CWE examples).
To address potential threats to construct validity, the \sensapi list was developed collaboratively with domain experts, ensuring that only APIs with recognized security impacts were included.
\rev{The CWE-example step used LLM-based tools, whose output can change between runs and can contain APIs that do not exist.
We reduced this threat in two ways: we kept only the APIs that both Gemini and ChatGPT returned, and we checked every candidate manually against the JDK documentation.
In addition, we checked the 201 JDK APIs of the final list against the JDK~11 class files; this check found seven misspelled names in the list in Appendix~\ref{sec:full-api-list}, which we corrected, and all 201 names now exist in JDK~11 (the other 26 APIs belong to the Servlet and WebSocket APIs of Java EE and Jakarta EE).
We also repeated the prompts five times for each of the 12 CWEs to which this step applied: both tools returned all 26 APIs of this step in every run, and the raw outputs contained no API that does not exist in the JDK.}

\begin{revblock}
\paragraph{Dependence on Endor Labs}
Two inputs of our study come from our industry partner, Endor Labs: the call graphs (Section~\ref{sec:cg-validation-method}) and the manual verification of the vulnerable functions of the 255 CVEs (Section~\ref{sec:cve-cwe-collection-mapping-impl}).
Both steps are proprietary, so we release their outputs: the replication package contains the \sensapi call-graph edges, the \sensapi call tables for each package version, and the verified vulnerable functions of the 255 CVEs.
The call graph validation in Section~\ref{sec:cg-validation} uses only open-source tools, and the replication package contains its OPAL configurations, scripts, and outputs, so others can run it from bytecode to results without Endor Labs tools.
\end{revblock}

\paragraph{Call Graph Construction and Analysis}
\rev{Our call graphs come from a third-party tool from Endor Labs~\cite{endor-labs}, which uses Class Hierarchy Analysis (CHA). At a virtual call site in Java bytecode, CHA creates an edge from the calling method to every matching method implementation in any Java archive of the package's dependency graph.
Like most program analyses used in practice, the analysis is neither sound nor complete~\cite{livshitsDefenseSoundinessManifesto2015,reifJudgeIdentifyingUnderstanding2019}, and it can both over-approximate and under-approximate runtime reachability.
First, CHA \emph{over-approximates}: it can include edges that no execution takes.
Our validation (Section~\ref{sec:cg-validation}) compared OPAL CHA with OPAL RTA: at the API level, RTA retained 96.9\% (intra-package) and 97.4\% (inter-package) of the package--API pairs found by CHA. This comparison estimates the effect of changing algorithms on our analyzed packages; the comparison does not measure net error against runtime behavior.
Second, the analysis \emph{under-approximates}: it may miss \sensapis that a package reaches only as the target of a reflective call, a dynamic proxy, or a class loaded at run time. This does not affect the calls \emph{to} the reflection APIs that we count, which are direct calls in the bytecode. Because the analysis can also add unreachable edges, the net direction of error in the overall counts is unknown; the reported results describe APIs found by our static analysis and may omit dynamically reached APIs.
Our analysis also does not check how a package uses a \sensapi, for example with which arguments or in which execution context, so a call does not by itself mean a security risk; a finer analysis, such as a data-flow analysis of the arguments, could separate benign calls from risky ones, and we leave it to future work.
We did not measure the number of omitted reflective targets. Measuring them would require dynamic analysis with client applications, which are not available for the library versions in our dataset. Prior work on application code found a median recall of 0.88 for static call graphs~\cite{suiRecallStaticCall2020}; that result should not be taken as an estimate for our packages.
The validation has three further limits.
The validation covers the latest versions of the 34 RQ3 packages, not a random sample of the 30,772 versions of RQ2.
We compared the Endor Labs tool and OPAL only in the intra-package scope, because the two tools define the inter-package scope differently; in this scope, they agree on which \sensapis each package calls (Spearman's $\rho = 0.99$), and we labeled every remaining difference.
Finally, OPAL 5.0.0 reported that some Java 21 class files exceed the class-file versions it supports; we used the same JDK and the same OPAL version for both algorithms.}
In addition, some package versions in our dataset lacked source code, leading the call graph generator to produce no output.

\paragraph{Survey Limitation}
The generalization of our findings beyond the pool of participants should be made with caution since this is a common limitation of works of this kind~\cite{millerWeFeelWere2023}.
The developers who participated in the survey might not have represented the entire developer population.
\rev{The response rate was 1.4\% (110 of 7,785 contacted developers), so the survey results describe our respondents only.}
Developers participating in an API study might be more knowledgeable/aware of APIs than the average (self-selection bias~\cite{rogelbergProfilingActivePassive2003,marcusWhoArePeople2005}).
Our study may be affected by social desirability bias~\cite{kreuterSocialDesirabilityBias2008}, similar to other studies~\cite{redmilesHowWellMy2019,wermke2022committed,gollaWhatWasThat2018}.
Participants' reported reactions might differ from what they would have done in real life.
Inspired by previous studies and survey best practices, we sought to minimize social desirability bias through iterative pre-testing and softening language.
The survey result may also be affected by how we explain the problem (survey clarity) and present the alternatives (visualization clarity).
To ensure our survey and visualization are clear, we conducted a pilot and modified our survey accordingly.

\section{Related Work}
\label{sec:related_works}
In this section, we discuss the related works to our study.

\paragraph{Capability Analysis, Sensitive API, and Permissions} 
Our study is inspired by previous capability analysis research, which aims to understand a program's capabilities (permissions) and detect any malicious behavior.
\textsc{bincapz}~\cite{bincapz} and \textsc{malcontent}~\cite{malcontent} by Chainguard, \textsc{capslock} by Google~\cite{capslock}, and \textsc{AppInspector} by Microsoft are tools *(rule-based) that can find the capabilities from binary or source code of a certain language.
Similarly, Gorla et al.~\cite{gorlaCheckingAppBehavior2014} use the term \emph{sensitive API} to find the capabilities of an Android application and possibly anomalous behavior. We built on a similar term as Gorla et al., but focused more on \emph{security}-sensitive APIs and the Maven ecosystem, as this ecosystem has also been found to have supply chain issues~\cite{log4shell,equifax-data-breach-article}.

The closest work to ours is that of Hermann et al.~\cite{hermann2015getting}. They looked at the Java Class Library (JCL) to identify usage of system capabilities, manually assigned capability markers, and used call graph analysis to determine whether a capability was reached. Our work presents a more comprehensive methodology for identifying these capabilities, which we call sensitive APIs, and analyzes their relationship with actual security vulnerabilities.
\rev{Table~\ref{tab:related-comparison} compares our work with the closest prior work and tools.
The main difference is the derivation of the \sensapi set: prior work uses a set selected by expert judgment, while we define a repeatable procedure that others can apply in any ecosystem and validate the result against real vulnerabilities.}

\begin{table*}[tbp]
\revcolor
\centering
\caption{Comparison of our work with prior work on capabilities and \sensapis.}
\label{tab:related-comparison}
\resizebox{\textwidth}{!}{%
\begin{tabular}{l l c c l c}
\toprule
\textbf{Work} & \makecell[l]{\textbf{Derivation of the \sensapi}\\\textbf{or capability set}} & \textbf{Repeatable} & \makecell{\textbf{Validated against}\\\textbf{vulnerabilities}} & \textbf{Scope} & \makecell{\textbf{Compares similar}\\\textbf{alternatives}} \\
\midrule
Hermann et al.~\cite{hermann2015getting} & \makecell[l]{Manual capability markers\\on native methods of the JCL} & \xmark & \xmark & \makecell[l]{JCL and Java libraries} & \xmark \\
\addlinespace
Ferreira et al.~\cite{ferreiraContainingMaliciousPackage2021} & \makecell[l]{Expert-defined permission categories\\(network, file system, process)} & \xmark & \xmark & \makecell[l]{npm packages and\\their dependencies} & \xmark \\
\addlinespace
\textsc{capslock}~\cite{capslock} & \makecell[l]{Curated mapping of Go standard\\library functions to capabilities} & \xmark & \xmark & \makecell[l]{Go packages and\\their dependencies} & \xmark \\
\addlinespace
\makecell[l]{\textsc{bincapz}/\textsc{malcontent}~\cite{bincapz,malcontent}} & \makecell[l]{Hand-written YARA rules} & \xmark & \xmark & \makecell[l]{Single binaries\\and scripts} & \xmark \\
\addlinespace
\textsc{ZTD}$_{JAVA}$~\cite{amusuoZTD$_JAVA$MitigatingSoftware2025} & \makecell[l]{Expert-defined permission types} & \xmark & \xmark & \makecell[l]{Java applications and\\their dependencies} & \xmark \\
\midrule
\textbf{This work} & \makecell[l]{Three-pronged procedure: ecosystem API\\documentation, CVE fixes, and CWE examples} & \cmark & \cmark & \makecell[l]{Maven packages and\\their dependencies} & \cmark \\
\bottomrule
\end{tabular}%
}
\end{table*}

Another way to control program capability is by enforcing permission to comply with certain policies~\cite{amusuoZTD$_JAVA$MitigatingSoftware2025, vasilakisPreventingDynamicLibrary2021,ferreiraContainingMaliciousPackage2021, ntousakisDemoDetectingThirdParty2021,ohmYouCanRun2023,wyssWolfDoorPreventing2022}.
Ferreira et al.~\cite{ferreiraContainingMaliciousPackage2021} proposed a lightweight permission system that protects applications from malicious updates of direct and transitive dependencies.
The permissions are categorized as \apicat{network access}, \apicat{file system access}, \apicat{process creation}, and \apicat{all}.
Our security-sensitive resource types are inspired by the first three. Our work complements these works on permissions by providing a list of \sensapis, analyzing their relationships with real security vulnerabilities, and examining their prevalence in open-source libraries.

\paragraph{Maven Supply Chain Security}
While there is a lack of capability studies on the Java/Maven ecosystem, there are several studies that analyze supply chain security in this ecosystem~\cite{wuUnderstandingThreatsUpstream2023, pashchenkoVuln4RealMethodologyCounting2022, mirEffectTransitivityGranularity2023}.
Wu et al.~\cite{wuUnderstandingThreatsUpstream2023} analyzed 50 million invocations from downstream to upstream projects and found that the most vulnerable functions (in the upstream) cannot be accessed from the downstream projects.
This result aligns with that of Mir et al.~\cite{mirEffectTransitivityGranularity2023}: less than 1\% of packages have a reachable call path to the vulnerable code in their dependencies.
Pashchenko et al.~\cite{pashchenkoVuln4RealMethodologyCounting2022} proposed a methodology to reduce false alerts of supply chain vulnerability named Vuln4Real. 
Zhang et al.~\cite{zhangMitigatingPersistenceOpenSource2023} \rev{studied why known vulnerabilities persist in the Maven ecosystem after a fix is available.}
While vulnerability analysis accounts for most research on supply chain security, breaking changes have also attracted research interest, as they are often associated with security patches~\cite{ochoaBreakingBadSemantic2022, raemaekersSemanticVersioningBreaking2014, jayasuriyaUnderstandingBreakingChanges2023}.
These studies highlight the importance of supply chain security in the Maven ecosystem, which becomes one of our motivations in this work.

\paragraph{Program Analysis for Security} 
We chose to use call graph analysis in our study as program analysis has been used in the SOTA for security purposes~\cite{hermann2015getting, keshaniFrankensteinFastLightweight2023, keshaniScalableCallGraph2021, mirEffectTransitivityGranularity2023}.
Call graph analysis is one of the most popular methods in program analysis.
Nielson et al.~\cite{nielsenModularCallGraph2021} presented JAM, a precise call-graph construction method for JavaScript programs used in security scanning.
In Java, there are Keshani et al.~\cite{keshaniFrankensteinFastLightweight2023, keshaniScalableCallGraph2021}
and Mir et al.~\cite{mirEffectTransitivityGranularity2023}, who are using call graph analysis for supply chain security.
Another common program analysis used for security purposes is taint analysis~\cite{staicuExtractingTaintSpecifications2020, muraleeARGUSFrameworkStaged2023, shcherbakovSilentSpringPrototype2023, schwartzAllYouEver2010}.
\rev{Staicu et al.~\cite{staicuExtractingTaintSpecifications2020} presented a scalable specification extraction technique for JavaScript libraries that builds on a dynamic taint analysis and uses existing test suites.
Muralee et al.~\cite{muraleeARGUSFrameworkStaged2023} presented a staged static taint analysis of GitHub workflows and actions.
Shcherbakov et al.~\cite{shcherbakovSilentSpringPrototype2023} showed how prototype pollution in Node.js can lead to remote code execution.
Schwartz et al.~\cite{schwartzAllYouEver2010} described dynamic taint analysis and forward symbolic execution for security.}
We acknowledge the possibility of unsoundness of our call graph in the threat to validity section (\S\ref{sec:threats}).
To avoid weaknesses of conventional program analysis, several studies prefer to use hybrid program analysis~\cite{duanMeasuringSupplyChain2021, liMiningNodeJs2022, vasilakisEfficientModulelevelDynamic2021, ntousakisDemoDetectingThirdParty2021}, anti-analysis techniques~\cite{jungDefeatingProgramAnalysis2021}, or large language models~\cite{liHitchhikerGuideProgram2023, xuLmPaImprovingDecompilation2023}.

\paragraph{Developers Study in Supply Chain}
While no study has explored the usefulness of understanding \sensapi for dependency selection, several studies have been done with developers to understand how they choose/update their dependencies and why~\cite{mujahidWhatAreCharacteristics2023, liExploringFactorsMetrics2022, lariosvargasSelectingThirdpartyLibraries2020,heAutomatingDependencyUpdates2023}.
From a security perspective, several studies have been conducted with developers to understand the relationship between dependency management and security~\cite{pashchenkoQualitativeStudyDependency2020a, wermkeAlwaysContributeBack2023}.
Among the SOTA studies, the closest to our developer study is that by Mujahid et al.~\cite{mujahidWhereGoNow2023}. They also presented some package alternatives to the developers and asked if they would consider them.
The difference with ours is that (1) they surveyed JavaScript/npm developers (we surveyed Java developers) and (2) they compared packages based on how they were declining, while we compared packages based on their \sensapi calls.

\section{Conclusion} %
\label{sec:concl}
\rev{In this paper, we present a repeatable methodology to construct a \sensapi list for an ecosystem, and we use the Java list to give visibility into the sensitive resources that a dependency can reach.
Our static analysis shows that functionally similar packages differ in the \sensapis and the capabilities that their call graphs reach.
This visibility can help developers compare alternatives, review the calls of a dependency, or restrict them with a permission system.
Our study does not show that this visibility reduces the attack surface of software or changes real-world dependency choices: the call graphs measure static reachability, and the survey measured the stated intentions of 110 respondents in a hypothetical scenario.
We envision that tool builders will integrate \sensapi information into dependency management tools, and future work can measure whether developers who use this information select different dependencies.}

\bibliographystyle{IEEEtran}
\bibliography{additional,websites,zotero}

\begin{thebibliography}{10}
\providecommand{\url}[1]{#1}
\csname url@samestyle\endcsname
\providecommand{\newblock}{\relax}
\providecommand{\bibinfo}[2]{#2}
\providecommand{\BIBentrySTDinterwordspacing}{\spaceskip=0pt\relax}
\providecommand{\BIBentryALTinterwordstretchfactor}{4}
\providecommand{\BIBentryALTinterwordspacing}{\spaceskip=\fontdimen2\font plus
\BIBentryALTinterwordstretchfactor\fontdimen3\font minus
  \fontdimen4\font\relax}
\providecommand{\BIBforeignlanguage}[2]{{%
\expandafter\ifx\csname l@#1\endcsname\relax
\typeout{** WARNING: IEEEtran.bst: No hyphenation pattern has been}%
\typeout{** loaded for the language `#1'. Using the pattern for}%
\typeout{** the default language instead.}%
\else
\language=\csname l@#1\endcsname
\fi
#2}}
\providecommand{\BIBdecl}{\relax}
\BIBdecl

\bibitem{cisa-log4j}
{CISA}, ``Review of the {D}ecember 2021 {L}og4j {E}vent,''
  \url{https://www.cisa.gov/sites/default/files/publications/CSRB-Report-on-Log4-July-11-2022_508.pdf},
  2022, last accessed: 11-Mar-2026.

\bibitem{eversonLog4shellRedefiningWeb2022}
D.~Everson, L.~Cheng, and Z.~Zhang, ``Log4shell: {{Redefining}} the {{Web
  Attack Surface}},'' in \emph{Proceedings 2022 {{Workshop}} on
  {{Measurements}}, {{Attacks}}, and {{Defenses}} for the {{Web}}}.\hskip 1em
  plus 0.5em minus 0.4em\relax San Diego, CA, USA: Internet Society, 2022.

\bibitem{cisa-log4j-mitigation}
{CISA}, ``{A}pache {L}og4j {V}ulnerability {G}uidance,''
  \url{https://www.cisa.gov/news-events/news/apache-log4j-vulnerability-guidance},
  2022, last accessed: 11-Mar-2026.

\bibitem{yuMaltrackerFineGrainedNPM2024}
Z.~Yu, M.~Wen, X.~Guo, and H.~Jin, ``Maltracker: {{A Fine-Grained NPM Malware
  Tracker Copiloted}} by {{LLM-Enhanced Dataset}},'' in \emph{Proceedings of
  the 33rd {{ACM SIGSOFT International Symposium}} on {{Software Testing}} and
  {{Analysis}}}, ser. {{ISSTA}} 2024.\hskip 1em plus 0.5em minus 0.4em\relax
  New York, NY, USA: Association for Computing Machinery, Sep. 2024, pp.
  1759--1771.

\bibitem{amusuoZTD$_JAVA$MitigatingSoftware2025}
P.~C. Amusuo, K.~A. Robinson, T.~Singla, H.~Peng, A.~Machiry,
  S.~{Torres-Arias}, L.~Simon, and J.~C. Davis, ``{ZTD}$_{JAVA}$: {Mitigating
  Software Supply Chain Vulnerabilities} via {Zero-Trust Dependencies},'' in
  \emph{International {{Conference}} on {{Software Engineering}}}, 2025.

\bibitem{bifolco2024accuracyGitHubDependencyGraph}
\BIBentryALTinterwordspacing
D.~Bifolco, S.~Nocera, S.~Romano, M.~Di~Penta, R.~Francese, and G.~Scanniello,
  ``On the accuracy of {GitHub}'s dependency graph,'' in \emph{Proceedings of
  the 28th International Conference on Evaluation and Assessment in Software
  Engineering}, ser. EASE '24.\hskip 1em plus 0.5em minus 0.4em\relax New York,
  NY, USA: Association for Computing Machinery, 2024, p. 242–251. [Online].
  Available: \url{https://doi.org/10.1145/3661167.3661175}
\BIBentrySTDinterwordspacing

\bibitem{lariosvargasSelectingThirdpartyLibraries2020}
E.~Larios~Vargas, M.~Aniche, C.~Treude, M.~Bruntink, and G.~Gousios,
  ``Selecting third-party libraries: The practitioners' perspective,'' in
  \emph{Proceedings of the 28th {{ACM Joint Meeting}} on {{European Software
  Engineering Conference}} and {{Symposium}} on the {{Foundations}} of
  {{Software Engineering}}}.\hskip 1em plus 0.5em minus 0.4em\relax Virtual
  Event USA: ACM, Nov. 2020, pp. 245--256.

\bibitem{hermann2015getting}
B.~Hermann, M.~Reif, M.~Eichberg, and M.~Mezini, ``Getting to know you: Towards
  a capability model for {Java},'' in \emph{Proceedings of the 2015 10th Joint
  Meeting on Foundations of Software Engineering}, 2015, pp. 758--769.

\bibitem{zenodo-us}
Anonymous, ``Replication package on {Zenodo},''
  \href{https://zenodo.org/records/15007760?token=eyJhbGciOiJIUzUxMiIsImlhdCI6MTc1MjI1NTg0NiwiZXhwIjoxOTA5OTU4Mzk5fQ.eyJpZCI6ImJjMTQxOTBmLWY1ZWItNDk1Ny1hMmZhLWMwMGRkNWE3OTgwMyIsImRhdGEiOnt9LCJyYW5kb20iOiIzNjgwYTBmYmM3ZDhiY2ZkMDNlYWFkOGI5ODliNmJjYyJ9.rc3gGZeQD_APrYhhTyOufM-DgsgUf2LWY1NIR24sSP1OTuXIlLfYIipEwUFIndroqp8mR4y5OOikLK0B3KuPfA}{Zenodo
  link}. If the link does not work, use the full link in the
  footnote\footnote{\url{https://zenodo.org/records/15007760?token=eyJhbGciOiJIUzUxMiIsImlhdCI6MTc1MjI1NTg0NiwiZXhwIjoxOTA5OTU4Mzk5fQ.eyJpZCI6ImJjMTQxOTBmLWY1ZWItNDk1Ny1hMmZhLWMwMGRkNWE3OTgwMyIsImRhdGEiOnt9LCJyYW5kb20iOiIzNjgwYTBmYmM3ZDhiY2ZkMDNlYWFkOGI5ODliNmJjYyJ9.rc3gGZeQD_APrYhhTyOufM-DgsgUf2LWY1NIR24sSP1OTuXIlLfYIipEwUFIndroqp8mR4y5OOikLK0B3KuPfA}}.,
  2025.

\bibitem{dunlapPairingSecurityAdvisories2024}
T.~Dunlap, J.~S. Meyers, B.~Reaves, and W.~Enck, ``Pairing {{Security
  Advisories}} with {{Vulnerable Functions Using Open-Source LLMs}},'' in
  \emph{Detection of {{Intrusions}} and {{Malware}}, and {{Vulnerability
  Assessment}}}, F.~Maggi, M.~Egele, M.~Payer, and M.~Carminati, Eds.\hskip 1em
  plus 0.5em minus 0.4em\relax Cham: Springer Nature Switzerland, 2024, vol.
  14828, pp. 350--369.

\bibitem{ferreiraContainingMaliciousPackage2021}
G.~Ferreira, L.~Jia, J.~Sunshine, and C.~K{\"a}stner, ``Containing {{Malicious
  Package Updates}} in npm with a {{Lightweight Permission System}},'' in
  \emph{2021 {{IEEE}}/{{ACM}} 43rd {{International Conference}} on {{Software
  Engineering}} ({{ICSE}})}, May 2021, pp. 1334--1346.

\bibitem{zhang2011empirical}
H.~Zhang, M.~A. Babar, X.~Bai, J.~Li, and L.~Huang, ``An empirical assessment
  of a systematic search process for systematic reviews,'' in \emph{15th Annual
  Conference on Evaluation \& Assessment in Software Engineering (EASE
  2011)}.\hskip 1em plus 0.5em minus 0.4em\relax IET, 2011, pp. 56--65.

\bibitem{zhang2011identifying}
H.~Zhang, M.~A. Babar, and P.~Tell, ``Identifying relevant studies in software
  engineering,'' \emph{Information and Software Technology}, vol.~53, no.~6,
  pp. 625--637, 2011.

\bibitem{tran2024good}
H.~K.~V. Tran, J.~B{\"o}rstler, N.~B. Ali, and M.~Unterkalmsteiner, ``How good
  are my search strings? reflections on using an existing review as a
  quasi-gold standard,'' \emph{arXiv preprint arXiv:2402.11041}, 2024.

\bibitem{kitchenham2013systematic}
B.~Kitchenham and P.~Brereton, ``A systematic review of systematic review
  process research in software engineering,'' \emph{Information and software
  technology}, vol.~55, no.~12, pp. 2049--2075, 2013.

\bibitem{maven}
``{Maven} {Central} {Repository},''
  \url{https://mvnrepository.com/open-source}, last accessed: 11-Mar-2026.

\bibitem{campbellCodingIndepthSemistructured2013}
J.~L. Campbell, C.~Quincy, J.~Osserman, and O.~K. Pedersen, ``Coding {{In-depth
  Semistructured Interviews}}: {{Problems}} of {{Unitization}} and {{Intercoder
  Reliability}} and {{Agreement}},'' \emph{Sociological Methods \& Research},
  vol.~42, no.~3, pp. 294--320, Aug. 2013.

\bibitem{owasp}
``{OWASP} {Top} {Ten},'' \url{https://owasp.org/www-project-top-ten/}, last
  accessed: 11-Mar-2026.

\bibitem{mitre_cwe}
{The MITRE Corporation}, ``Common weakness enumeration,''
  \url{https://cwe.mitre.org/index.html}, 2024, last accessed: 11-Mar-2026.

\bibitem{zimmermannCardsortingTextThemes2016}
T.~Zimmermann, ``Card-sorting: {{From}} text to themes,'' in \emph{Perspectives
  on {{Data Science}} for {{Software Engineering}}}, T.~Menzies, L.~Williams,
  and T.~Zimmermann, Eds.\hskip 1em plus 0.5em minus 0.4em\relax Boston: Morgan
  Kaufmann, Jan. 2016, pp. 137--141.

\bibitem{basakWhatChallengesDevelopers2023}
S.~K. Basak, L.~Neil, B.~Reaves, and L.~Williams, ``What {{Challenges Do
  Developers Face About Checked-in Secrets}} in {{Software Artifacts}}?'' in
  \emph{2023 {{IEEE}}/{{ACM}} 45th {{International Conference}} on {{Software
  Engineering}} ({{ICSE}})}, May 2023, pp. 1635--1647.

\bibitem{rahmanWhatQuestionsProgrammers2018}
A.~Rahman, A.~Partho, P.~Morrison, and L.~Williams, ``What questions do
  programmers ask about configuration as code?'' in \emph{Proceedings of the
  4th {{International Workshop}} on {{Rapid Continuous Software Engineering}}},
  ser. {{RCoSE}} '18.\hskip 1em plus 0.5em minus 0.4em\relax New York, NY, USA:
  Association for Computing Machinery, May 2018, pp. 16--22.

\bibitem{acmdigitallib}
{Association for Computing Machinery}, ``{ACM} digital library,''
  \url{https://dl.acm.org/}, 2025, last accessed: 11-Mar-2026.

\bibitem{dunlapVFCFinderPairingSecurity2024}
T.~Dunlap, E.~Lin, W.~Enck, and B.~Reaves, ``{{VFCFinder}}: {{Pairing Security
  Advisories}} and {{Patches}},'' in \emph{Proceedings of the 19th {{ACM Asia
  Conference}} on {{Computer}} and {{Communications Security}}}.\hskip 1em plus
  0.5em minus 0.4em\relax Singapore Singapore: ACM, Jul. 2024, pp. 1128--1142.

\bibitem{gisevInterraterAgreementInterrater2013}
N.~Gisev, J.~S. Bell, and T.~F. Chen, ``Interrater agreement and interrater
  reliability: {{Key}} concepts, approaches, and applications,'' \emph{Research
  in Social and Administrative Pharmacy}, vol.~9, no.~3, pp. 330--338, May
  2013.

\bibitem{krugerSecuringYourCryptoAPI2023}
S.~Kr{\"u}ger, M.~Reif, A.-K. Wickert, S.~Nadi, K.~Ali, E.~Bodden, Y.~Acar,
  M.~Mezini, and S.~Fahl, ``Securing {{Your Crypto-API Usage Through Tool
  Support}} - {{A Usability Study}},'' in \emph{2023 {{IEEE Secure Development
  Conference}} ({{SecDev}})}, Oct. 2023, pp. 14--25.

\bibitem{cwe-246}
{MITRE}, ``{CWE-246},'' \url{https://cwe.mitre.org/data/definitions/246.html},
  last accessed: 11-Mar-2026.

\bibitem{cwe-577}
------, ``{CWE-577},'' \url{https://cwe.mitre.org/data/definitions/577.html},
  last accessed: 11-Mar-2026.

\bibitem{cwe-470}
------, ``{CWE-470},'' \url{https://cwe.mitre.org/data/definitions/470.html},
  last accessed: 11-Mar-2026.

\bibitem{cve-2024-8048}
``{CVE-2024-8048},'' \url{https://www.cve.org/CVERecord?id=CVE-2024-8048},
  2024, last accessed: 11-Mar-2026.

\bibitem{cve-2018-1000613}
``{CVE-2018-1000613},''
  \url{https://www.cve.org/CVERecord?id=CVE-2018-1000613}, last accessed:
  11-Mar-2026.

\bibitem{agresti2016statistics}
\BIBentryALTinterwordspacing
A.~Agresti, C.~Franklin, and B.~Klingenberg, \emph{Statistics: The Art and
  Science of Learning from Data}.\hskip 1em plus 0.5em minus 0.4em\relax
  Pearson Education, 2016. [Online]. Available:
  \url{https://books.google.it/books?id=Vql5CwAAQBAJ}
\BIBentrySTDinterwordspacing

\bibitem{bui2024apr4vul}
Q.-C. Bui, R.~Paramitha, D.-L. Vu, F.~Massacci, and R.~Scandariato,
  ``{APR4Vul}: an empirical study of automatic program repair techniques on
  real-world {Java} vulnerabilities,'' \emph{Empirical software engineering},
  vol.~29, no.~1, p.~18, 2024.

\bibitem{stackoverflow-2021}
StackOverflow, ``{Stack} {Overflow} {Developer} {Survey},''
  \url{https://insights.stackoverflow.com/survey/2021}, 2021, last accessed:
  11-Mar-2026.

\bibitem{deanOptimizationObjectorientedPrograms1995}
J.~Dean, D.~Grove, and C.~Chambers, ``Optimization of object-oriented programs
  using static class hierarchy analysis,'' in \emph{ECOOP'95 ---
  Object-Oriented Programming, 9th European Conference}, ser. LNCS, vol. 952,
  1995, pp. 77--101.

\bibitem{mirEffectTransitivityGranularity2023}
A.~M. Mir, M.~Keshani, and S.~Proksch, ``On the {{Effect}} of {{Transitivity}}
  and {{Granularity}} on {{Vulnerability Propagation}} in the {{Maven
  Ecosystem}},'' in \emph{{{IEEE International Conference}} on {{Software
  Analysis}}, {{Evolution}} and {{Reengineering}} ({{SANER}})}.\hskip 1em plus
  0.5em minus 0.4em\relax IEEE, 2023.

\bibitem{pontaDetectionAssessmentMitigation2020}
S.~E. Ponta, H.~Plate, and A.~Sabetta, ``Detection, assessment and mitigation
  of vulnerabilities in open source dependencies,'' \emph{Empirical Software
  Engineering}, vol.~25, no.~5, pp. 3175--3215, Sep. 2020.

\bibitem{helmUnimocgModularCallgraph2024}
D.~Helm, T.~Roth, S.~Keidel, M.~Reif, and M.~Mezini, ``Unimocg: Modular
  call-graph algorithms for consistent handling of language features,'' in
  \emph{Proceedings of the 33rd ACM SIGSOFT International Symposium on Software
  Testing and Analysis (ISSTA)}, 2024.

\bibitem{baconFastStaticAnalysis1996}
D.~F. Bacon and P.~F. Sweeney, ``Fast static analysis of {C++} virtual function
  calls,'' in \emph{Proceedings of the 11th ACM SIGPLAN Conference on
  Object-Oriented Programming, Systems, Languages, and Applications (OOPSLA)},
  1996, pp. 324--341.

\bibitem{reifCallGraphConstruction2016}
M.~Reif, M.~Eichberg, B.~Hermann, J.~Lerch, and M.~Mezini, ``Call graph
  construction for {Java} libraries,'' in \emph{Proceedings of the 2016 24th
  ACM SIGSOFT International Symposium on Foundations of Software Engineering
  (FSE)}, 2016, pp. 474--486.

\bibitem{depsdev}
``{Open} {Source} {Insights}: Understand your dependencies,''
  \url{https://deps.dev/}, last accessed: 11-Mar-2026.

\bibitem{gerosaShiftingSandsMotivation2021}
M.~Gerosa, I.~Wiese, B.~Trinkenreich, G.~Link, G.~Robles, C.~Treude,
  I.~Steinmacher, and A.~Sarma, ``The {{Shifting Sands}} of {{Motivation}}:
  {{Revisiting What Drives Contributors}} in {{Open Source}},'' in \emph{2021
  {{IEEE}}/{{ACM}} 43rd {{International Conference}} on {{Software
  Engineering}} ({{ICSE}})}, May 2021, pp. 1046--1058.

\bibitem{guerraHowAnnotationsAffect2024}
E.~Guerra, E.~Gomes, J.~Ferreira, I.~Wiese, P.~Lima, M.~Gerosa, and
  P.~Meirelles, ``How do annotations affect {{Java}} code readability?''
  \emph{Empirical Software Engineering}, vol.~29, no.~3, p.~62, May 2024.

\bibitem{mujahidWhereGoNow2023}
S.~Mujahid, D.~E. Costa, R.~Abdalkareem, and E.~Shihab, ``Where to {{Go Now}}?
  {{Finding Alternatives}} for {{Declining Packages}} in the npm
  {{Ecosystem}},'' in \emph{2023 38th {{IEEE}}/{{ACM International Conference}}
  on {{Automated Software Engineering}} ({{ASE}})}, 2023, pp. 1628--1639.

\bibitem{charmazConstructingGroundedTheory2014}
K.~Charmaz, \emph{Constructing {{Grounded Theory}}}.\hskip 1em plus 0.5em minus
  0.4em\relax SAGE Publications Ltd, 2014.

\bibitem{corbinGroundedTheoryPractice1997}
J.~M. Corbin, \emph{Grounded {{Theory}} in {{Practice}}}.\hskip 1em plus 0.5em
  minus 0.4em\relax SAGE Publications, Mar. 1997.

\bibitem{corbinGroundedTheoryResearch1990}
J.~M. Corbin and A.~Strauss, ``Grounded theory research: {{Procedures}},
  canons, and evaluative criteria,'' \emph{Qualitative Sociology}, vol.~13,
  no.~1, pp. 3--21, Mar. 1990.

\bibitem{birksGroundedTheoryPractical2022}
M.~Birks and J.~Mills, \emph{Grounded {{Theory}}: {{A Practical Guide}}}.\hskip
  1em plus 0.5em minus 0.4em\relax SAGE Publications, 2022.

\bibitem{urquhartGroundedTheoryQualitative2022}
C.~Urquhart, \emph{Grounded {{Theory}} for {{Qualitative Research}}: {{A
  Practical Guide}}}.\hskip 1em plus 0.5em minus 0.4em\relax SAGE Publications,
  2022.

\bibitem{mcdonaldReliabilityInterraterReliability2019}
N.~McDonald, S.~Schoenebeck, and A.~Forte, ``Reliability and {{Inter-rater
  Reliability}} in {{Qualitative Research}}: {{Norms}} and {{Guidelines}} for
  {{CSCW}} and {{HCI Practice}},'' \emph{Proc. ACM Hum.-Comput. Interact.},
  vol.~3, no. CSCW, pp. 72:1--72:23, Nov. 2019.

\bibitem{wermke2022committed}
D.~Wermke, N.~W{\"o}hler, J.~H. Klemmer, M.~Fourn{\'e}, Y.~Acar, and S.~Fahl,
  ``Committed to trust: A qualitative study on security \& trust in open source
  software projects,'' in \emph{2022 IEEE symposium on Security and Privacy
  (SP)}.\hskip 1em plus 0.5em minus 0.4em\relax IEEE, 2022, pp. 1880--1896.

\bibitem{wermkeAlwaysContributeBack2023}
D.~Wermke, J.~H. Klemmer, N.~W{\"o}hler, J.~Schm{\"u}ser, H.~S. Ramulu,
  Y.~Acar, and S.~Fahl, ``"{{Always Contribute Back}}": {{A Qualitative Study}}
  on {{Security Challenges}} of the {{Open Source Supply Chain}},'' in
  \emph{2023 {{IEEE Symposium}} on {{Security}} and {{Privacy}} ({{SP}})}, May
  2023, pp. 1545--1560.

\bibitem{http2_rfc}
``{HTTP/2} {RFC},'' \url{https://datatracker.ietf.org/doc/html/rfc9113}, last
  accessed: 11-Mar-2026.

\bibitem{zahanOpenSSFScorecardPath2023}
N.~Zahan, P.~Kanakiya, B.~Hambleton, S.~Shohan, and L.~Williams, ``{{OpenSSF
  Scorecard}}: {{On}} the {{Path Toward Ecosystem-Wide Automated Security
  Metrics}},'' \emph{IEEE Security \& Privacy}, vol.~21, no.~6, pp. 76--88,
  Nov. 2023.

\bibitem{acm-code-of-ethics}
\BIBentryALTinterwordspacing
{ACM Code 2018 Task Force}, ``{ACM} code of ethics and professional conduct.''
  [Online]. Available: \url{https://www.acm.org/code-of-ethics}
\BIBentrySTDinterwordspacing

\bibitem{github_tos}
``{GitHub} {TOS},''
  \url{https://docs.github.com/en/site-policy/acceptable-use-policies/github-acceptable-use-policies},
  2024, last accessed: 11-Mar-2026.

\bibitem{endor-labs}
{Endor Labs}, ``{Endor Labs},'' \url{https://www.endorlabs.com/}, 2025, last
  accessed: 11-Mar-2026.

\bibitem{livshitsDefenseSoundinessManifesto2015}
B.~Livshits, M.~Sridharan, Y.~Smaragdakis, O.~Lhot{\'a}k, J.~N. Amaral,
  B.-Y.~E. Chang, S.~Z. Guyer, U.~P. Khedker, A.~M{\o}ller, and D.~Vardoulakis,
  ``In defense of soundiness: A manifesto,'' \emph{Communications of the ACM},
  vol.~58, no.~2, pp. 44--46, Jan. 2015.

\bibitem{reifJudgeIdentifyingUnderstanding2019}
M.~Reif, F.~K{\"u}bler, M.~Eichberg, D.~Helm, and M.~Mezini, ``Judge:
  Identifying, understanding, and evaluating sources of unsoundness in call
  graphs,'' in \emph{Proceedings of the 28th {{ACM SIGSOFT International
  Symposium}} on {{Software Testing}} and {{Analysis}}}, ser. {{ISSTA}}
  2019.\hskip 1em plus 0.5em minus 0.4em\relax New York, NY, USA: Association
  for Computing Machinery, Jul. 2019, pp. 251--261.

\bibitem{suiRecallStaticCall2020}
L.~Sui, J.~Dietrich, A.~Tahir, and G.~Fourtounis, ``On the recall of static
  call graph construction in practice,'' in \emph{Proceedings of the ACM/IEEE
  42nd International Conference on Software Engineering (ICSE)}, 2020, pp.
  1049--1060.

\bibitem{millerWeFeelWere2023}
C.~Miller, C.~K{\"a}stner, and B.~Vasilescu, ````{{We Feel Like We}}'re
  {{Winging It}}:'' {{A Study}} on {{Navigating Open-Source Dependency
  Abandonment}},'' in \emph{Proceedings of the 31st {{ACM Joint European
  Software Engineering Conference}} and {{Symposium}} on the {{Foundations}} of
  {{Software Engineering}}}.\hskip 1em plus 0.5em minus 0.4em\relax San
  Francisco CA USA: ACM, Nov. 2023, pp. 1281--1293.

\bibitem{rogelbergProfilingActivePassive2003}
S.~G. Rogelberg, J.~M. Conway, M.~E. Sederburg, C.~Spitzm{\"u}ller, S.~Aziz,
  and W.~E. Knight, ``Profiling {{Active}} and {{Passive Nonrespondents}} to an
  {{Organizational Survey}},'' \emph{Journal of Applied Psychology}, vol.~88,
  no.~6, pp. 1104--1114, Dec. 2003.

\bibitem{marcusWhoArePeople2005}
B.~Marcus and A.~Sch{\"u}tz, ``Who {{Are}} the {{People Reluctant}} to
  {{Participate}} in {{Research}}? {{Personality Correlates}} of {{Four
  Different Types}} of {{Nonresponse}} as {{Inferred}} from {{Self-}} and
  {{Observer Ratings}},'' \emph{Journal of Personality}, vol.~73, no.~4, pp.
  959--984, 2005.

\bibitem{kreuterSocialDesirabilityBias2008}
F.~Kreuter, S.~Presser, and R.~Tourangeau, ``Social {{Desirability Bias}} in
  {{CATI}}, {{IVR}}, and {{Web Surveys}}: {{The Effects}} of {{Mode}} and
  {{Question Sensitivity}},'' \emph{Public Opinion Quarterly}, vol.~72, no.~5,
  pp. 847--865, Dec. 2008.

\bibitem{redmilesHowWellMy2019}
E.~M. Redmiles, S.~Kross, and M.~L. Mazurek, ``How {{Well Do My Results
  Generalize}}? {{Comparing Security}} and {{Privacy Survey Results}} from
  {{MTurk}}, {{Web}}, and {{Telephone Samples}},'' in \emph{2019 {{IEEE
  Symposium}} on {{Security}} and {{Privacy}} ({{SP}})}, May 2019, pp.
  1326--1343.

\bibitem{gollaWhatWasThat2018}
M.~Golla, M.~Wei, J.~Hainline, L.~Filipe, M.~D{\"u}rmuth, E.~Redmiles, and
  B.~Ur, ``"{{What}} was that site doing with my {{Facebook}} password?":
  {{Designing Password-Reuse Notifications}},'' in \emph{Proceedings of the
  2018 {{ACM SIGSAC Conference}} on {{Computer}} and {{Communications
  Security}}}.\hskip 1em plus 0.5em minus 0.4em\relax Toronto Canada: ACM, Oct.
  2018, pp. 1549--1566.

\bibitem{bincapz}
Chainguard, ``bincapz,'' \url{https://github.com/chainguard-dev/bincapz}, last
  accessed: 11-Mar-2026.

\bibitem{malcontent}
------, ``Malcontent,'' \url{https://github.com/chainguard-dev/malcontent},
  2025, last accessed: 11-Mar-2026.

\bibitem{capslock}
Google, ``Capslock,'' \url{https://github.com/google/capslock}, 2023, last
  accessed: 11-Mar-2026.

\bibitem{gorlaCheckingAppBehavior2014}
A.~Gorla, I.~Tavecchia, F.~Gross, and A.~Zeller, ``Checking app behavior
  against app descriptions,'' in \emph{Proceedings of the 36th {{International
  Conference}} on {{Software Engineering}}}, ser. {{ICSE}} 2014.\hskip 1em plus
  0.5em minus 0.4em\relax New York, NY, USA: Association for Computing
  Machinery, May 2014, pp. 1025--1035.

\bibitem{log4shell}
A.~Berged, ``What is log4shell? the log4j vulnerability explained (and what to
  do about it),'' \url{https://www.dynatrace.com/news/blog/what-is-log4shell/},
  2023, last accessed: 11-Mar-2026.

\bibitem{equifax-data-breach-article}
``Equifax data breach {FAQ}: {What} happened, who was affected, what was the
  impact?''
  \url{https://www.csoonline.com/article/567833/equifax-data-breach-faq-what-happened-who-was-affected-what-was-the-impact.html},
  2017, last accessed: 11-Mar-2026.

\bibitem{vasilakisPreventingDynamicLibrary2021}
N.~Vasilakis, C.-A. Staicu, G.~Ntousakis, K.~Kallas, B.~Karel, A.~DeHon, and
  M.~Pradel, ``Preventing {{Dynamic Library Compromise}} on {{Node}}.js via
  {{RWX-Based Privilege Reduction}},'' in \emph{Proceedings of the 2021 {{ACM
  SIGSAC Conference}} on {{Computer}} and {{Communications Security}}}.\hskip
  1em plus 0.5em minus 0.4em\relax Virtual Event Republic of Korea: ACM, Nov.
  2021, pp. 1821--1838.

\bibitem{ntousakisDemoDetectingThirdParty2021}
G.~Ntousakis, S.~Ioannidis, and N.~Vasilakis, ``Demo: {{Detecting Third-Party
  Library Problems}} with {{Combined Program Analysis}},'' in \emph{Proceedings
  of the 2021 {{ACM SIGSAC Conference}} on {{Computer}} and {{Communications
  Security}}}, ser. {{CCS}} '21.\hskip 1em plus 0.5em minus 0.4em\relax New
  York, NY, USA: Association for Computing Machinery, Nov. 2021, pp.
  2429--2431.

\bibitem{ohmYouCanRun2023}
M.~Ohm, T.~Pohl, and F.~Boes, ``You {{Can Run But You Can}}'t {{Hide}}:
  {{Runtime Protection Against Malicious Package Updates For Node}}.js,'' arXiv
  preprint arXiv:2305.19760, May 2023.

\bibitem{wyssWolfDoorPreventing2022}
E.~Wyss, A.~Wittman, D.~Davidson, and L.~De~Carli, ``Wolf at the {{Door}}:
  {{Preventing Install-Time Attacks}} in npm with {{Latch}},'' in
  \emph{Proceedings of the 2022 {{ACM}} on {{Asia Conference}} on {{Computer}}
  and {{Communications Security}}}.\hskip 1em plus 0.5em minus 0.4em\relax
  Nagasaki Japan: ACM, May 2022, pp. 1139--1153.

\bibitem{wuUnderstandingThreatsUpstream2023}
Y.~Wu, Z.~Yu, M.~Wen, Q.~Li, D.~Zou, and H.~Jin, ``Understanding the
  {{Threats}} of {{Upstream Vulnerabilities}} to {{Downstream Projects}} in the
  {{Maven Ecosystem}},'' in \emph{International {{Conference}} on {{Software
  Engineering}}}, 2023.

\bibitem{pashchenkoVuln4RealMethodologyCounting2022}
I.~Pashchenko, H.~Plate, S.~E. Ponta, A.~Sabetta, and F.~Massacci,
  ``{{Vuln4Real}}: {{A Methodology}} for {{Counting Actually Vulnerable
  Dependencies}},'' \emph{IEEE Transactions on Software Engineering}, vol.~48,
  no.~5, pp. 1592--1609, May 2022.

\bibitem{zhangMitigatingPersistenceOpenSource2023}
L.~Zhang, C.~Liu, S.~Chen, Z.~Xu, L.~Fan, L.~Zhao, Y.~Zhang, and Y.~Liu,
  ``Mitigating {{Persistence}} of {{Open-Source Vulnerabilities}} in {{Maven
  Ecosystem}},'' in \emph{2023 38th {{IEEE}}/{{ACM International Conference}}
  on {{Automated Software Engineering}} ({{ASE}})}, Sep. 2023, pp. 191--203.

\bibitem{ochoaBreakingBadSemantic2022}
L.~Ochoa, T.~Degueule, J.-R. Falleri, and J.~Vinju, ``Breaking bad?
  {{Semantic}} versioning and impact of breaking changes in {{Maven
  Central}},'' \emph{Empirical Software Engineering}, vol.~27, no.~3, p.~61,
  Mar. 2022.

\bibitem{raemaekersSemanticVersioningBreaking2014}
S.~Raemaekers, A.~{van Deursen}, and J.~Visser, ``Semantic {{Versioning}}
  versus {{Breaking Changes}}: {{A Study}} of the {{Maven Repository}},'' in
  \emph{2014 {{IEEE}} 14th {{International Working Conference}} on {{Source
  Code Analysis}} and {{Manipulation}}}, Sep. 2014, pp. 215--224.

\bibitem{jayasuriyaUnderstandingBreakingChanges2023}
D.~Jayasuriya, V.~Terragni, J.~Dietrich, S.~Ou, and K.~Blincoe, ``Understanding
  {{Breaking Changes}} in the {{Wild}},'' in \emph{Proceedings of the 32nd
  {{ACM SIGSOFT International Symposium}} on {{Software Testing}} and
  {{Analysis}}}.\hskip 1em plus 0.5em minus 0.4em\relax Seattle WA USA: ACM,
  Jul. 2023, pp. 1433--1444.

\bibitem{keshaniFrankensteinFastLightweight2023}
M.~Keshani, G.~Gousios, and S.~Proksch, ``Frankenstein: Fast and lightweight
  call graph generation for software builds,'' \emph{Empirical Software
  Engineering}, vol.~29, no.~1, p.~1, Nov. 2023.

\bibitem{keshaniScalableCallGraph2021}
M.~Keshani, ``Scalable {{Call Graph Constructor}} for {{Maven}},'' in
  \emph{2021 {{IEEE}}/{{ACM}} 43rd {{International Conference}} on {{Software
  Engineering}}: {{Companion Proceedings}} ({{ICSE-Companion}})}, May 2021, pp.
  99--101.

\bibitem{nielsenModularCallGraph2021}
B.~B. Nielsen, M.~T. Torp, and A.~M{\o}ller, ``Modular call graph construction
  for security scanning of {{Node}}.js applications,'' in \emph{Proceedings of
  the 30th {{ACM SIGSOFT International Symposium}} on {{Software Testing}} and
  {{Analysis}}}, ser. {{ISSTA}} 2021.\hskip 1em plus 0.5em minus 0.4em\relax
  New York, NY, USA: Association for Computing Machinery, Jul. 2021, pp.
  29--41.

\bibitem{staicuExtractingTaintSpecifications2020}
C.-A. Staicu, M.~T. Torp, M.~Sch{\"a}fer, A.~M{\o}ller, and M.~Pradel,
  ``Extracting taint specifications for {{JavaScript}} libraries,'' in
  \emph{Proceedings of the {{ACM}}/{{IEEE}} 42nd {{International Conference}}
  on {{Software Engineering}}}, ser. {{ICSE}} '20.\hskip 1em plus 0.5em minus
  0.4em\relax New York, NY, USA: Association for Computing Machinery, Oct.
  2020, pp. 198--209.

\bibitem{muraleeARGUSFrameworkStaged2023}
S.~Muralee, I.~Koishybayev, A.~Nahapetyan, G.~Tystahl, B.~Reaves, A.~Bianchi,
  W.~Enck, A.~Kapravelos, and A.~Machiry, ``{{ARGUS}}: {{A Framework}} for
  {{Staged Static Taint Analysis}} of {{GitHub Workflows}} and {{Actions}},''
  in \emph{{{USENIX}} Security Symposium}, 2023.

\bibitem{shcherbakovSilentSpringPrototype2023}
M.~Shcherbakov, M.~Balliu, and C.-A. Staicu, ``Silent {{Spring}}: {{Prototype
  Pollution Leads}} to {{Remote Code Execution}} in {{Node}}.js,'' in
  \emph{{{USENIX Security}}}, 2023.

\bibitem{schwartzAllYouEver2010}
E.~J. Schwartz, T.~Avgerinos, and D.~Brumley, ``All {{You Ever Wanted}} to
  {{Know}} about {{Dynamic Taint Analysis}} and {{Forward Symbolic Execution}}
  (but {{Might Have Been Afraid}} to {{Ask}}),'' in \emph{2010 {{IEEE
  Symposium}} on {{Security}} and {{Privacy}}}.\hskip 1em plus 0.5em minus
  0.4em\relax Oakland, CA, USA: IEEE, 2010, pp. 317--331.

\bibitem{duanMeasuringSupplyChain2021}
R.~Duan, O.~Alrawi, R.~P. Kasturi, R.~Elder, B.~Saltaformaggio, and W.~Lee,
  ``Towards {{Measuring Supply Chain Attacks}} on {{Package Managers}} for
  {{Interpreted Languages}},'' in \emph{Proceedings 2021 {{Network}} and
  {{Distributed System Security Symposium}}}.\hskip 1em plus 0.5em minus
  0.4em\relax Virtual: Internet Society, 2021.

\bibitem{liMiningNodeJs2022}
S.~Li, M.~Kang, J.~Hou, and Y.~Cao, ``Mining {{Node}}.js {{Vulnerabilities}}
  via {{Object Dependence Graph}} and {{Query}},'' in \emph{31st {{USENIX
  Security Symposium}} ({{USENIX Security}} 22)}, 2022.

\bibitem{vasilakisEfficientModulelevelDynamic2021}
N.~Vasilakis, G.~Ntousakis, V.~Heller, and M.~C. Rinard, ``Efficient
  module-level dynamic analysis for dynamic languages with module
  recontextualization,'' in \emph{Proceedings of the 29th {{ACM Joint Meeting}}
  on {{European Software Engineering Conference}} and {{Symposium}} on the
  {{Foundations}} of {{Software Engineering}}}.\hskip 1em plus 0.5em minus
  0.4em\relax Athens Greece: ACM, Aug. 2021, pp. 1202--1213.

\bibitem{jungDefeatingProgramAnalysis2021}
C.~Jung, D.~Kim, W.~Wang, Y.~Zheng, K.~H. Lee, and Y.~Kwon, ``Defeating
  {{Program Analysis Techniques}} via {{Ambiguous Translation}},'' in
  \emph{2021 36th {{IEEE}}/{{ACM International Conference}} on {{Automated
  Software Engineering}} ({{ASE}})}, Nov. 2021, pp. 1382--1387.

\bibitem{liHitchhikerGuideProgram2023}
H.~Li, Y.~Hao, Y.~Zhai, and Z.~Qian, ``The {{Hitchhiker}}'s {{Guide}} to
  {{Program Analysis}}: {{A Journey}} with {{Large Language Models}},'' arXiv
  preprint arXiv:2308.00245, Jul. 2023.

\bibitem{xuLmPaImprovingDecompilation2023}
X.~Xu, Z.~Zhang, S.~Feng, Y.~Ye, Z.~Su, N.~Jiang, S.~Cheng, L.~Tan, and
  X.~Zhang, ``{{LmPa}}: {{Improving Decompilation}} by {{Synergy}} of {{Large
  Language Model}} and {{Program Analysis}},'' arXiv preprint arXiv:2306.02546,
  Jun. 2023.

\bibitem{mujahidWhatAreCharacteristics2023}
S.~Mujahid, R.~Abdalkareem, and E.~Shihab, ``What are the characteristics of
  highly-selected packages? {{A}} case study on the npm ecosystem,''
  \emph{Journal of Systems and Software}, vol. 198, p. 111588, Apr. 2023.

\bibitem{liExploringFactorsMetrics2022}
X.~Li, S.~Moreschini, Z.~Zhang, and D.~Taibi, ``Exploring factors and metrics
  to select open source software components for integration: {{An}} empirical
  study,'' \emph{Journal of Systems and Software}, vol. 188, p. 111255, Jun.
  2022.

\bibitem{heAutomatingDependencyUpdates2023}
R.~He, H.~He, Y.~Zhang, and M.~Zhou, ``Automating {{Dependency Updates}} in
  {{Practice}}: {{An Exploratory Study}} on {{GitHub Dependabot}},'' \emph{IEEE
  Transactions on Software Engineering}, vol.~49, no.~8, pp. 4004--4022, Aug.
  2023.

\bibitem{pashchenkoQualitativeStudyDependency2020a}
I.~Pashchenko, D.-L. Vu, and F.~Massacci, ``A {{Qualitative Study}} of
  {{Dependency Management}} and {{Its Security Implications}},'' in
  \emph{Proceedings of the 2020 {{ACM SIGSAC Conference}} on {{Computer}} and
  {{Communications Security}}}.\hskip 1em plus 0.5em minus 0.4em\relax Virtual
  Event USA: ACM, Oct. 2020, pp. 1513--1531.

\end{thebibliography}

\clearpage
\appendix
\section{Difference of \Sensapi usage in \texorpdfstring{\rev{the Other}}{the Other} Maven Package Categories}
In this section, we show the rest of the heatmaps from Section~\ref{sec:sen-api-diff}.
Each heatmap shows how many times security-sensitive resources and capabilities are used by functionally similar packages.
\rev{The last heatmap shows the JDBC Driver category, which we report only as an exploratory case (Section~\ref{sec:jdbc-exploratory}).}

\begin{figure}[hbp]
    \centering
    \includegraphics[width=\linewidth]{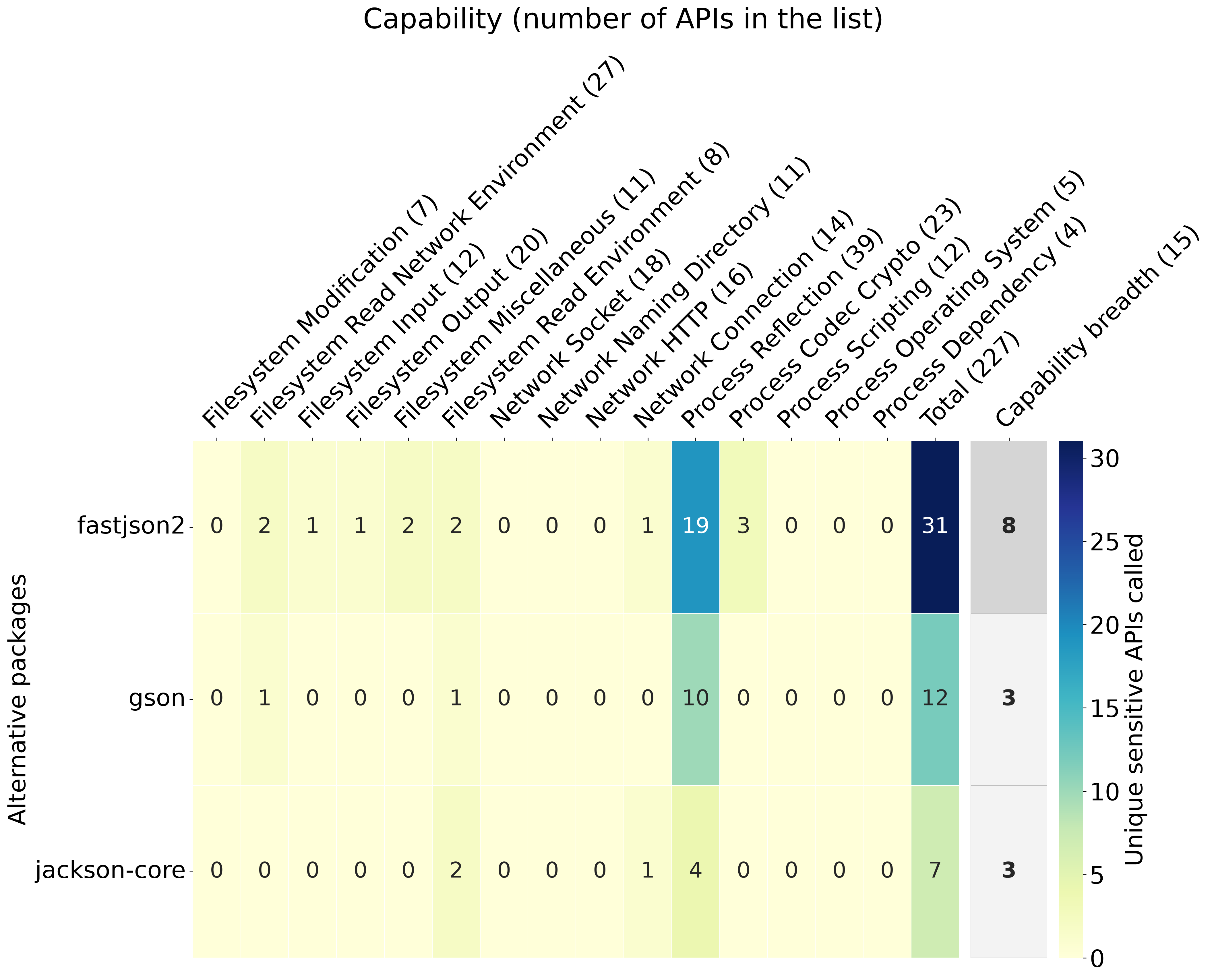}
    \caption{\rev{Security-sensitive API usage in ``JSON Libraries'' package category, at the API level. Each cell is the number of unique \sensapis of a capability that the package calls in its inter-package call graph; the number in parentheses in a column header is the number of APIs of that capability in our list. The last column shows the capability breadth (capability level).}}
    \label{fig:json-libraries-heatmap}
\end{figure}
\begin{figure}[hbp]
    \centering
    \includegraphics[width=\linewidth]{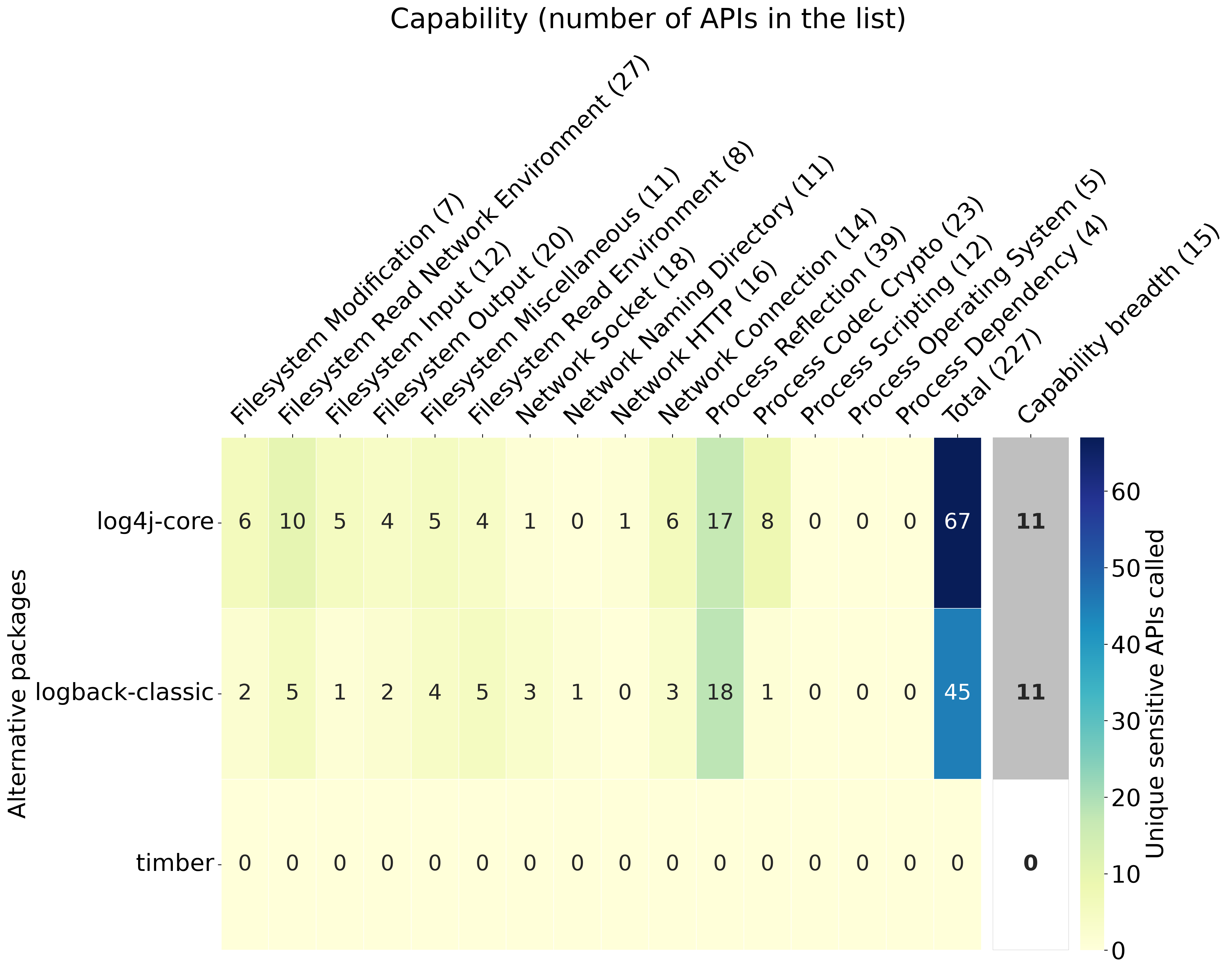}
    \caption{\rev{Security-sensitive API usage in ``Logging Frameworks'' package category, at the API level. Each cell is the number of unique \sensapis of a capability that the package calls in its inter-package call graph; the number in parentheses in a column header is the number of APIs of that capability in our list. The last column shows the capability breadth (capability level).}}
    \label{fig:logging-frameworks-heatmap}
\end{figure}
\begin{figure}[hbp]
    \centering
    \includegraphics[width=\linewidth]{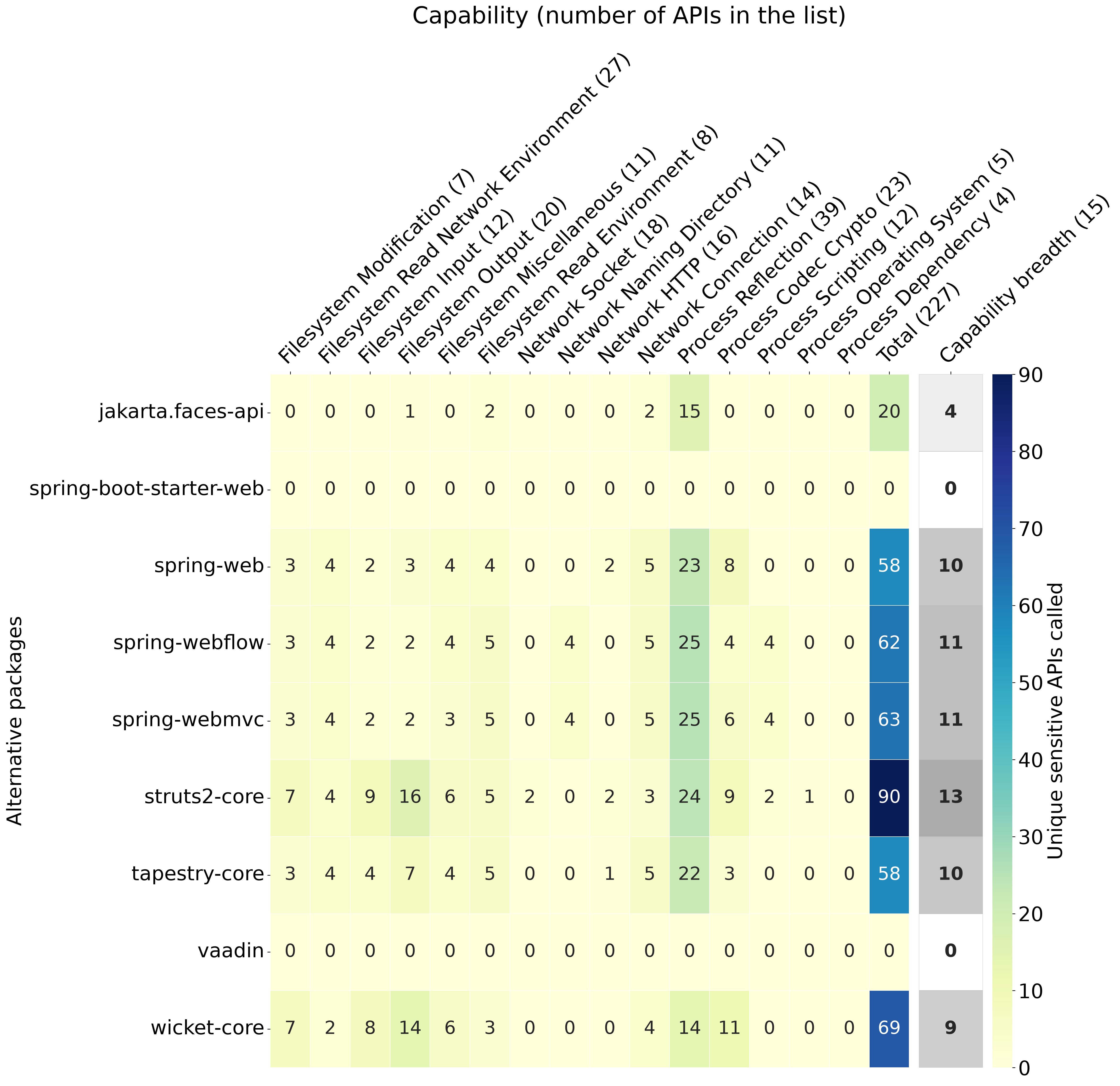}
    \caption{\rev{Security-sensitive API usage in ``Web Frameworks'' package category, at the API level. Each cell is the number of unique \sensapis of a capability that the package calls in its inter-package call graph; the number in parentheses in a column header is the number of APIs of that capability in our list. The last column shows the capability breadth (capability level).}}
    \label{fig:web-frameworks-heatmap}
\end{figure}
\begin{figure}[hbp]
    \centering
    \includegraphics[width=\linewidth]{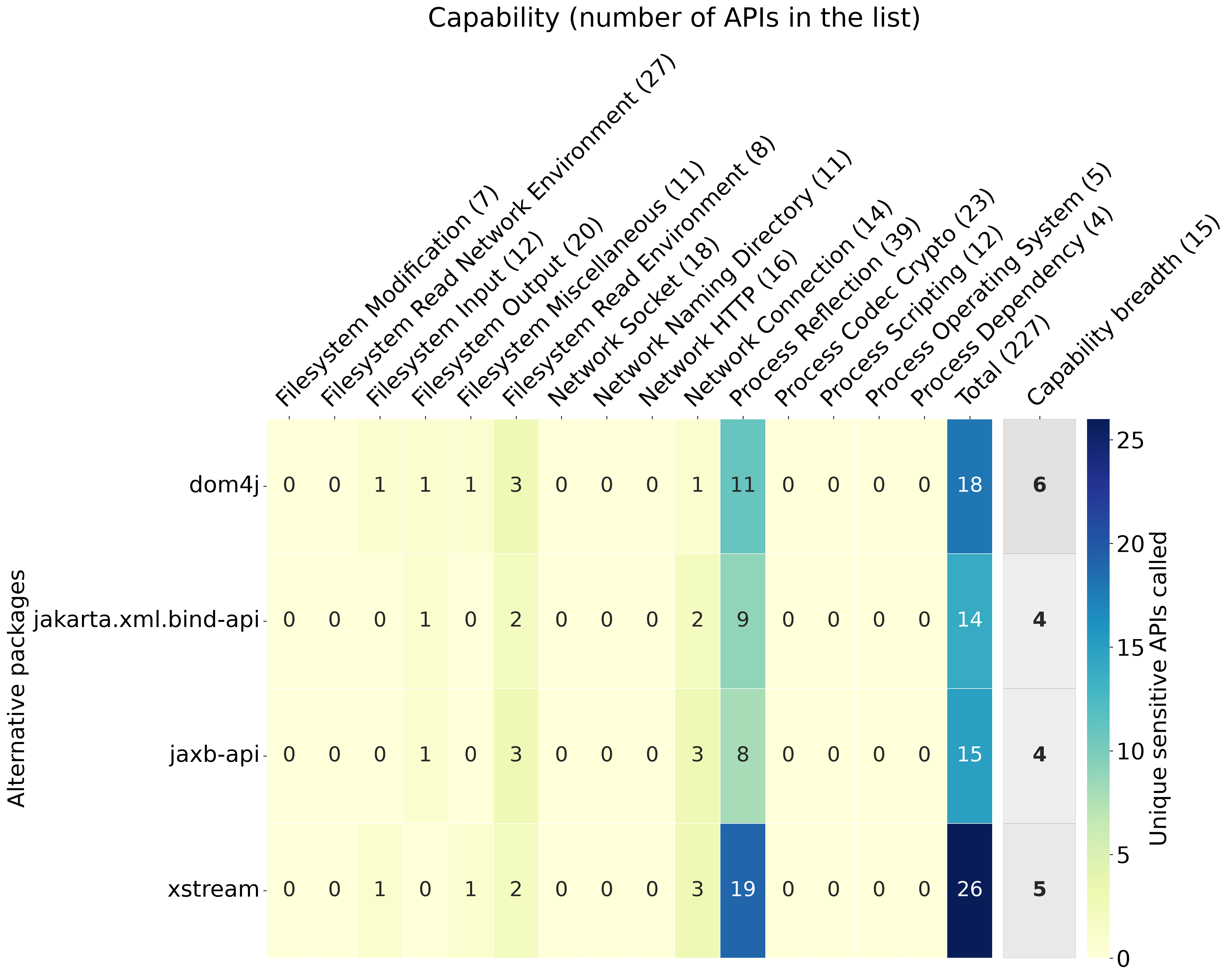}
    \caption{\rev{Security-sensitive API usage in ``XML Processing'' package category, at the API level. Each cell is the number of unique \sensapis of a capability that the package calls in its inter-package call graph; the number in parentheses in a column header is the number of APIs of that capability in our list. The last column shows the capability breadth (capability level).}}
    \label{fig:xml-processing-heatmap}
\end{figure}

\begin{figure}[hbp]
    \centering
    \includegraphics[width=\linewidth]{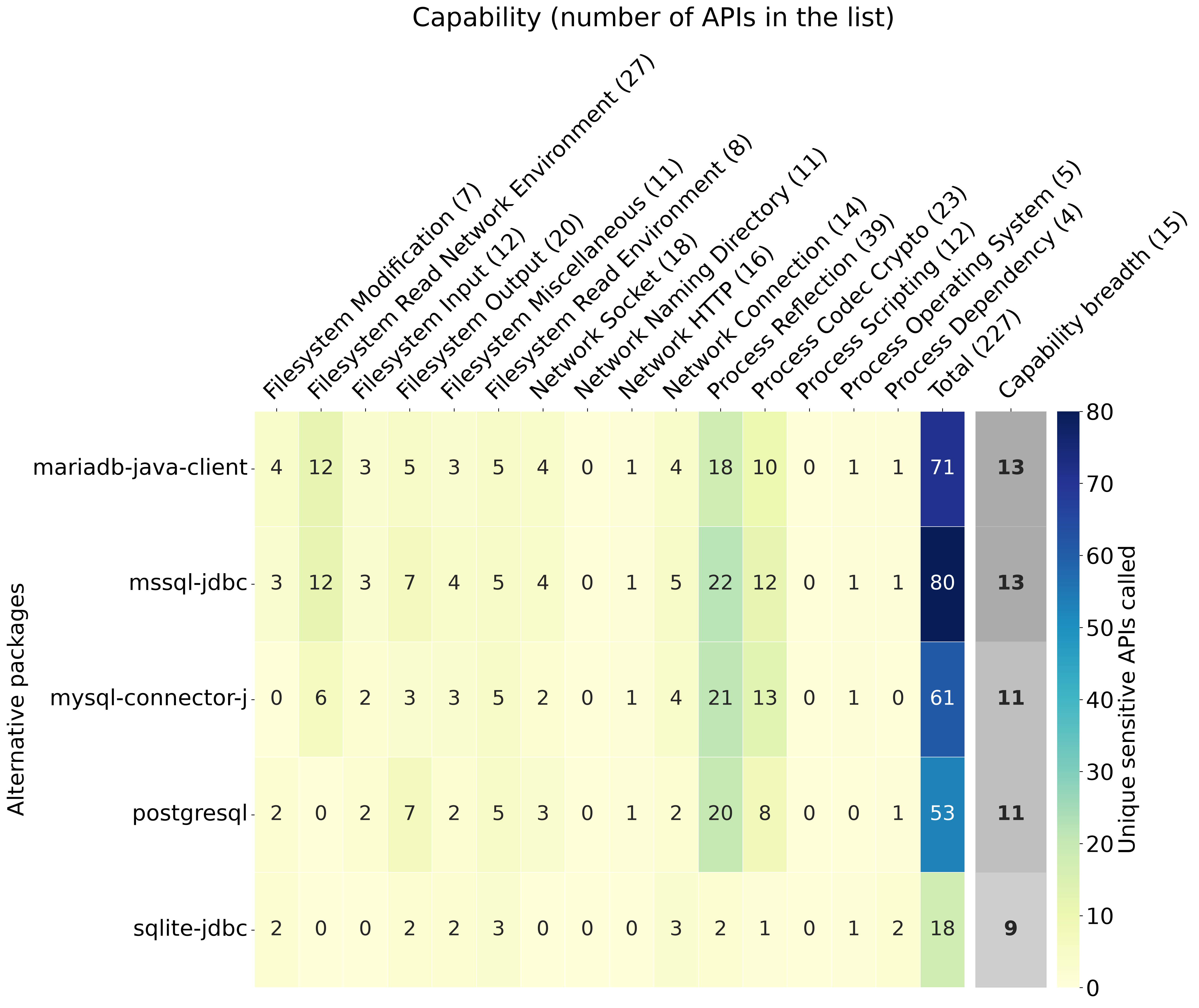}
    \caption{\rev{Security-sensitive API usage in ``JDBC Driver'' package category (exploratory case), at the API level. Each cell is the number of unique \sensapis of a capability that the package calls in its inter-package call graph; the number in parentheses in a column header is the number of APIs of that capability in our list. The last column shows the capability breadth (capability level).}}
    \label{fig:jdbc-driver-heatmap}
\end{figure}

\begin{revblock}
\subsection{Exploratory Case: JDBC Driver}
\label{sec:jdbc-exploratory}
The JDBC Driver category does not meet inclusion criterion 2: the database that a project uses determines its driver, so the packages are not alternatives to one another.
We therefore do not use this category in \rqref{rq:sen-api-diff} and show its heatmap only as an exploratory case (Figure~\ref{fig:jdbc-driver-heatmap}).
At the API level, the five drivers call between 18 (\pkgname{sqlite-jdbc}) and 80 (\pkgname{mssql-jdbc}) unique \sensapis and reach 9 to 13 of the 15 capabilities.
They are the only packages in our sample that call \apicat{process/dependency} (four of the five drivers).
These differences reflect different databases and protocols, not a choice between alternatives that a developer can make.

We keep this exploratory case because a project that uses a database must use the driver for that database. The developer cannot replace the driver, but can still see which sensitive resources it reaches, review those calls, or restrict them with a permission system such as \textsc{ZTD}$_{JAVA}$~\cite{amusuoZTD$_JAVA$MitigatingSoftware2025}.
\end{revblock}

\clearpage
\newpage
\section{Full API list}
\label{sec:full-api-list}

\begin{xltabular}{\linewidth}{lll}
  \caption{Security-sensitive Java API List}\\
  \toprule \textbf{Resource} & \textbf{Capability} & \textbf{APIs}\\ \midrule
  \endfirsthead
  \endhead
  \multicolumn{3}{r}{{Continued on next page}} \\ 
  \endfoot
  \multicolumn{3}{r}{{End}} \\
  \endlastfoot
  \label{tab:full_list}
        & & java.nio.channels.FileChannel.read() \\
        & & java.nio.file.Files.lines() \\
        & & java.nio.file.Files.newBufferedReader() \\
        & & java.nio.file.Files.newInputStream() \\ 
        & & java.nio.file.Files.readAllBytes() \\ 
        & & java.nio.file.Files.readAllLines() \\
        & & java.nio.file.Files.readString() \\
        & & java.nio.file.Files.readSymbolicLink() \\
        & & javax.xml.parsers.DocumentBuilder.parse() \\
        & & javax.xml.parsers.SAXParser.parse() \\
        & & java.util.jar.JarFile.getInputStream() \\
        & \multirow{-13}{*}{\apicat{Input}} & java.util.jar.JarFile.stream()\\
        \hhline{~--}
        & & java.nio.file.Files.find() \\
        & & java.nio.file.Files.getPosixFilePermissions() \\
        & & java.nio.file.Files.list() \\
        & & java.io.File.exists() \\
        & & java.io.File.getAbsolutePath() \\
        & & java.io.File.getCanonicalFile() \\ 
        & & java.io.File.getPath() \\
        & & java.io.File.toPath() \\
        & & java.io.File() \\
        & & java.nio.channels.FileChannel() \\
        & \multirow{-11}{*}{\apicat{Miscellaneous}} & java.nio.file.Paths.get() \\
        \hhline{~--}
        & & java.io.File.delete() \\
        & & java.io.File.deleteOnExit() \\
        & & java.nio.file.Files.copy() \\
        & & java.nio.file.Files.delete() \\
        & & java.nio.file.Files.deleteIfExists() \\
        & & java.nio.file.Files.move() \\
        \apicat{Filesystem} & \multirow{-5}{*}{\apicat{Modification}} & java.nio.file.Files.newByteChannel() \\ \hhline{~--}
        & & java.io.File.createNewFile() \\
        & & java.io.File.createTempFile() \\
        & & java.io.File.mkdir() \\
        & & java.io.FileOutputStream() \\
        & & java.io.FileOutputStream.write() \\
        & & java.io.RandomAccessFile() \\ 
        & & java.io.RandomAccessFile.write() \\
        & & java.nio.file.Files.createFile() \\
        & & java.nio.file.Files.createLink() \\
        & & java.nio.file.Files.createSymbolicLink() \\
        & & java.nio.file.Files.createTempFile() \\
        & & java.nio.file.Files.newBufferedWriter() \\
        & & java.nio.file.Files.newOutputStream() \\
        & & java.nio.file.Files.setLastModifiedTime() \\
        & & java.nio.file.Files.setOwner() \\
        & & java.nio.file.Files.setPosixFilePermissions() \\
        & & java.nio.file.Files.write() \\
         & & java.nio.file.Files.writeString() \\
        & & java.util.logging.Logger.info()\\
        & \multirow{-17}{*}{\apicat{Output}}  & java.util.logging.Logger.log() \\
        \hhline{~--}
        & & java.lang.ProcessBuilder.environment() \\
        & & java.lang.System.getenv() \\
        & & java.lang.System.getProperties() \\
        & & java.lang.System.getProperty() \\
        & & java.lang.System.getSecurityManager() \\
        & & java.util.Properties.getProperty() \\
        & & java.util.Properties.setProperty() \\
        & \multirow{-10}{*}{\apicat{Read\_env}} & java.util.Properties() \\
        \hhline{~--}
        & & java.net.InetAddress.getAllByName() \\
        & & \rev{java.net.InetAddress.getHostAddress()} \\
        & & \rev{java.net.InetAddress.getAddress()} \\
        & & java.net.InetAddress.getHostName() \\
        & & java.net.InetAddress.getLocalHost() \\
        & & java.net.InetAddress.getLoopbackAddress() \\
        & & java.net.InetAddress.isReachable() \\
        & & java.net.InetSocketAddress() \\
        & & java.net.InetSocketAddress.getHostName() \\
        & & java.net.InetSocketAddress.getPort() \\
        & & java.net.NetworkInterface.getInetAddresses() \\
        & & \rev{java.net.Socket.getInetAddress()} \\
        & & java.net.URI() \\
        & & java.net.URI.getAuthority()  \\
        & & java.net.URI.getHost() \\
        & & java.net.URI.getPort()  \\
        & & java.net.URI.getRawSchemeSpecificPart() \\
        & & javax.servlet.http.HttpServletRequest.getContextPath() \\
        & & javax.servlet.http.HttpServletRequest.getLocalAddr() \\
        & & javax.servlet.http.HttpServletRequest.getLocalName() \\
        & & javax.servlet.http.HttpServletRequest.getLocalPort() \\
        & & javax.servlet.http.HttpServletRequest.getRemoteAddr() \\
        & & javax.servlet.http.HttpServletRequest.getRemotePort() \\
        & & javax.servlet.http.HttpServletRequest.getQueryString() \\
        & & javax.servlet.http.HttpServletRequest.getRequestURI() \\
        & & javax.servlet.http.HttpServletRequest.getServletPath() \\
        & \multirow{-27}{*}{\apicat{Read\_network\_env}} &  javax.servlet.http.HttpServletResponse.setStatus() \\
        \midrule \midrule
        & & jakarta.servlet.ServletContext.getRequestDispatcher() \\
        & & java.net.HttpURLConnection() \\
        & & java.net.HttpURLConnection.connect() \\
        & & java.net.HttpURLConnection.getInputStream() \\
        & & java.net.HttpURLConnection.getOutputStream() \\
        & & java.net.URL() \\
        & & java.net.URL.openConnection() \\
        & & java.net.URL.openStream() \\
        & & java.net.URL.getContent() \\
        & & java.net.URLConnection.getInputStream() \\
        & & java.net.URLConnection.getOutputStream() \\
        & & java.sql.DriverManager.getConnection()\\
        & & javax.servlet.http.HttpServletRequest.getRequestDispatcher() \\
        & \multirow{-16}{*}{\apicat{Connection}} & javax.sql.DataSource.getConnection() \\ \hhline{~--}
        & & jakarta.servlet.http.HttpServletRequest.getParameter()\\
        & & jakarta.servlet.http.HttpServletRequest.getRequestDispatcher() \\
        \apicat{Network} & & jakarta.servlet.http.HttpServletRequest.setAttribute() \\
        & & jakarta.servlet.http.HttpServletResponse.getWriter()\\
        & & jakarta.servlet.http.HttpServletResponse.setStatus() \\
        & & java.net.http.HttpClient.Builder.build() \\
        & & java.net.http.HttpClient.newBuilder() \\
        & & java.net.http.HttpClient.newHttpClient() \\
        & & javax.net.ssl.HttpsURLConnection.setHostnameVerifier()\\
        & & javax.security.auth.login.LoginContext.login()\\
        & & \rev{javax.servlet.http.HttpServletResponse.addHeader()} \\ 
        & & javax.servlet.http.HttpServletResponse.getWriter()\\
        & & javax.servlet.http.HttpServletResponse.sendRedirect()\\
        & & javax.servlet.http.HttpServletRequest.getParameter()\\
        & & javax.servlet.http.HttpSession.getAttribute()\\
        & \multirow{-21}{*}{\apicat{Http}} & javax.servlet.http.HttpSession.setAttribute()\\
        \hhline{~--}
        & & javax.naming.Context.bind() \\
        & & javax.naming.Context.list() \\
        & & javax.naming.Context.listBindings() \\
        & & javax.naming.Context.lookup() \\
        & & javax.naming.Context.lookupLink() \\
        & & javax.naming.Context.rebind() \\
        & & javax.naming.Context.rename() \\
        & & javax.naming.Context.unbind() \\
        & & javax.naming.InitialContext() \\
        & & javax.naming.InitialContext.doLookup() \\
        & \multirow{-13}{*}{\apicat{Naming\_directory}} & javax.naming.InitialContext.lookup() \\\hhline{~--}
        & & java.net.DatagramSocket()\\
        & & java.net.DatagramSocket.connect()\\
        & & java.net.http.HttpClient.newWebSocketBuilder()\\
        & & java.net.http.WebSocket.Builder.buildAsync()\\
        & & java.net.http.WebSocket.sendBinary()\\
        & & java.net.http.WebSocket.sendPing()\\
        & & java.net.http.WebSocket.sendPong()\\
        & & java.net.http.WebSocket.sendText()\\
        & & java.net.ServerSocket()\\
        & & java.net.ServerSocket.accept()\\
        & & java.net.Socket()\\
        & & java.net.Socket.connect()\\
        & & java.net.Socket.getInputStream()\\
        & & javax.net.ServerSocketFactory.createServerSocket() \\
        & & javax.net.SocketFactory.createSocket() \\
        & & javax.rmi.ssl.SslRMIClientSocketFactory.createSocket()\\
        & & javax.rmi.ssl.SslRMIServerSocketFactory.createServerSocket()\\
        & \multirow{-22}{*}{\apicat{Socket}} & javax.websocket.WebSocketContainer()\\
         \midrule \midrule
        & & java.net.URLDecoder.decode() \\
        & & java.net.URLEncoder.encode() \\
        & & java.nio.charset.CharsetDecoder.charset() \\
        & & java.nio.charset.CharsetEncoder.charset() \\
        & & java.security.MessageDigest.getInstance() \\
        & & java.security.MessageDigest() \\
        & & java.util.Base64.Decoder.decode() \\
        & & java.util.Base64.Decoder.wrap() \\
        & & java.util.Base64.Encoder.encode() \\
        & & java.util.Base64.Encoder.encodeToString() \\
        & & java.util.Base64.Encoder.wrap() \\
        & & java.util.Base64.getDecoder() \\
        & & java.util.Base64.getEncoder() \\
        & & javax.crypto.Cipher.doFinal() \\
        & & javax.crypto.Cipher.getInstance() \\
        & & javax.crypto.Cipher.init() \\
        & & javax.crypto.Cipher.update() \\
        & & javax.crypto.Cipher.updateAAD() \\
        & & javax.crypto.Cipher.wrap() \\
        & & javax.crypto.Cipher.unwrap() \\
        & & javax.crypto.Cipher.getIV() \\
        & & javax.websocket.Decoder.Binary.decode() \\
        & \multirow{-21}{*}{\apicat{Codec\_crypto}} & javax.websocket.Decoder.Text.decode() \\\hhline{~--}  
        & & java.lang.Runtime.load()\\
        & & java.lang.Runtime.loadLibrary()\\
        & & java.lang.System.load()\\
        & \multirow{-4}{*}{\apicat{Dependency}} & java.lang.System.loadLibrary()\\\hhline{~--} 
        & & java.awt.Desktop.open()\\
        & & java.lang.ProcessBuilder.start()\\
        & & java.lang.ProcessBuilder.startPipeline()\\
        & & java.lang.ProcessBuilder()\\
        \apicat{Process} & \multirow{-4}{*}{\apicat{Operating\_system}}& java.lang.Runtime.exec()\\\hhline{~--} 
        & & java.beans.Introspector.getBeanInfo() \\
        & & java.beans.PropertyDescriptor() \\
        & & java.beans.PropertyDescriptor.getWriteMethod() \\
        & & java.io.ObjectInputStream.readObject() \\
        & & java.lang.Class.forName() \\
        & & java.lang.Class.getClasses() \\
        & & java.lang.Class.getClassLoader() \\
        & & java.lang.Class.getDeclaredClasses() \\
        & & \rev{java.lang.Class.getDeclaringClass()} \\
        & & java.lang.Class.getEnclosingClass() \\
        & & java.lang.Class.getEnclosingMethod() \\
        & & java.lang.Class.getConstructor() \\
        & & java.lang.Class.getConstructors() \\
        & & java.lang.Class.getDeclaredConstructor() \\
        & & java.lang.Class.getDeclaredConstructors() \\
        & & java.lang.Class.getDeclaredMethod() \\
        & & java.lang.Class.getDeclaredMethods() \\
        & & java.lang.Class.getMethod() \\
        & & java.lang.Class.getMethods() \\
        & & java.lang.Class.getSuperclass() \\
        & \apicat{Reflection} & java.lang.Class.newInstance() \\
        & & java.lang.ClassLoader() \\
        & & java.lang.ClassLoader.defineClass() \\
        & & java.lang.ClassLoader.findClass() \\
        & & \rev{java.lang.ClassLoader.findLoadedClass()} \\
        & & \rev{java.lang.ClassLoader.findSystemClass()} \\
        & & java.lang.ClassLoader.getSystemClassLoader() \\
        & & java.lang.ClassLoader.loadClass() \\
        & & java.lang.reflect.AccessibleObject.setAccessible() \\
        & & java.lang.reflect.Constructor.newInstance() \\
        & & java.lang.reflect.Method.getDeclaringClass() \\
        & & java.lang.reflect.Method.invoke() \\
        & & java.net.URLClassLoader() \\
        & & java.net.URLClassLoader.newInstance() \\
        & & java.net.URLClassLoader.findClass() \\
        & & java.rmi.server.RMIClassLoader.getClassLoader() \\
        & & java.rmi.server.RMIClassLoader.loadClass() \\
        & & java.util.ServiceLoader.load() \\
        & & java.util.ServiceLoader.loadInstalled() \\\hhline{~--} 
        & & javax.script.ScriptEngine.eval() \\
        & & javax.script.ScriptEngineFactory.getScriptEngine() \\
        & & javax.script.ScriptEngineManager() \\
        & & javax.script.ScriptEngineManager.getEngineByExtension() \\
        & & javax.script.ScriptEngineManager.getEngineByMimeType() \\
        & & javax.script.ScriptEngineManager.getEngineByName() \\
        & & javax.script.ScriptEngineManager.getEngineFactories() \\
        & & jdk.jshell.JShell.Builder.build() \\
        & & jdk.jshell.JShell.builder() \\
        & & jdk.jshell.JShell.create() \\
        & & jdk.jshell.JShell.eval() \\
        & \multirow{-13}{*}{\apicat{Scripting}} & jdk.nashorn.api.scripting.NashornScriptEngine.eval() \\
         \bottomrule
\end{xltabular}

\clearpage
\newpage
\section{Survey Protocol}
\addcontentsline{toc}{section}{Appendix: Survey Protocol}

\noindent This appendix presents the complete survey instrument used to investigate developers' perception towards security-sensitive API calls in open-source packages, as administered to practitioners via Qualtrics.

\subsection*{Preamble and Definition Block}

\noindent A survey by WSPR lab researchers to investigate the usefulness of knowledge
about security-sensitive API calls in open-source packages.

\medskip
\noindent Thanks for your interest in our survey! Before we start, we would like to
define our definition of security-sensitive API for this survey.

\medskip
\noindent \textbf{Security-sensitive API:} We consider the APIs that directly interact
with the \textit{FILESYSTEM}, the \textit{NETWORK}, or the \textit{PROCESS} as security-sensitive APIs. For example,

\begin{itemize}
  \item \texttt{java.io.FileReader()} is used to take input from a file and is
        categorized into \textit{FILESYSTEM (Input)},
  \item \texttt{java.net.HttpURLConnection.connect()} is used to open a communication
        link and is categorized into the \textit{NETWORK (Http)} category,
  \item \texttt{javax.script.ScriptEngine.eval()} is used to execute a script and is
        categorized into the \textit{PROCESS (Scripting)} category.
\end{itemize}

\noindent We have 15 such categories in our analysis --- 5 in \textit{FILESYSTEM}, 4
in \textit{NETWORK}, and 5 in \textit{PROCESS}.

\subsection*{Block 1: Participant and Project Context}

\medskip
\paragraph{Q4.} When you adopted \textit{[selected package]} in your project \textit{[project name]} as a \textit{[selected category]} library, which security-sensitive API categories did you \textbf{expect} to be called from \textit{[selected package]}? (Hover over to get more information about the category and an example.)

\begin{itemize}
  \item[$\square$] Filesystem (Read Environment)
  \item[$\square$] Filesystem (Read Network Environment)
  \item[$\square$] Filesystem (Input)
  \item[$\square$] Filesystem (Miscellaneous)
  \item[$\square$] Filesystem (Modification)
  \item[$\square$] Filesystem (Output)
  \item[$\square$] Network (Connection)
  \item[$\square$] Network (HTTP)
  \item[$\square$] Network (Naming Directory)
  \item[$\square$] Network (Socket)
  \item[$\square$] Process (Codec Crypto)
  \item[$\square$] Process (Dependency)
  \item[$\square$] Process (Operating System)
  \item[$\square$] Process (Reflection)
  \item[$\square$] Process (Scripting)
\end{itemize}

\subsection*{Block 2: Heatmap Presentation and Expectation Gap}

\noindent Participants were shown a package-category heatmap corresponding to their
selected package category. The heatmap was presented with the following explanatory
text:

\medskip
\begin{quote}
In our analysis, this is the security-sensitive API calls of \textit{[selected
package category]}.

This heatmap shows the relationship between sensitive API categories and the packages
that use them. The columns are sensitive API categories, while the rows are packages
with similar functionality. Each number indicates how many times a package (row) calls
a sensitive API category (column). Darker colors indicate more calls, and lighter
colors indicate fewer calls.
\end{quote}

\noindent The following paragraphs describe the heatmaps shown for each package
category. Each heatmap is a matrix whose rows are alternative packages within the
category and whose columns are the 15 sensitive API categories (Filesystem: Read
Environment, Read Network Environment, Input, Miscellaneous, Modification, Output;
Network: Connection, HTTP, Naming Directory, Socket; Process: Codec Crypto,
Dependency, Operating System, Object Oriented, Scripting), plus a total call count
column. Cell values represent the raw count of calls from that package to that API
category; a rightmost \textit{total} column sums all category counts per package.

\medskip
\paragraph{Q5.} \textit{[Project name]}'s used dependency \textit{[selected package]}
is using these security-sensitive API categories which you did not expect.

\medskip
\noindent Do you think the information of security-sensitive API calls can aid when
comparing/choosing alternative packages for a given functionality?

\begin{itemize}
  \item[$\circ$] Yes
  \item[$\circ$] No
\end{itemize}

\medskip
\paragraph{Q6 (if Yes).} Why do you think the security-sensitive API calls
\textbf{is} useful?

\medskip
\noindent \textit{[Free-text entry]}

\medskip
\paragraph{Q7 (if No).} Why do you think the security-sensitive API calls
\textbf{is not} useful?

\medskip
\noindent \textit{[Free-text entry]}

\subsection*{Block 3: Decision Impact}

\paragraph{Q8.} If you had known this security-sensitive API information before
choosing the dependency, how would it affect your decision?

\begin{itemize}
  \item[$\circ$] I would keep using my chosen dependency in this project
  \item[$\circ$] I would keep using my chosen dependency in this project but I will
                 factor sensitive API calls in future choices
  \item[$\circ$] I would adopt the dependency with the lowest sensitive API usage
  \item[$\circ$] I would adopt the dependency based on its calls to a specific category
  \item[$\circ$] Other, please specify: \underline{\hspace{5cm}}
\end{itemize}

\medskip
\paragraph{Q9.} Please elaborate your choice from the previous question (e.g., why).

\medskip
\noindent \textit{[Free-text entry]}

\subsection*{Block 4: Sensitivity Rating}

\paragraph{Q10.} In your opinion, which of the following API categories should be
considered security-sensitive for any package? (Hover over to get more information
about the category and an example.)

\begin{itemize}
  \item[$\square$] Filesystem (Read Environment)
  \item[$\square$] Filesystem (Read Network Environment)
  \item[$\square$] Filesystem (Input)
  \item[$\square$] Filesystem (Miscellaneous)
  \item[$\square$] Filesystem (Modification)
  \item[$\square$] Filesystem (Output)
  \item[$\square$] Network (Connection)
  \item[$\square$] Network (HTTP)
  \item[$\square$] Network (Naming Directory)
  \item[$\square$] Network (Socket)
  \item[$\square$] Process (Codec Crypto)
  \item[$\square$] Process (Dependency)
  \item[$\square$] Process (Operating System)
  \item[$\square$] Process (Object Oriented)
  \item[$\square$] Process (Scripting)
\end{itemize}

\medskip
\paragraph{Q11.} What degree of sensitivity would you personally assign to these APIs
(1 = least sensitive, 10 = most sensitive)? Hover to get the description and an
example of the security-sensitive API category.

\medskip
\noindent \textit{[Slider scale 1--10 for each of the 15 categories listed in Q10]}

\subsection*{Block 5: Closing}

\paragraph{Q12.} Please provide your email address if you want to enter the draw for
an Amazon gift card. Five participants will receive a \$20 gift card each from a
random draw. (Optional)

\medskip
\noindent \textit{[Free-text entry]}

\medskip
\paragraph{Q13.} Do you want to be notified about the results and the outcome of our
research? (Optional)

\begin{itemize}
  \item[$\circ$] Yes
  \item[$\circ$] No
\end{itemize}

\end{document}